\documentclass[amsmath,preprintnumbers,showpacs,aps,nofootinbib,tightenlines,floatfix,superscriptaddress]{revtex4} 

\usepackage{amsmath}
\usepackage{amssymb}
\usepackage{braket}
\usepackage[utf8]{inputenc}
\usepackage{graphicx}
\usepackage{tej}
\usepackage{todonotes}
\usepackage{esint}

\usepackage{amssymb,latexsym}
\usepackage{slashed,array,mathtools,amsfonts,multirow,hyperref,bm,color,subfigure}

\newcommand{\avg}[1]{\left< #1 \right>} 

\newcommand{\eff}{\text{eff}}
\newcommand{\R}{\mathcal{R}}
\newcommand{\UWTheta}{\widetilde{\Theta}}

\renewcommand{\TODO}[1]{{\color{red} #1}}
\renewcommand{\setminus}[0]{\backslash}

\begin{document}

\title{Phase Unwrapping and One-Dimensional Sign Problems}
\author{William Detmold }  \thanks{wdetmold@mit.edu}
\affiliation{Center for Theoretical Physics, Massachusetts Institute of Technology, Cambridge, MA 02139, USA}
\author{Gurtej Kanwar} \thanks{gurtej@mit.edu}
\affiliation{Center for Theoretical Physics, Massachusetts Institute of Technology, Cambridge, MA 02139, USA}
\author{Michael L. Wagman } \thanks{mlwagman@mit.edu}
\affiliation{Center for Theoretical Physics, Massachusetts Institute of Technology, Cambridge, MA 02139, USA}

\begin{abstract}
  Sign problems in path integrals arise when different field configurations contribute with different signs or phases.
  Phase unwrapping describes a family of signal processing techniques in which phase differences between elements of a time series are integrated to construct noncompact  unwrapped phase differences.
  By combining phase unwrapping with a cumulant expansion, path integrals with sign problems arising from phase fluctuations can be systematically approximated as linear combinations of path integrals without sign problems.
  This work explores phase unwrapping in zero-plus-one-dimensional complex scalar field theory.
  Results with improved signal-to-noise ratios for the spectrum of scalar field theory can be obtained from unwrapped phases, but the size of cumulant expansion truncation errors is found to be undesirably sensitive to the parameters of the phase unwrapping algorithm employed. 
  It is argued that this numerical sensitivity arises from discretization artifacts that become large when phases fluctuate close to singularities of a complex logarithm in the definition of the unwrapped phase.
 \end{abstract}

\preprint{MIT-CTP/5023}
\pacs{11.15.Ha, 
      12.38.Gc 
}

\maketitle

\section{Introduction}

If the properties of quantum states with large baryon number could be calculated directly from the Standard Model, open questions could be answered regarding the boundaries of the periodic table, the composition of neutron stars, and the interpretation of low-energy experimental searches for beyond-the-Standard-Model interactions in nuclear targets.
Many electroweak effects can be calculated accurately in perturbation theory, but at low energies relevant to nuclear systems the effects of the strong nuclear force described by quantum chromodynamics (QCD) can only be accurately computed nonperturbatively.
Lattice Quantum Field Theory (LQFT) provides a method for nonperturbatively calculating path integrals in many QFTs in which ultraviolet divergences have been regularized by replacing spacetime with a discrete lattice of points and infrared divergences have been regularized by restricting spacetime to a finite volume.
Renormalized QFT observables are obtained from the continuum and infinite-volume limits of LQFT results.
Monte Carlo (MC) methods can be used to calculate path integrals representing observables in lattice QCD (LQCD) and other LQFTs by stochastically sampling field configurations from appropriately chosen probability distributions and averaging observables over quantum fluctuations.
If the probability distribution used for  MC  sampling is proportional to the contribution of each field configuration to the thermal partition function, then equilibrium thermodynamic observables can be computed from MC ensemble averages.

Sign problems arise in MC calculations when contributions of different field configurations to path integrals have different signs or more generally when path integral contributions are complex and have different phases.
For example, the partition function for QCD at nonzero baryon chemical potential has a sign problem and cannot be used to define a probability distribution for MC simulations of nonzero baryon density systems.
When a partition function has a sign problem, one can instead MC sample according to a different probability distribution and then attempt to reweight the contribution of each field configuration by the ratio of the desired complex weight to the positive weight used for MC importance sampling.
In reweighting approaches to the baryon chemical potential sign problem, the signal-to-noise (StN) ratio of the reweighting factor, that is the average reweighting factor divided by the square root of its variance, vanishes exponentially quickly as the spacetime volume is taken to infinity~\cite{Gibbs:1986ut,Cohen:2003kd,Cohen:2003ut,Splittorff:2006fu,Splittorff:2006vj,Splittorff:2007ck,deForcrand:2010ys,Alexandru:2014hga}.
Sign problems can arise for particular observables even when the partition function does not have a sign problem if different contributions to the path integrals representing these observables have different phases.
Ensemble averages of the observable may then have small StN ratios analogous to the StN ratios of complex reweighting factors for nonzero baryon density partition functions.

Baryon and nuclear correlation functions have StN ratios that decrease exponentially at a rate predicted by the moment analysis of Parisi~\cite{Parisi:1983ae} and Lepage~\cite{Lepage:1989hd} when the total baryon number integrated over spacetime is increased.
Many LQCD calculations of baryon and nuclear correlation functions rely on a golden window of intermediate source/sink separations where signals are consistent with single-state behavior but this StN problem is not too severe~\cite{Beane:2009kya,Beane:2009gs,Beane:2009py,Beane:2010em,Beane:2011pc,Beane:2014oea,Detmold:2014hla}.
The baryon correlation function StN problem arises from phase fluctuations that lead to sign problems in correlation functions and is absent when phase fluctuations are ignored~\cite{Wagman:2016bam}.
The average values of baryon and nuclear correlation functions decay exponentially faster in the limit of large source/sink separation than their average magnitudes, which demonstrates that exponentially precise cancellations must occur between different contributions to MC ensemble averages and signals an exponentially severe StN problem.
LQCD simulations of mesons and baryons with the quark masses tuned to reproduce experimental observables are being used today to predict increasingly complex observables directly from the Standard Model.
However, LQCD simulations of nuclei are still generally limited to simulations of small (baryon number $B=1-5$) nuclei with quark masses tuned to unphysically heavy values~\cite{Beane:2006mx,Yamazaki:2009ua,Doi:2011gq,Beane:2015yha,Berkowitz:2015eaa,Chang:2015qxa,Doi:2015oha,Detmold:2015daa,Orginos:2015aya,Yamazaki:2015asa,Yamazaki:2015vjn,Parreno:2016fwu,Savage:2016kon,Chang:2017eiq,Doi:2017cfx,Doi:2017zov,Gongyo:2017fjb,Nemura:2017bbw,Shanahan:2017bgi,Tiburzi:2017iux,Wagman:2017tmp,Winter:2017bfs,Francis:2018qch,Iritani:2018zbt}.
Although physical quark mass simulations of small nuclei are becoming computationally feasible, current methods for calculating nuclear correlation functions require computational resources that grow exponentially with baryon number. 
These problems motivate the study of alternative approaches to calculating path integrals with sign problems arising from phase fluctuations.

A zero-plus-one-dimensional ($(0+1)D$)  complex scalar field StN problem arising from phase fluctuations in correlation functions with nonzero $U(1)$ charge is studied in this work as a toy model of the baryon correlation function StN problem in LQCD.\footnote{Other interesting LQCD observables face distinct StN problems.
For instance, isoscalar meson correlation functions are uncharged under $U(1)$ symmetries and possess exponential StN problems but not $U(1)$ phase fluctuations. 
Excited-state energies are extracted from differences of correlation functions with the same quantum numbers and face StN problems arising from the exponentially precise cancellations needed to project out ground-state contributions and leave exponentially faster decaying excited-state contributions.
In  large nuclei, StN problems associated with MeV excitation energies are negligible compared to the phase fluctuation StN problem associated with the multi-GeV rest mass of the nucleus.
LQCD calculations of isoscalar mesons and exotic hadrons conversely face StN problems primarily from sources besides $U(1)$ phase fluctuations, and phase unwrapping is not immediately applicable to these systems.}
Scalar field phase fluctuations are found to qualitatively resemble the LQCD baryon correlation function phase fluctuations described in Ref.~\cite{Wagman:2016bam} and in particular are shown to be wrapped normally distributed and have exponential StN problems in an analytically tractable approximation where magnitude fluctuations are neglected and phase fluctuations are assumed to be small.
Analytically integrating over phase fluctuations using a change of variables similar to the dual lattice variables employed in Ref.~\cite{Endres:2006xu} enables calculations of correlation functions that avoid sign and StN problems in $(0+1)D$ complex scalar field theory, but it remains challenging to extend similar analytic integration methods to more complicated theories such as LQCD~\cite{Chandrasekharan:2008gp,deForcrand:2010ys,Gattringer:2012df,Gattringer:2012ap,Gattringer:2014nxa,Gattringer:2016kco,Giuliani:2017qeo}.
Instead, a new method is explored in this work in which phase differences are ``unwrapped,'' or numerically integrated over a series of spacetime separations.
The resulting unwrapped phases are noncompact random variables rather than circular random variables defined modulo $2\pi$.
Moments of unwrapped phase differences can be calculated from positive-definite path integrals that do not have sign problems and do not generically require computational resources that increase exponentially with increasing $U(1)$ charge.
Correlation functions can be calculated from moments of unwrapped phases using cumulant expansion techniques similar to those of Ref.~\cite{Endres:2011jm}, although beyond leading order in the cumulant expansion sign and StN problems can reemerge from differences of cumulants.\footnote{Cumulant expansions of noncompact ``extensive phases'' have also been applied to sign problems in QCD and other theories at nonzero chemical potential~\cite{Ejiri:2007ga,Nakagawa:2011eu,Ejiri:2012wp,Greensite:2013gya,Garron:2017fta,Bloch:2018yhu}.}
Phase unwrapping in conjunction with this cumulant expansion allows generic complex correlation functions with nonzero $U(1)$ charge to be represented by series of path integrals without sign problems; however, it is shown below that finding a robust numerical implementation of phase unwrapping is challenging even in $(0+1)D$ complex scalar field theory.

The phase unwrapping techniques used here to map series of random phases to series of noncompact random variables are analogous to phase unwrapping techniques used  in signal processing, radar interferometry, x-ray crystallography, magnetic resonance imaging, and other areas of science and engineering~\cite{Judge:1994,Ghiglia:98,Ying:2006,Kitahara:2015}.
The idea of phase unwrapping correlation functions in LQFT was briefly mentioned in Ref.~\cite{Wagman:2016bam} but its numerical implementation faces challenges arising from large phase fluctuations and was not pursued in detail.
This work presents a detailed study of phase unwrapping in $(0+1)D$ complex scalar field theory in order to explore its potential applicability in physically relevant higher-dimensional LQFTs.
The $1D$ phase unwrapping algorithms studied here can be immediately applied to three-momentum-projected correlation functions in LQCD and other higher-dimensional LQFTs; however, $1D$ phase unwrapping algorithms generically suffer from numerical instabilities and do not provide a robust solution to LQFT sign and StN problems.
These numerical instabilities are argued to arise from an accumulation of phase unwrapping ambiguities related to large phase jumps that occur with non-negligible probability even on lattices with very fine levels of discretization in $(0+1)D$ complex scalar field theory.
Multidimensional phase unwrapping algorithms are known to avoid analogous numerical instabilities, and more robust phase unwrapping algorithms might be achieved in future investigations of phase unwrapping in multidimensional LQFTs.

The remainder of this work is organized as follows.
LQFT for $(0+1)D$ complex scalar fields is reviewed in Sec.~\ref{sec:scalarstats}.
After discussing sign and StN problems in free complex scalar field theory in Sec.~\ref{sec:freestats},
analytic integration over phase fluctuations and its effect on sign and StN problems is discussed in Sec.~\ref{sec:dual} and is used to derive
wrapped phase statistics in Sec.~\ref{sec:wrappedstats} that confirm the phase fluctuation origin of the sign and StN problems.
Phase unwrapping is introduced in Sec.~\ref{sec:unwrappedstats}.
A cumulant expansion method for relating wrapped and unwrapped phase distributions is introduced in Sec.~\ref{sec:cumulant}.
Numerical studies comparing 1D phase unwrapping and the cumulant expansion, analytic phase integration, and standard MC methods are presented in Sec.~\ref{sec:unwrap1D}.
After brief remarks on applications to higher-dimensional LQFTs in Sec.~\ref{subsec:cho-results},
conclusions are presented in Sec.~\ref{sec:conclusions}.

\section{Complex scalar field statistics}\label{sec:scalarstats}

\subsection{Sign and signal-to-Noise problems}\label{sec:freestats}

Consider a complex scalar field $\varphi(t)$ defined on a uniform lattice of points $t=0,\dots,L-1$ representing a discretized $(0+1)D$ flat Euclidean spacetime.
Units where the lattice spacing is set to unity are used throughout.
Periodic boundary conditions (PBCs) are imposed on $\varphi$, and $L$ is assumed to be even for simplicity.
The free complex scalar field action is then given by
\begin{equation}
  \begin{split}
    S(\varphi)     &\equiv \sum_{t=0}^{L-1} \left( \varphi^*(t+1) - \varphi^*(t) \right)\left( \varphi(t+1) - \varphi(t) \right) + M^2 |\varphi(t)|^2 \\
    &= \sum_{n=-L/2}^{L/2-1} |\varphi(n)|^2\left( 4 \sin^2\left( \frac{n \pi}{L} \right) + M^2 \right),
  \end{split}\label{eq:Sdef}
\end{equation}
where the $L$ dependence of the action and other quantities below is left implicit, $\varphi(L) \equiv \varphi(0)$ by PBCs,  and the last expression introduces a discrete Fourier transform 
\begin{equation}
  \begin{split}
    \varphi(n) \equiv \frac{1}{\sqrt{L}}\sum_{t=0}^{L-1} \varphi(t)e^{-2\pi i n t /L}, \hspace{20pt} \varphi(t) = \frac{1}{\sqrt{L}}\sum_{n=-L/2}^{L/2-1} \varphi(n)e^{2\pi i n t /L}.
  \end{split}\label{eq:phidft}
\end{equation}
The partition function can be evaluated by Gaussian integration in polar coordinates $\varphi(n) = |\varphi(n)|e^{i\theta(n)}$,
\begin{equation}
  \begin{split}
    Z &\equiv \int \mathcal{D}\varphi\; e^{-S(\varphi)} \equiv \int \prod_{t=0}^{L-1} \left[ \frac{1}{\pi}\; d\text{Re}\varphi(t)\; d\text{Im}\varphi(t) \right] e^{-S(\varphi)} \\
    &= \prod_{n=-L/2}^{L/2-1} \int_{-\pi}^\pi \frac{1}{\pi}d\theta(n) \int_0^\infty |\varphi(n)|d|\varphi(n)| e^{-|\varphi(n)|^2\left( 4 \sin^2\left(  \frac{n\pi}{L} \right) + M^2 \right)} \\ 
    &=  \exp\left\lbrace  - \sum_{n=-L/2}^{L/2-1} \ln\left[ 4 \sin^2\left( \frac{n \pi}{L} \right) +  M^2 \right]  \right\rbrace.
  \end{split}\label{eq:Zdef}
\end{equation}
The scalar field propagator is given by
\begin{equation}
  \begin{split}
    G(t) &\equiv  \avg{\varphi(t)\varphi^*(0)} \equiv \frac{1}{Z}\int \mathcal{D}\varphi\; e^{-S} \varphi(t)\varphi^*(0) \\
    &= \frac{1}{L}\sum_{n=-L/2}^{L/2-1} \frac{e^{2 \pi i n t/L}}{4 \sin^2\left( \frac{n\pi}{L} \right) + M^2}.
  \end{split}\label{eq:Gprop}
\end{equation}
In the limit $L\rightarrow\infty$ the sum can be replaced by an integral that can be evaluated by contour integral methods~\cite{Smit:2002}. 
Corrections to this form are exponentially suppressed in $L$ and can be evaluated analytically if the LQFT action in Eq.~\eqref{eq:Sdef} is replaced by its continuum counterpart.
This provides a spectral representation for the propagator,
\begin{equation}
  \begin{split}
    G(t)  &= Z_{1;0,1} e^{-E t}\left[ 1 + O\left(e^{-E (L-t)}\right) \right], 
  \end{split}\label{eq:Gspec}
\end{equation}
where the $L\rightarrow \infty$ complex scalar propagator pole $E$ and residue $Z_{1;0,1}$ are given by
\begin{equation}
  \begin{split}
    E = 2\; \text{arcsinh}\left( \frac{M}{2} \right), \hspace{20pt} Z_{1;0,1} = \avg{|\varphi|^2} = \frac{1}{M \sqrt{4 + M^2}}.
  \end{split}\label{eq:lattpole}
\end{equation}

The free complex scalar field action Eq.~\eqref{eq:Sdef} has a $U(1)$ symmetry
\begin{equation}
  \begin{split}
    \varphi \rightarrow e^{-i\alpha}\varphi,\hspace{20pt} \varphi^* \rightarrow e^{i\alpha}\varphi^*.
  \end{split}\label{eq:U1def}
\end{equation}
This $U(1)$ symmetry can be used to classify sectors of states in the LQFT Hilbert space that do not mix under (Euclidean) time evolution.
The charge of the vacuum is defined to be $Q=0$.
Field products of the form
\begin{equation}
  \begin{split}
    \mathcal{O}_{Q,2P}(t) \equiv \begin{cases} \varphi(t)^{|Q|} |\varphi(t)|^{2P}, & Q \geq 0 \\  {\varphi^*(t)}^{|Q|} |\varphi(t)|^{2P} , & Q < 0 \end{cases}
  \end{split}\label{eq:Odef}
\end{equation}
transform under $U(1)$ in the charge $Q$ representation,
\begin{equation}
  \begin{split}
    \mathcal{O}_{Q,2P} \rightarrow e^{-iQ \alpha}\mathcal{O}_{Q,2P}.
  \end{split}\label{eq:OU1}
\end{equation}
In operator language, the $\mathcal{O}_{Q,2P}$ for all $Q, P \in \mathbb{Z}$ with $P \geq 0$ form a complete basis of interpolating operators for Hilbert space states connected to the Euclidean vacuum by polynomial functions of field operators.
Two-point correlation functions involving these operators can be evaluated in terms of the scalar propagator by Wick's theorem as
\begin{equation}
  \begin{split}
    G_{Q,2P}(t) &\equiv \avg{ \mathcal{O}_{Q,2P}(t) \mathcal{O}_{Q,2P}^*(0) } \\
    &= P!(|Q|+P)!\sum_{k=0}^P \binom{P}{k} \binom{|Q| + P}{k} G(t)^{|Q| + 2P - 2k} G(0)^{2k}  \\
    &= P!(|Q|+P)!\avg{|\varphi|^2}^{|Q| + 2P} \sum_{k=0}^P \binom{P}{k} \binom{|Q| + P}{k} e^{-(|Q|+2P-2k)Et}\left[ 1 + O\left(e^{-E (L-t)}\right) \right].
  \end{split}\label{eq:GQP}
\end{equation}
Comparing to the general spectral representation guaranteed by unitarity,
\begin{equation}
  \begin{split}
    G_{Q,2P}(t) \equiv \sum_n Z_{n;Q,2P} e^{-E_n t}\left[ 1 + O\left(e^{-E (L-t)}\right) \right],
  \end{split}\label{eq:GQPspec}
\end{equation}
the energies of states produced by $\mathcal{O}_{Q,2P}$ are seen to be $E_n \in \{|Q|E,\  (|Q|+2)E,\ \dots,\ (|Q|+2P)E\}$, and the overlaps $Z_{n;Q,2P}$ of the $n$th energy eigenstate can be determined straightforwardly.
The full spectrum of the theory includes energies $E_m = m E$ for all integers $m\geq 0$.
Each eigenvalue $E_m$ appears in $(m+1)$ different charge sectors $Q = -m,\ -m+2,\ \dots,\ m-2,\ m$, signaling an $(m+1)$-fold degeneracy of energy eigenstates.

Since the action in Eq.~\eqref{eq:Sdef} is real, $e^{-S}$ is a positive-definite function that can be interpreted as a probability distribution
\begin{equation}
  \begin{split}
    \mathcal{P}(\varphi) \equiv \frac{1}{Z} e^{-S(\varphi)}.
  \end{split}\label{eq:CHOprob}
\end{equation}
MC methods can be used to produce stochastic samples of complex scalar fields distributed by Eq.~\eqref{eq:CHOprob} using for example the Metropolis algorithm.\footnote{See Ref.~\cite{Lepage:1998dt} for a pedagogical review of MC methods for the simple harmonic oscillator that can be readily applied to $(0+1)D$ complex scalar field theory.}
After generating a MC ensemble of field configurations $\varphi_i$
with $i = 1,\dots,N$ and $N\gg 1$, the scalar field propagator can be determined by approximating the path integral with the ensemble average
\begin{equation}
  \begin{split}
     \overline{G}(t)  =  \frac{1}{N} \sum_{i=1}^N \varphi_i(t){\varphi_i}^*(0) = G(t)\left[ 1 + O\left( \sqrt{\frac{\text{Var}(G(t))}{N}} \right) \right].
  \end{split}\label{eq:sampleG}
\end{equation}
At fixed $t$ and asymptotically large $N$, convergence of the sample mean to the true average with $1/\sqrt{N}$ scaling is guaranteed (neglecting MC autocorrelations) by the Berry–Esseen theorem since all moments of $\mathcal{P}(\varphi)$ are finite.
The $1/\sqrt{N}$ prefactor of $\text{Var}(G(t))$ describes the size of statistical errors at asymptotically large $N$ according to the central limit theorem but may only provide a rough guide to the error scaling for arbitrary $\mathcal{P}(\varphi)$ and finite $N$.
It is noteworthy that $G(t)$ is real by Eq.~\eqref{eq:GQP}, but individual samples $\varphi_i(t){\varphi_i}^*(0)$ and $\overline{G}(t)$ at finite $N$ are complex.
At large $N$ the real part of $\overline{G}(t)$ converges to $G(t)$ as in Eq.~\eqref{eq:sampleG} and the imaginary part of $\overline{G}(t)$ converges to zero with similar $1/\sqrt{N}$ scaling.
Ensemble average estimators can similarly be constructed for general correlation functions,
\begin{equation}
  \begin{split}
     \overline{G}_{Q,2P}(t)  &= \frac{1}{N}\sum_{i=1}^N \mathcal{O}^i_{Q,2P}(t)\mathcal{O}^i_{-Q,2P}(0) \\
    &= G_{Q,2P}(t)\left[ 1 + O\left( \sqrt{\frac{\text{Var}(G_{Q,2P}(t))}{N}} \right) \right],
  \end{split}\label{eq:sampleGQP}
\end{equation}
where $\mathcal{O}^i_{Q,2P}$ is defined by Eq.~\eqref{eq:Odef} with $\varphi$ replaced by $\varphi_i$ and $1/\sqrt{N}$ error scaling follows from the Berry-Esseen and central limit theorems. 
Ground-state energies can be determined from the large-$t$ behavior of the effective masses derived from ensemble average correlation functions,
\begin{equation}
  \begin{split}
    E_{Q,2P}(t) \equiv - \partial_t \ln \left(\overline{G}_{Q,2P}(t)\right) \equiv - \ln\left( \overline{G}_{Q,2P}(t+1) \right) + \ln\left( \overline{G}_{Q,2P}(t) \right).
  \end{split}\label{eq:EQ2P}
\end{equation}

\begin{figure}
\centering
\includegraphics[width=\textwidth]{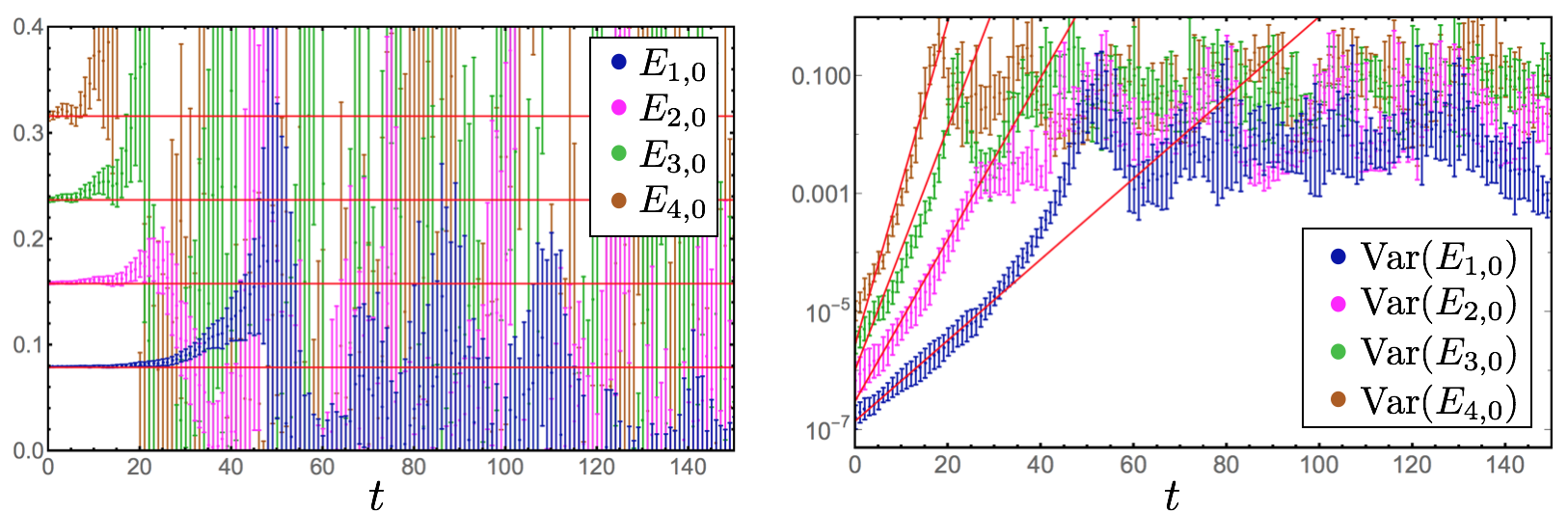}
\caption{ The left plot shows MC ensemble average ground-state energies $E_{Q,0}$ of $U(1)$ charge sectors $Q=1,\dots,4$ in $(0+1)D$ complex scalar field theory. Analytic results valid in the $L\rightarrow \infty$ limit from Eq.~\eqref{eq:lattpole} are shown as red lines. The right plot shows the variance in these ground-state energies. The red lines show the theoretically predicted $e^{2E_{Q,0} t}$ scaling. Error bars denote $68\%$ confidence intervals determined by bootstrap resampling correlation functions calculated from $L$ source points on $N = 5000$ MC field configurations of complex scalar fields with $M^2 = 0.00625$ and $L=512$ generated using the Metropolis algorithm. This ensemble is denoted $C_0$ below, see Sec.~\ref{sec:unwrap1D} and Appendix~\ref{app:MC} for more details.}
\label{fig:freeQ}
\end{figure}

Following standard Parisi-Lepage arguments~\cite{Parisi:1983ae,Lepage:1989hd}, the variance of $G_{Q,2P}$ can be described by a linear combination of correlation functions
\begin{equation}
  \begin{split}
    \text{Var}(\text{Re}[G_{Q,2P}(t)]) &\equiv \frac{1}{4}\avg{ \Big( \mathcal{O}_{Q,2P}(t)\mathcal{O}_{Q,2P}(0)^* + \mathcal{O}_{Q,2P}(t)^*\mathcal{O}_{Q,2P}(0) \Big) ^2 } - G_{Q,2P}(t)^2 \\
    &= \frac{1}{2} G_{0,2|Q| + 4P}(t) + \frac{1}{2}G_{2Q,4P}(t) - G_{Q,2P}(t)^2,
  \end{split}\label{eq:GQPvar}
\end{equation}
where $G_{-Q,2P} = G_{Q,2P}$  has been used following Eq.~\eqref{eq:GQP}.
The variance of $\overline{G}_{Q,2P}$ is related to the variance of $G_{Q,2P}$ by $1/\sqrt{N}$ in the large-$N$ limit by the central limit theorem, giving at large $N$
\begin{equation}
  \begin{split}
    \text{StN}(\text{Re}[\overline{G}_{Q,2P}(t)]) &\equiv \frac{G_{Q,2P}(t)}{\sqrt{\text{Var}(\text{Re}[\overline{G}_{Q,2P}(t)])}} \\
    &= \sqrt{N} \frac{G_{Q,2P}(t)}{\sqrt{ \frac{1}{2} G_{0,2Q + 4P}(t) + \frac{1}{2}G_{2Q,4P}(t) - G_{Q,2P}(t)^2 }} + O(N^0) \\
    &= \sqrt{2 N} e^{-Q E t} \left[ 1 + O\left( e^{-2 E t} \right) + O(N^{-1/2}) \right].
  \end{split}\label{eq:GPQStN}
\end{equation}
Correlation functions describing sectors with $U(1)$ charge $Q\neq 0$ face an exponentially severe StN problem where the exponent is proportional to the $U(1)$ charge of the system.
This result is confirmed numerically in MC results shown in Fig.~\ref{fig:freeQ}.

In LQCD, baryon correlation functions similarly face an exponentially severe StN problem whose exponent is proportional to $U(1)_B$ baryon number charge.
MC studies indicate that this StN problem is related to the sign problem caused by correlation function phase fluctuations~\cite{Wagman:2016bam}.
Analogous features can be seen analytically in free complex scalar field theory correlation functions.
A magnitude-phase decomposition of the complex scalar field
\begin{equation}
  \begin{split}
    \varphi(t) = e^{R(t) + i\theta(t)}, \hspace{20pt} R(t) \equiv \ln|\varphi(t)|, \hspace{20pt} \theta(t) \equiv \text{arg}(\varphi(t)),
  \end{split}\label{eq:phimagphase}
\end{equation}
and an analogous decomposition of the scalar boson propagator and correlation functions
\begin{equation}
  \begin{split}
    G(t) \equiv \avg{C(t)} &= \avg{e^{\mathcal{R}(t) + i \Theta(t)}}, \hspace{20pt} \mathcal{R}(t) \equiv R(t) + R(0), \hspace{20pt} \Theta(t) \equiv \theta(t) - \theta(0),
  \end{split}\label{eq:Gmagphasedef}
\end{equation}
can be inserted into the path integral representation of the  propagator, Eq.~\eqref{eq:Gprop}, to give
\begin{equation}
  \begin{split}
    G(t) = \frac{1}{Z}\int \mathcal{D} \varphi\; e^{-S + \mathcal{R}(t) + i\Theta(t)}.
  \end{split}\label{eq:Gmagphase}
\end{equation}
Fluctuations of the scalar field phase give scalar boson propagators a sign problem. 
The path integrand in Eq.~\eqref{eq:Gmagphase} is not positive-definite and cannot be interpreted as a probability distribution in MC simulations.\footnote{Note that the magnitude of the integrand of Eq.~\eqref{eq:Gmagphase} is positive-definite but not properly normalized as a probability distribution. A suitably normalized positive-definite probability distribution can be found by replacing $\frac{1}{Z}$ by $\frac{1}{\widetilde{Z}}$ where $\widetilde{Z} = \int \mathcal{D} \varphi\; e^{-S + \R}$. The scalar field propagator can be computed from field configurations sampled from the integral $\frac{1}{\widetilde{Z}}\int \mathcal{D} \varphi\; e^{-S + \R}$ by taking the ensemble average of $e^{i\Theta}$ and including an additional factor of $\frac{\widetilde{Z}}{Z}$ that can be computed from propagator magnitudes generated with standard MC sampling of Eq.~\eqref{eq:CHOprob}.}

\begin{figure}
\centering
\includegraphics[width=\textwidth]{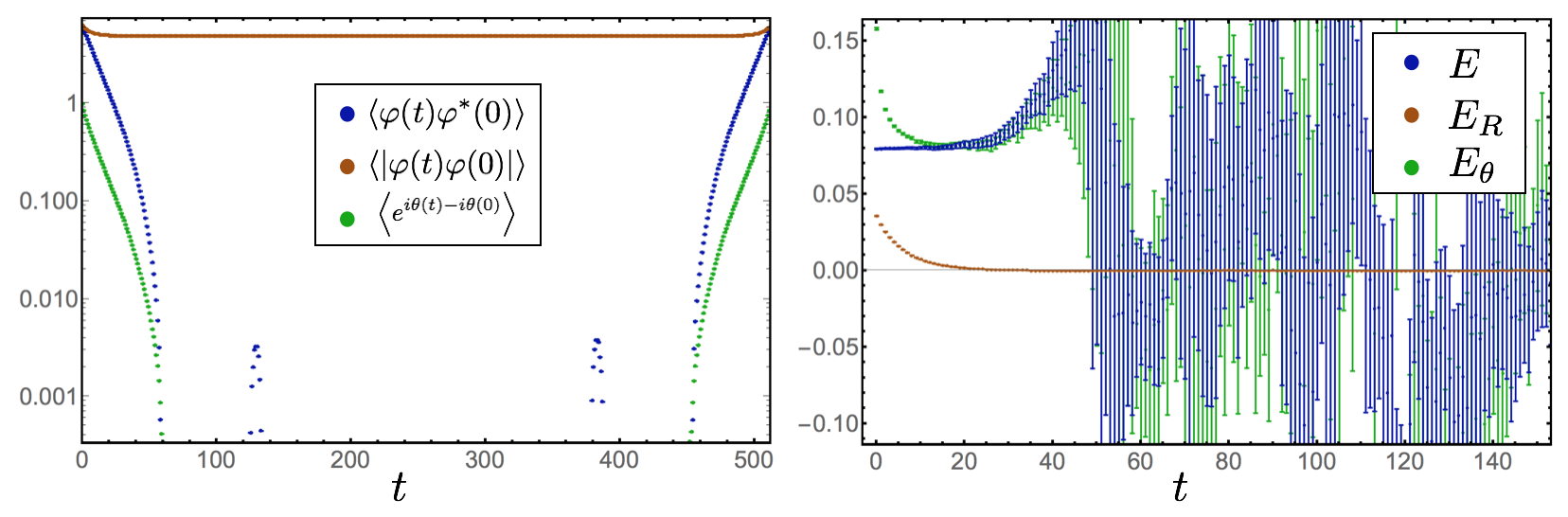}
\caption{The left plot shows a magnitude-phase decomposition of the scalar boson propagator using the same  MC ensemble $C_0$ of free $(0+1)D$ complex scalar field configurations as in Fig.~\ref{fig:freeQ}. Ensemble average calculations of the propagator $\avg{\varphi(t)\varphi^*(0)}$ are shown in blue, calculations of the propagator magnitudes $\avg{|\varphi(t)\varphi(0)|}$ are shown in orange, and calculations of the propagator phase factor $\avg{e^{i\theta(t) - i\theta(0)}}$ are shown in green. The right plot shows the corresponding effective masses $E = -\partial_t \ln \avg{\varphi(t)\varphi^*(0)}$ in blue, $E_R = -\partial_t \ln \avg{|\varphi(t)\varphi(0)|}$ in orange, and $E_\theta = -\partial_t \ln \avg{e^{i\theta(t) - i \theta(0)}}$ in green.}
\label{fig:freeStN}
\end{figure}

Applying a similar decomposition to generic correlation functions gives
\begin{equation}
  \begin{split}
    G_{Q,2P}(t) = \frac{1}{Z}\int \mathcal{D}\varphi\; e^{-S + (2P+|Q|)\R(t) + iQ\Theta(t)}.
  \end{split}\label{eq:GQPmagphase}
\end{equation}
Free complex scalar field correlation functions have a sign problem if they describe states with $U(1)$ charge $Q \neq 0$.
Identical considerations apply to the StN problem in Eq.~\eqref{eq:GPQStN}, demonstrating that sign and StN problems for complex scalar field correlation functions both arise in the presence of nonzero $U(1)$ charge.
The average of the correlation function magnitude,
\begin{equation}
  \begin{split}
    \avg{ |\mathcal{O}_{Q,2P}(t)\mathcal{O}_{Q,2P}^*(0)| } &= \avg{ e^{(|Q|+2P)\R (t)} } = \avg{ |\varphi(t)|^{|Q|+2P}|\varphi(0)|^{|Q|+2P}} ,
  \end{split}\label{eq:magprop}
\end{equation}
depends nonanalytically on the Fourier modes $\varphi(n)$ but can be calculated simply in MC studies of  $(0+1)D$ complex scalar field theory.
As shown in Fig.~\ref{fig:freeStN}, the scalar boson propagator magnitude is $O(1)$ both sample-by-sample and in expectation with no severe StN problem.
Analogous behavior is seen for the magnitudes of generic correlation functions.
The phase of the scalar boson propagator is $O(1)$ sample-by-sample by definition but $O(e^{-Et})$ in expectation with a severe $O(e^{-Et})$ StN problem as shown in Fig.~\ref{fig:freeStN}.
The phase of a general correlation function depends only on the $U(1)$ charge of the correlation function
\begin{equation}
  \begin{split}
    \Theta_Q(t) \equiv \text{arg}\left[ \mathcal{O}_{Q,2P}(t)\mathcal{O}_{Q,2P}^*(0) \right] = iQ \Theta(t),
  \end{split}\label{eq:phaseprop}
\end{equation}
and $\avg{e^{i\Theta_Q}}$ has both an expectation value and a StN problem of $O(e^{-Q E t})$ as shown in Fig.~\ref{fig:freeStN}.

Additional StN degradation is present in calculations of excited-state energies.
Correlation functions $G_{0,2P}$, which include both vacuum and $Q=0$ excited-state contributions, have qualitatively similar behavior to $G_{0,1} = \avg{ |\varphi(t)\varphi(0)| }$ in Fig.~\ref{fig:freeStN} with $O(1)$ signal and root-mean-square variance independent of $t$.
After subtracting vacuum contributions, connected $Q=0$ correlation functions $G^{conn}_{0,2P}$ are given by
\begin{equation}
  \begin{split}
    G^{conn}_{0,2P} &\equiv \avg{|\varphi(t)\varphi(0)|^{2P}} - \avg{|\varphi(t)|^{2P}}\avg{|\varphi(0)|^{2P}} \\
  &=  (P!)^2\avg{|\varphi|^2}^{2P} \sum_{k=1}^P { P \choose P-k } e^{-2k E t} \\
    &\sim e^{-2 E t},
  \end{split}\label{eq:GPconn}
\end{equation}
where $\sim$ denotes proportionality at large $t$.
Connected $Q=0$ correlation functions are $O(e^{-2 E t})$ in expectation but $O(1)$ sample-by-sample and have a StN problem identical to $Q=2$ charged correlation functions.
Eq.~\eqref{eq:GPconn} provides a simple example of a sign problem: expectation values of $O(1)$ random variables must cancel to exponentially increasing precision in order to achieve constant precision in calculations of $G^{conn}_{0,2P}$ at increasing $t$.
Interpreting the vacuum-subtracted correlation functions $G^{conn}_{0,2P}$ as belonging to the $E_n \neq 0$ ``charge sector,'' excited-state spectroscopy in free $(0+1)D$ scalar field theory can be interpreted as possessing a sign problem and StN problem associated with exponentially small differences of averages that is associated with the presence of nonzero energy ``charge.''
This appearance of a sign problem and StN problem for interpolating operators overlapping on to excited-states of the $Q=0$ vacuum sector is analogous to the exponentially worse StN problem faced by excited-state correlation functions than ground-state correlation functions in LQCD when excited-state correlation functions are constructed from nonpositive linear combinations of correlation functions with the same quantum numbers.

\subsection{Dual lattice variable sign problem solution}\label{sec:dual}

Generalizing Eq.~\eqref{eq:Sdef} to include an arbitrary $U(1)$ invariant potential $V(|\varphi|)$ and inserting the magnitude-phase decomposition of Eq.~\eqref{eq:phimagphase} gives
\begin{equation}
  \begin{split}
    S(\varphi) = \sum_{t=0}^{L-1} \left\lbrace 2|\varphi(t)|^2 - \kappa(t) \cos(\Delta(t)) + V(|\varphi(t)|) \right\rbrace
  \end{split}\label{eq:actiondecomp}
\end{equation}
with
\begin{equation}
  \begin{split}
    \kappa(t) \equiv 2 |\varphi(t)||\varphi(t-1)|
    \qquad \text{and} \qquad
    \Delta(t) \equiv \theta(t) - \theta(t-1).
  \end{split}\label{eq:kappaDeltaDefn}
\end{equation}
The partition function can be factored into magnitude and phase contributions as
\begin{equation}
  \begin{split}
    Z &= \int_0^\infty \prod_{t= 0}^{L-1} \left[
      d|\varphi(t)|\; |\varphi(t)| \; e^{-2|\varphi(t)|^2 - V(|\varphi(t)|)} \right]
    \int_{-\pi}^\pi \prod_{t= 0}^{L-1} \left[
      \frac{1}{\pi }d \theta(t) \; e^{\kappa(t) \cos(\Delta(t))} \right].
  \end{split}\label{eq:Zdecomp}
\end{equation}
It is shown in Appendix~\ref{app:dual} that the integral over phase fluctuations in Eq.~\eqref{eq:Zdecomp} can be performed analytically using a change of variables analogous to the dual lattice variables used in Ref.~\cite{Endres:2006xu}.
Analytically integrating over phase fluctuations provides a positive-definite representation for correlation functions, 
\begin{equation} \begin{aligned}
    G_{Q,2P}(t) &= \sum_{q \in \mathbb{Z}} \int \mathcal{D}|\varphi| \mathcal{P}_0(|\varphi|) \;  |\varphi(t)|^{|Q|+2P} |\varphi(0)|^{|Q| + 2P}  \prod_{t^\prime = 1}^t \frac{I_{|Q + q|}(\kappa(t))}{I_0(\kappa(t))}  \prod_{t^\prime = t+1}^L \frac{I_{|q|}(\kappa(t))}{I_0(\kappa(t))},
  \end{aligned}\label{eq:GQPdualrep}
\end{equation}
where the $I_q$ are modified Bessel functions,
$\mathcal{P}_0$ is a probability distribution suitable for MC sampling of scalar
field magnitudes after phase fluctuations have been analytically integrated out,
\begin{equation}
  \begin{split}
    \mathcal{D}|\varphi| \mathcal{P}_0(|\varphi|) = \frac{2}{Z_0}   \prod_{t=0}^{L-1} \left[  |\varphi(t)| d|\varphi(t)| e^{-2|\varphi(t)|^2 - V(|\varphi(t)|)} \;  2\;  I_{0}\left( \kappa(t) \right) \right],
  \end{split}\label{eq:Pprimedef}
\end{equation}
and $Z_0$ is a normalization coefficient defined by $\int \mathcal{D}|\varphi| \mathcal{P}_0(|\varphi|) = 1$.
The explicit sum over scalar field phase winding numbers $q$ is included in Eq.~\eqref{eq:GQPdualrep} in order to avoid ``topological freezing'' arising with stochastic sampling over winding numbers; see Appendix~\ref{app:dual} for details.
Significant contributions to Eq.~\eqref{eq:GQPdualrep} arise for $q = -Q,\dots,+Q$ but topological charge sectors with $|q| > |Q|$ make subdominant contributions that rapidly converge to zero and allow the sum over topological charge sector to be truncated in practical calculations.
Given a finite MC ensemble of scalar field magnitude $|\varphi_i|$, $i=1,\dots,N$ sampled from  Eq.~\eqref{eq:Pprimedef}, correlation functions can be estimated from the corresponding ensemble averages
\begin{equation}\begin{aligned}
    \overline{G}_{Q,2P}^{dual}(t) &= \frac{1}{N} \sum_{i=1}^{N} 
    \sum_{q \in \mathbb{Z}} \Bigg\{
    |\varphi_i(t)|^{|Q|+2P} |\varphi_i(0)|^{|Q| + 2P}  \prod_{t^\prime = 1}^t \frac{I_{|Q + q|}(\kappa(t))}{I_0(\kappa(t))}  \prod_{t^\prime = t+1}^L \frac{I_{|q|}(\kappa(t))}{I_0(\kappa(t))} \Bigg\},
  \end{aligned}\label{eq:GQ2Pdual}
\end{equation}
where $\overline{G}_{Q,2P}^{dual}$ denotes ensemble average calculations of $G_{Q,2P}$ in this dual-variable approach.

\begin{figure}
\centering
\includegraphics[width=\textwidth]{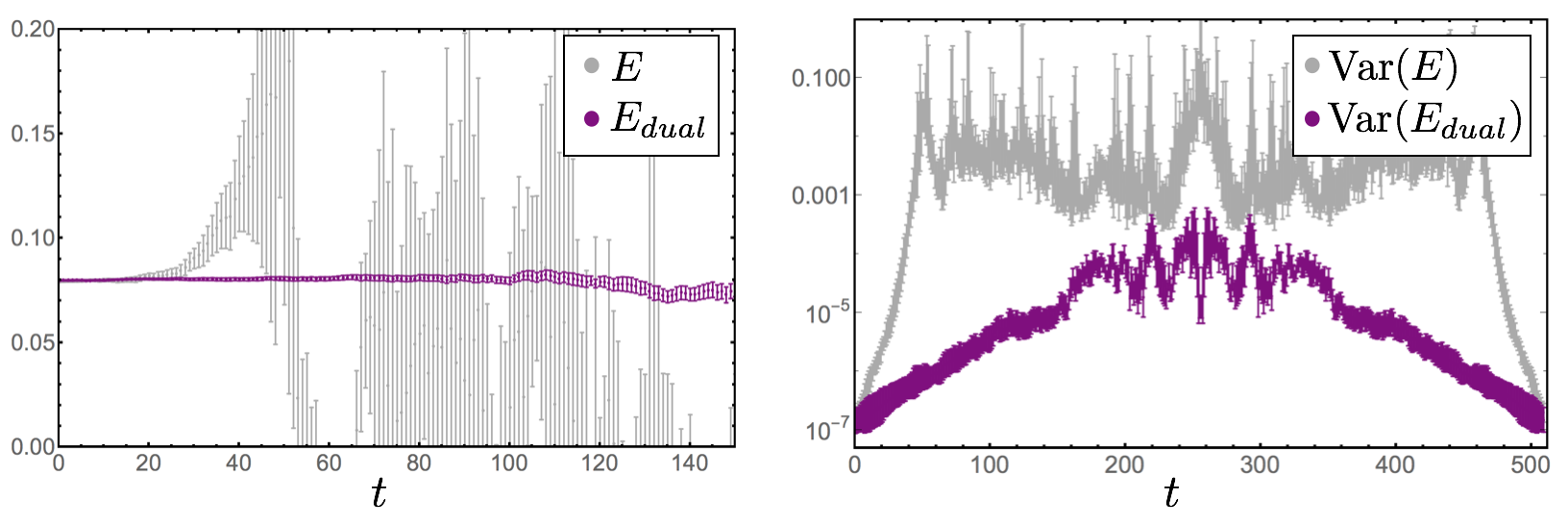}
\caption{ The left plot shows the standard effective mass of the scalar boson propagator $E = -\partial_t \ln \overline{G}_{1,0}$ in gray as well as the dual effective mass $E_{dual} = -\partial_t \ln \overline{G}^{dual}_{1,0}$ in purple
calculated with MC sampling of the dual representation defined in Eq.~\eqref{eq:Pprimedef} - Eq.~\eqref{eq:GQ2Pdual} where phase fluctuations have been integrated out analytically. The right plot shows bootstrap estimates of the variance of the effective mass for the same dual MC estimate of the propagator in purple as well as estimates of the variance of the standard MC propagator shown in Fig.~\ref{fig:freeStN} in gray. For both standard and dual representations, error bars denote $68\%$ confidence intervals determined by bootstrap resampling correlation functions calculated using $5000$ field configurations with $M^2 = 0.00625$ and $L=512$. The same MC ensemble $C_0$ as in Fig.~\ref{fig:freeStN} is used to determine $\overline{G}_{1,0}$.}
\label{fig:dualStN}
\end{figure}

The variance of correlation functions after integrating over dual lattice phase variables is given by
\begin{equation}
  \begin{split}
    \frac{1}{\sqrt{N}}\text{Var}\left(  \overline{G}_{Q,2P}^{dual} \right) &=  -G_{Q,2P}^2 + \sum_{q \in \mathbb{Z}} \int \mathcal{D}|\varphi| \mathcal{P}_0(|\varphi|) \;  |\varphi(t)|^{2|Q|+4P} |\varphi(0)|^{2|Q| + 4P}  \prod_{t^\prime = 1}^t \frac{I_{|Q + q|}(\kappa(t))^2}{I_0(\kappa(t))^2}  \prod_{t^\prime = t+1}^L \frac{I_{|q|}(\kappa(t))^2}{I_0(\kappa(t))^2} \\
    &\sim e^{-2|Q|Et},
  \end{split}\label{eq:Gdualvar}
\end{equation}
where the scaling estimate arises from counting each positive-definite factor of $I_{|Q|}/I_0$ as $O(e^{-|Q|E})$ for $t \ll L$ and $1/\sqrt{N}$ corrections have been neglected.
This suggests that charged scalar correlation functions with analytically integrated dual phase variables avoid both the sign problem for importance sampling correlation functions and the $O(e^{-QEt})$ StN degradation associated with correlation functions from standard MC methods where the phase is stochastically sampled. 
One still expects StN degradation arising from numerically estimating the average product of an increasingly large number of variables as $t$ is increased, but this residual StN problem is not associated with estimating a signal that vanishes in the large-$t$ limit and should therefore be much less severe.\footnote{
Multilevel hierarchical integration~\cite{Luscher:2001up} can be used to exponentially reduce the StN problem associated with sampling products of increasingly many factors in Eqs.~\eqref{eq:CHOZdecomp}-\eqref{eq:CHOpropdecomp} as in Refs.~\cite{DellaMorte:2007zz,DellaMorte:2008jd,DellaMorte:2010yp,Ce:2016idq,Ce:2016ajy}. This approach has been explored, and for instance a two-level hierarchical integration  scheme for calculating correlation functions from Eqs.~\eqref{eq:Pprimedef}-\eqref{eq:GQ2Pdual}  achieves the expected $N_1^{-1}N_0^{-1/2}$ error scaling at moderately large $N_1 \ll N_0$.
}
For numerical verification, the free scalar boson propagator for ensemble $C_0$ is compared to the propagator determined by MC sampling of the dual representation Eq.~\eqref{eq:Pprimedef} with identical parameters $M^2 =0.00625$ and $L=512$ in Fig.~\ref{fig:dualStN}.
The dual representation provides calculations of the propagator and effective mass with slower StN degradation than the standard representation.

The integrand in Eq.~\eqref{eq:Zdecomp} includes products of $I_{|q|}$ functions reminiscent of transfer matrix products appearing in symmetry-projected path integral constructions of Ref.~\cite{DellaMorte:2007zz,DellaMorte:2008jd,DellaMorte:2010yp}. Times between the scalar source and sink are associated with factors of $I_{|Q+q|}$ Bessel functions orthogonal to the $I_{|q|}$ Bessel functions associated with the partition function and with the propagator for times outside the source and sink.
This suggests that integration over the phase projects the transfer matrix to sectors of definite scalar $U(1)$ charge.
Formally, group-theoretical projectors are constructed for path integrals by integration over all elements of a symmetry group.
Integration over decoupled $e^{i\Delta}$ factors associated with each link is equivalent to integration over the group of  local $U(1)$ transformations $\varphi(t) \rightarrow e^{i\alpha(t)}\varphi(t)$, and so dual lattice phase integration acts as a projector to sectors of the Hilbert space with definite $U(1)$ charge.
This projection provides the essential mechanism by which dual lattice phase integration avoids the $U(1)$ charged correlation function sign and StN problem.

\subsection{Wrapped phase statistics}\label{sec:wrappedstats}

The magnitude-phase decomposition of the partition function in Eq.~\eqref{eq:Zdecomp}
shows that for a given scalar field magnitude the phase differences $\Delta(t)$
are independent in the $L\rightarrow \infty$ limit where the PBC constraint $\sum_{t=0}^{L-1} \Delta(t) = 2\pi w$ can be neglected.
The $L\rightarrow \infty$ distribution for $\Delta(t)$ is given from Eq.~\eqref{eq:Zdecomp} in terms of $\kappa(t)$ by
\begin{equation}
  \begin{split}
    \mathcal{P}(\Delta(t)) =  \frac{1}{2\pi I_0(\kappa(t))}\; e^{\kappa(t)\cos(\Delta(t))}.
  \end{split}\label{eq:VM}
\end{equation}
This distribution is known as a von Mises distribution and is well studied in circular statistics~\cite{Fisher:1995,Mardia:2009}.
The resulting probability distribution describing phase differences $\Theta(t) = \theta(t) - \theta(0)$
as sums of independent von Mises random variables can be expressed as
\begin{equation}
  \begin{split}
    \mathcal{P}(\Theta) = \frac{1}{2\pi}\sum_{n\in\mathbb{Z}} e^{-in\Theta} \prod_{t^\prime = 1}^t \left[ \frac{I_{|n|}(\kappa(t^\prime))}{I_0(\kappa(t^\prime))} \right].
  \end{split}\label{eq:PthetaVM}
\end{equation}
It is difficult to calculate many properties of this probability distribution analytically.
The remainder of Sec.~\ref{sec:scalarstats} studies a simpler approximation to Eq.~\eqref{eq:PthetaVM} where StN ratios can be calculated analytically for correlation function estimators using phase unwrapping introduced in Sec.~\ref{sec:unwrappedstats}.

\begin{figure}
\centering
\includegraphics[width=\textwidth]{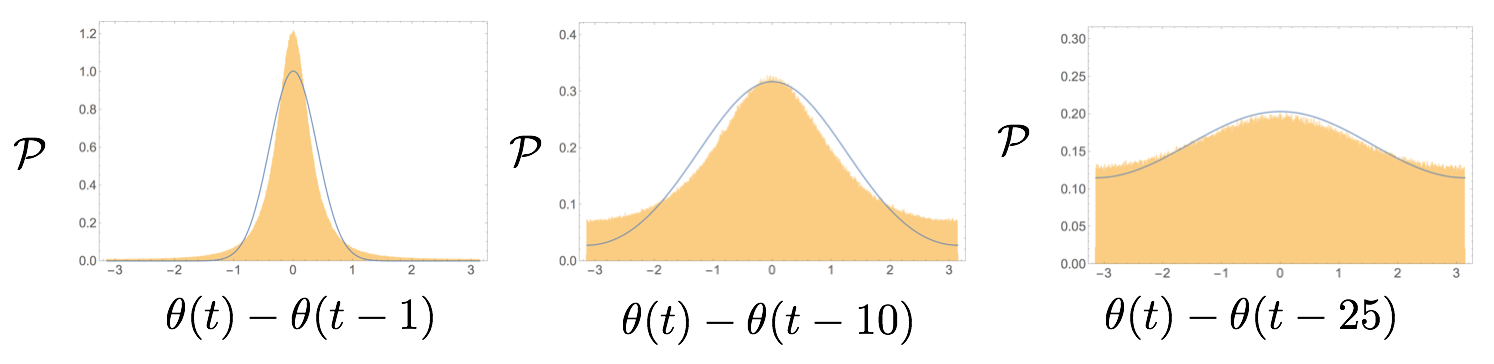}
\caption{ Histograms of differences of scalar field phases separated by 1, 10, and 25 lattice sites from left to right for ensemble $C_0$. These phase difference correspond to the phases of $G_{1,0}(1)$, $G_{1,0}(10)$, and $G_{1,0}(25)$ respectively.  The 2,560,000 samples resulting from $L=512$ differences on each of $N=5000$ field configurations are grouped into 500 bins and normalized so that the histograms represent the empirical probability distribution functions $\mathcal{P}(\theta(t) - \theta(t-1))$, $\mathcal{P}(\theta(t) - \theta(t-10))$ and $\mathcal{P}(\theta(t) - \theta(t-25))$ respectively. The blue curves show the predictions for these distributions from Eq.~\eqref{eq:WNtheta}-Eq.~\eqref{eq:kappa}, which assume that magnitude fluctuations do not affect the phase distribution and are only exact to leading order in Eq.~\eqref{eq:small}.}
\label{fig:phasehist}
\end{figure}

A simpler approximation to Eq.~\eqref{eq:PthetaVM} can be derived under the assumptions
\begin{equation}
  \begin{split}
    \frac{|\varphi(t)||\varphi(t^\prime)| - \avg{|\varphi(t)||\varphi(t^\prime)|}}{\avg{|\varphi(t)||\varphi(t^\prime)|}} \ll 1    
	\qquad \text{and} \qquad
	\Delta(t) \ll 1.
  \end{split}\label{eq:small}
\end{equation}
For fine discretizations with $M^2 \ll 1$, the gradient term provides the dominant contribution to the action and Eq.~\eqref{eq:small} should approximately hold for generic neighborhoods of generic field configurations.
Note however that Eq.~\eqref{eq:small} is not exact in any limit of complex scalar field theory.
The non-trivial consequences of relaxing Eq.~\eqref{eq:small} are explored below by comparing numerical MC results to analytic expressions derived to leading order in Eq.~\eqref{eq:small}, see in particular Figs.~\ref{fig:phasehist} and \ref{fig:phase-jumps}.
Throughout the remainder of this section $\approx$ will be used to denote equality to leading order in the small quantities indicated in Eq.~\eqref{eq:small} and ignoring terms that vanish in the large $t$ limit below Eq~\eqref{eq:kappa}.
In this approximation, phase differences between adjacent lattice sites are identically distributed as well as independent since
\begin{equation}
  \begin{split}
    \kappa(t) \approx \kappa \equiv \frac{1}{L} \sum_{t} \avg{\kappa(t)}.
  \end{split}\label{eq:kappaapprox}
\end{equation}
The assumption $\Delta \ll 1$ can be used to further simplify Eq.~\eqref{eq:VM}.
Expanding the cosine to second order in $\Delta$, restoring invariance under $\Delta \rightarrow \Delta + 2\pi k$ shifts through explicit summation, and adjusting the overall normalization to enforce $\int d\Delta \;\mathcal{P}(\Delta) = 1$ exactly at this order gives
\begin{equation}
  \begin{split}
    \mathcal{P}(\Delta) &\approx \sqrt{\frac{\kappa}{2\pi}}\sum_{k\in\mathbb{Z}}  e^{-\kappa (\Delta + 2\pi k)^2 / 2} \\
    &= \frac{1}{2\pi}\sum_{n\in\mathbb{Z}} e^{i n \Delta-n^2 /(2\kappa)}
  \end{split}\label{eq:WN}
\end{equation}
where the second line can be obtained using Poisson summation.
Eq.~\eqref{eq:WN} defines the wrapped normal probability describing a normally distributed random variable defined modulo $2\pi$.
The wrapped normal and von Mises distributions both approach normal distributions in the limit of small width $\kappa \rightarrow \infty$ and uniform distribution in the limit of large width $\kappa \rightarrow 0$, but the distributions differ at intermediate $\kappa$.

The wrapped normal characteristic function is identical to the normal characteristic function,
\begin{equation}
  \begin{split}
    \Phi_{\mathcal{P}(\Delta)}(n) \equiv \avg{e^{in\Delta}} = \int d\Delta \; e^{in\Delta} \mathcal{P}(\Delta) \approx   e^{-n^2 /2\kappa}.
  \end{split}\label{eq:Deltachar}
\end{equation}
The characteristic function of $\Theta$ can be described as a product of characteristic functions of $\Delta$,
\begin{equation}
  \begin{split}
    \Phi_{\mathcal{P}(\Theta)}(n) &\equiv \avg{ e^{i n \Theta} }
    = \avg{\prod_{t^\prime=1}^t e^{i n \Delta(t^\prime)} }
    = \prod_{t^\prime=1}^t \Phi_{\mathcal{P}(\Delta(t^\prime))}(n).
  \end{split}\label{eq:thetachar}
\end{equation}
The probability distribution of $\Theta$ is given by a Fourier transform of this characteristic function,
\begin{equation}
  \begin{split}
    \mathcal{P}(\Theta) &= \frac{1}{2\pi}\sum_{n\in\mathbb{Z}} e^{-in\Theta} \Phi_{\mathcal{P}(\Theta)}(n)\\
    &= \frac{1}{2\pi} \sum_{n\in\mathbb{Z}}  e^{-in \Theta} \prod_{t^\prime =1}^t \Phi_{\mathcal{P}(\Delta(t^\prime))}(n) \\
  &\approx \frac{1}{2\pi}\sum_{n\in\mathbb{Z}} e^{-in \Theta} e^{ - t n^2 /2\kappa}.
  \end{split}\label{eq:WNtheta}
\end{equation}
Under the assumption of Eq.~\eqref{eq:small}, the scalar boson propagator is given by
\begin{equation}
  \begin{split}
    G(t) &\approx \avg{|\varphi(t)\varphi(0)|} \avg{e^{i\Theta}} = \avg{|\varphi(t)\varphi(0)|} \Phi_{\mathcal{P}(\Theta)}(1) \\
    &\approx \avg{|\varphi(t)\varphi(0)|}  e^{-t/(2\kappa)}.
  \end{split}\label{eq:propapprox}
\end{equation}
Comparing this to the large-time spectral representation Eq.~\eqref{eq:Gspec}, the correct ground-state energy and overlap factor are reproduced if
\begin{equation}
  \begin{split}
    \kappa \approx \frac{1}{2E}, \hspace{20pt} \avg{|\varphi(t)\varphi(0)|} \approx  Z_{1;0,1},
  \end{split}\label{eq:kappa}
\end{equation}
where $t$ is assumed to be large. 
The expectation value of the ensemble average correlation function can be calculated in this approximation as
\begin{equation}
  \begin{split}
    \avg{\overline{G}}  &= \frac{1}{N}\sum_{i=1}^N \avg{|\varphi_i(t)||\varphi_i(0)|\cos(\Theta_i) }\\
    &\approx Z_{1;0,1} \frac{1}{N}\sum_{i=1}^N \avg{\cos(\Theta_i) } \\
    &\approx Z_{1;0,1} e^{-E t}.
  \end{split}\label{eq:Gbardef}
\end{equation}
Its variance is given by
\begin{equation}
  \begin{split}
    \text{Var}(\overline{G})  &\approx Z_{1;0,1}^2 \left\lbrace \avg{\left( \frac{1}{N}\sum_i \cos(\Theta_i) \right)^2} - \avg{\left( \frac{1}{N}\sum_i \cos(\Theta_i) \right)}^2 \right\rbrace \\
    &= \frac{Z_{1;0,1}^2}{2N}\left(1 + \avg{\cos(2 \Theta_i)} - \avg{\cos(\Theta_i)}^2 \right) \\
    &\approx \frac{Z_{1;0,1}^2}{2N}\left(1 - e^{-2Et} \right),
  \end{split}\label{eq:GbarVar}
\end{equation}
and its StN ratio is
\begin{equation}
  \begin{split}
    \text{StN}(\overline{G}) &= \frac{\avg{\overline{G}}}{\sqrt{\text{Var}(\overline{G})}} \approx \sqrt{2N}  \frac{e^{-Et}}{\sqrt{1 - e^{-2Et}}}.
  \end{split}\label{eq:GbarStN}
\end{equation}
It is noteworthy that the full StN problem for the scalar propagator arises at leading order in Eq.~\eqref{eq:small} where magnitude fluctuations are neglected and phase differences are wrapped normally distributed.
Determination of the  scalar propagator pole mass from MC sampling phases distributed according to Eq.~\eqref{eq:WNtheta} is equivalent to parameter inference for a wrapped normal distribution with variance $1/\kappa \approx 2E$.
Avoiding large finite sample size errors in wrapped normal parameter inference requires~\cite{Fisher:1995}
\begin{equation}
  \begin{split}
    \frac{1}{\sqrt{N}} \lesssim \avg{\cos(\Theta)} \approx e^{-E t}
  \end{split}\label{eq:circstatsbound}
\end{equation}
indicating the window of time in which reliable parameter inference is possible has size scaling only as $\log{N}$.

As shown in Fig.~\ref{fig:phasehist}, Eq.~\eqref{eq:WNtheta} roughly captures the $t$ dependence of phase difference distributions for MC ensemble $C_0$ but does not provide a precise fit to MC results.
The empirical distribution of $\Delta$ is better described by a heavy-tailed wrapped stable distribution than by a wrapped normal distribution.
Similar heavy-tailed phase derivatives were seen to arise for baryon correlation functions in Ref.~\cite{Wagman:2016bam}, where it was conjectured that these heavy tails arose from non-perturbative strong interaction physics.\footnote{The real parts of baryon correlation functions are also heavy-tailed, as pointed out in Ref.~\cite{davidkaplanLuschertalk}.}
The appearance of heavy tails in free scalar field theory suggests that they have a generic origin.
The von Mises distribution describing phase derivatives for a fixed field magnitude does not have heavy tails, and so the heavy tails present in MC distributions of phase derivatives must arise from integration over magnitude fluctuations.
Large phase jumps leading to deviations from Eq.~\eqref{eq:WNtheta} and their relation to magnitude fluctuations are discussed further below.
It is also noteworthy that correlation functions for higher charge sector can be computed under the assumptions of Eq.~\eqref{eq:small} by Eq.~\eqref{eq:WNtheta} as
\begin{equation}
  \begin{split}
    G_{Q,2P} \approx \avg{|\varphi(t)\varphi(0)|}^{|Q|+2P}\avg{e^{iQ\Theta}} \approx Z_{1;0,1}^{|Q|+2P} e^{-t Q^2 /(2\kappa)},
  \end{split}\label{eq:GQ2Papprox}
\end{equation}
which does not reproduce the linear spectrum of free-field theory in Eq.~\eqref{eq:GQP}.
These deficiencies are addressed in Sec.~\ref{sec:unwrap1D} with numerical MC studies not relying on the assumptions of Eq.~\eqref{eq:small}.

\subsection{Unwrapped phase statistics}\label{sec:unwrappedstats}

The results of the last section demonstrate that exponential StN degradation appears in calculations of the average cosines of wrapped normal phase differences.
This suggests that to avoid sign and StN problems, one needs to avoid numerical sampling of circular random variables.
For $(0+1)D$ complex scalar field theory phase fluctuations can be integrated analytically and the resulting path integrals describing charged scalar correlation functions have positive-definite weights and a less severe StN problem.
Similar methods can be applied to more complicated scalar field theories in more dimensions~\cite{Endres:2006xu} and also have a long history of application to lattice gauge theory~\cite{Ukawa:1979yv}.
The search for a more general dual representation of LQCD where properties of finite-density matter can be computed with path integrals with positive-definite weights is an active area of ongoing research, see for instance Refs.~\cite{Chandrasekharan:2008gp,deForcrand:2010ys,Gattringer:2012df,Gattringer:2012ap,Gattringer:2014nxa,Gattringer:2016kco,Giuliani:2017qeo}.
It is also possible to look for path integral deformations or changes of variables that reduce the severity of phase fluctuations.
Methods based on Lefschetz thimbles and more general classes of complex path integral deformations have successfully transformed path integrals with phase fluctuations in a variety of LQFTs into (sums of) path integrals where the phase is exactly fixed or at least fluctuating less severely than in the original theory~\cite{Witten:2010cx,Cristoforetti:2012su,Aarts:2013fpa,Fujii:2013sra,Cristoforetti:2013wha,Tanizaki:2014tua,Fujii:2015bua,Alexandru:2015sua,Alexandru:2015xva,Tanizaki:2016xcu,Alexandru:2016san,Alexandru:2016gsd,Mori:2017pne,Alexandru:2018fqp}.
Still, it is an open challenge to find an efficient representation for computing finite-density observables or multibaryon correlation functions in LQCD that avoids sign and StN problems.

The problems inherent to numerical sampling of circular random variables could be avoided if one could instead numerically sample a noncompact real random variable.
Intuitively, one may imagine stochastically sampling a real random variable representing the angular displacement accumulated by the scalar field phase in the interval $[0,t]$ including any $2\pi$ revolutions around the unit circle.
Other works have explored accessing distributions of an
analogous ``extensive phase''
in the context of QCD and other theories at nonzero chemical potential~\cite{Ejiri:2007ga,Nakagawa:2011eu,Ejiri:2012wp,Greensite:2013gya,Garron:2017fta,Bloch:2018yhu}.
This alternative has also been explored in other areas of science and engineering where circular random variables appear.
A variety of ``phase unwrapping'' techniques have been developed to extract noncompact variables representing angular displacement from numerical samples of compact phases, see Refs.~\cite{Judge:1994,Ghiglia:98,Ying:2006,Kitahara:2015} for reviews.
An unwrapped phase difference describes a difference between phases at opposite ends of a parametrized path plus $2\pi$ times the ``winding number'' counting the number of full revolutions of the unit circle accumulated along the path.
A formal definition of the unwrapped phase dating back to the 1975 homomorphic signal processing of Oppenheim and Schafer~\cite{Oppenheim:1975}  can be used algebraically to compute the unwrapped phase of a complex polynomial~\cite{McGowan:1982,Steiglitz:1983,AlNashi:1989,Yamada:1998,Yamada:2002,Yamada:2011,Kitahara:2015}, while numerical techniques can be used to approximately compute the unwrapped phase from sufficiently finely sampled time series of phases~\cite{Oppenheim:1975,Tribolet:1977,Itoh:1982} and higher-dimensional arrays of wrapped phases~\cite{Goldstein:1988,Huntley:1989,Bone:91,Huntley:01,Jenkinson:2003,Hooper:07,Abdul-Rahman:09}.

A continuous function denoted Arg that describes accumulated phase differences along a $1D$ path is defined in Appendix~\ref{app:unwrapping} and when applied to $1D$ correlation functions gives
\begin{equation}
  \begin{split}
    \text{Arg}(C(t)) = \text{Arg}(|C(t)|e^{i\theta(t) - i\theta(0)}) = \int_{\theta(0)}^{\theta(t)} d\theta^\prime = \int_0^t \frac{d\theta}{dt^\prime} dt^\prime.
  \end{split}\label{eq:ArgGdef}
\end{equation}
Since $\Theta(t) = \text{arg}(C(t))$ is not continuous at branch cut crossings, the fundamental theorem of calculus cannot be directly applied to Eq.~\eqref{eq:ArgGdef} if there is a branch cut crossing in the interval $[0,t]$.
By deforming the integration contour to replace branch cut crossings by integrals encircling a neighborhood of the origin (see Appendix~\ref{app:unwrapping}), Eq.~\eqref{eq:ArgGdef} can be transformed into an integral over a domain where $\theta(t)$ is analytic plus $2\pi$ times the total number of oriented branch cut crossings to give
\begin{equation}
  \begin{split}
    \text{Arg}(C(t)) = \Theta(t) + 2\pi \nu(t)
    = \text{arg}(C(t)) + 2\pi \nu(t)
    \equiv \widetilde{\theta}(t) - \widetilde{\theta}(0)
    \equiv \widetilde{\Theta}(t).
  \end{split}\label{eq:ArgGnu}
\end{equation}
The unwrapped phase difference $\widetilde{\Theta}(t)$ associated with a LQFT propagator
therefore differs from the principal-valued or ``wrapped'' phase difference by $2\pi$
times an integer winding number $\nu(t)$ that counts the number of oriented branch
cut crossings of the propagator in $[0,t]$.

\begin{figure}
\centering
\includegraphics[width=\textwidth]{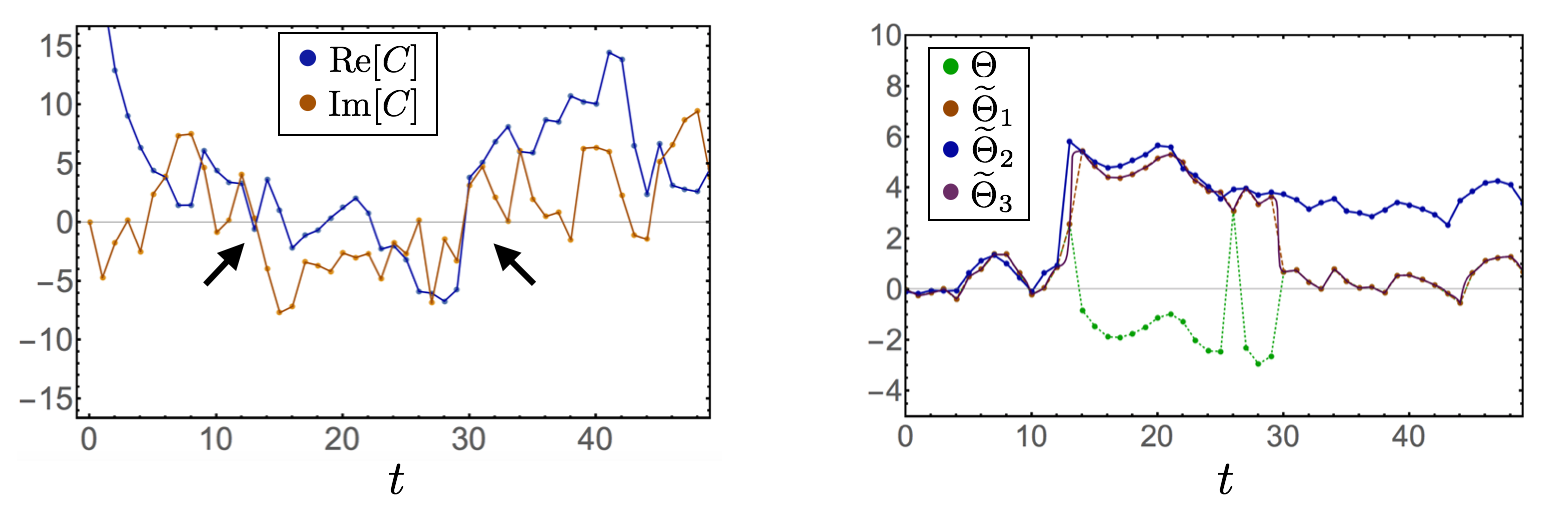}
\caption{The left plot shows real and imaginary parts of a particular $(0+1)D$ free complex scalar field propagator $C(t)$ generated with $M^2 = 0.01$ exhibiting near-zeros of the magnitude indicated by the arrows. The right plot shows the wrapped correlator phase $\Theta(t) = \text{arg}C(t)$ defined with $-\pi < \Theta \leq \pi$ and three calculations of the unwrapped phase $\text{Arg}C(t) \in \mathbb{R}$ obtained by numerical integration of Eq.~\eqref{eq:ArgGdef} for $\widetilde{\Theta}_1$, determination of winding numbers by assuming Eq.~\eqref{eq:1Dbound} for $\widetilde{\Theta}_2$, and algebraic phase unwrapping of a linear polynomial interpolation according to Ref.~\cite{Kitahara:2015} for $\widetilde{\Theta}_3$. The numerical and algebraic winding number methods agree exactly at all lattice sites and only differ in their interpolation between lattice sites. Numerical integration of Eq.~\eqref{eq:ArgGdef} leads to $O(\pi)$ deviations from both winding number methods for all $t > 30$.}
\label{fig:hardunwrapping}
\end{figure}

In LQFT as well as in other applications of complex time series, wrapped phases $\theta(t)$ are determined directly from ``data'' by applying $\text{arg}$ to complex random variables.
The phase unwrapping problem is to determine winding numbers $\nu(t)$ that make the unwrapped phase $\widetilde{\theta}(t)$ a continuous function of $t$ across the branch cuts of $\theta(t)$.
For a complex time series that samples a smooth function with sufficiently fine resolution, one expects that branch cut discontinuities of $\theta(t)$ can be identified and winding numbers can be assigned to keep $\widetilde{\theta}(t)$ continuous across these branch cuts of $\theta(t)$.
It was explicitly demonstrated by Itoh in Ref.~\cite{Itoh:1982} that the assumption
\begin{equation}
  \begin{split}
    |\widetilde{\theta}(t) - \widetilde{\theta}(t-1)| < \pi,
  \end{split}\label{eq:1Dbound}
\end{equation}
is sufficient to uniquely define winding numbers for a time series of wrapped phases
\begin{equation}
  \begin{split}
    \nu(0) = 0,\hspace{20pt} \nu(t) = \nu(t-1) + \begin{cases}
      1, & -2\pi < \theta(t) - \theta(t-1) \leq -\pi \\
      0, & -\pi < \theta(t) - \theta(t-1) \leq \pi \\
      -1, & \pi < \theta(t) - \theta(t-1) \leq 2\pi \end{cases}.
  \end{split}\label{eq:nu1Ddef}
\end{equation}
In LQFT, one might hope that Eq.~\eqref{eq:1Dbound} is valid in generic field configurations when the lattice spacing is much smaller than all physical correlation lengths.
However, Eq.~\eqref{eq:ArgGdef} shows that points with $|\varphi(t)|=0$ have infinite $d\widetilde{\theta}/dt$ even in the continuum.
As demonstrated by example in Fig.~\ref{fig:hardunwrapping}, near-zeros of $|C(t)|$ can occur for $(0+1)D$ free complex scalar field theory with $M^2 = 0.01$.
The wrapped phase of the same correlation function is also shown in Fig.~\ref{fig:hardunwrapping} along with results for three different $(0+1)D$ phase unwrapping schemes:
\begin{enumerate}
  \item Numerical integration of $d\widetilde{\theta}/dt$ according to a linear discretization of Eq.~\eqref{eq:ArgGdef}.
  \item Numerical integration of $\theta(t) - \theta(t-1)$ with winding numbers $\nu(t)$ assigned by Eq.~\eqref{eq:nu1Ddef} to satisfy $|\partial_t \widetilde{\theta}| < \pi$.
  \item Algebraic phase unwrapping of a linear polynomial interpolation of $\varphi$ using the numerically stabilized Strum sequence method of Kitahara and Yamada~\cite{Kitahara:2015}.  \end{enumerate} 
The unwrapped phase defined by numerically integrating Eq.~\eqref{eq:ArgGdef} does not satisfy Eq.~\eqref{eq:nu1Ddef} at $t=30$, which coincides with a near-zero of the magnitude as indicated in Fig.~\ref{fig:hardunwrapping}.
This violation of Eq.~\eqref{eq:nu1Ddef} leads to $O(\pi)$ discrepancies between the results of unwrapping based on numerical integration of Eq.~\eqref{eq:ArgGdef} and both the numerical and algebraic winding number determination methods.
Notably, $O(\pi)$ discrepancies occur at all lattices points with $t$ larger than the point where Eq.~\eqref{eq:1Dbound} is violated.
Near-zeros therefore produce an accumulating $O(\pi)$ sensitivity in the unwrapped phase
which increases with increasing $t$ unless Eq.~\eqref{eq:1Dbound} holds at all points.
The accumulation-of-errors problem will present numerical difficulties for the simple phase unwrapping schemes explored in Sec.~\ref{sec:unwrap1D}.
These difficulties become more tractable in higher-dimensional phase unwrapping problems~\cite{Goldstein:1988,Huntley:1989,Ghiglia:98}, essentially because redundancies in the multidimensional gradient of a smooth function provide additional information that can be used to guide unwrapping across regions where Eq.~\eqref{eq:1Dbound} is violated.
The theory of multidimensional phase unwrapping is much richer than the $1D$ theory explored below and deferred to future work.

The remainder of this section addresses the StN behavior of the unwrapped phase and how properties of its distribution can be used to usefully estimate the average phase cosine.
The small fluctuation assumptions of Eq.~\eqref{eq:small} are used to leading order and phase differences are therefore wrapped normally distributed.
It is also assumed that the unwrapped phase $\widetilde{\theta}$ differs from $\theta$ by $2\pi$ times an integer winding number $\nu$ and therefore that 
\begin{equation}
  \begin{split}
    W[\widetilde{\theta}] \equiv \widetilde{\theta} \text{ mod } 2\pi = \theta ,
  \end{split}\label{eq:Wdef}
\end{equation}
where the wrapping operator $W$ restricts the unwrapped phase to the interval $(-\pi,\pi]$.
Wrapped normal phase differences, Eq.~\eqref{eq:WN}, can be generated by applying $W$ to a normally distributed unwrapped phase difference,
\begin{equation}
  \begin{split}
    \mathcal{P}(\widetilde{\Theta}) = \sqrt{\frac{\kappa }{2\pi t}} e^{-\kappa  \widetilde{\Theta}^2/(2t)}.
  \end{split}\label{eq:N}
\end{equation}
By construction the average cosine of the wrapped and unwrapped phases are identical
\begin{equation}
  \begin{split}
    \avg{\cos(\widetilde{\Theta})} &= \avg{\cos(\Theta + 2\pi\nu )} \\
    &=  \avg{\cos(\Theta)} \\
    &=  e^{-t/(2\kappa)} \approx e^{-E t}.
  \end{split}\label{eq:cosUW}
\end{equation}
The sample mean cosine of an ensemble of unwrapped phases could be used to estimate the ground-state energy with identical results and identical StN degradation as calculations based on the sample mean of the wrapped phase cosine.
However, the boson mass can be estimated more efficiently from a MC ensemble of normally distributed unwrapped phase differences by
\begin{equation}
  \begin{split}
    \widetilde{E}(t) \equiv   \frac{1}{2N}\sum_{i=1}^N  \left[ \widetilde{\Theta}_i(t)^2 - \widetilde{\Theta}_i(t+1)^2 \right],
  \end{split}\label{eq:Mbartildedef}
\end{equation}
where $\widetilde{\Theta} \rightarrow - \widetilde{\Theta}$ symmetry has been assumed on the basis of $\Theta \rightarrow - \Theta$ symmetry (which follows in the infinite-statistics $N\rightarrow \infty$ limit from the reality of correlation functions guaranteed by unitarity).
The corresponding estimate of the correlation function is
\begin{equation}
  \begin{split}
    \widetilde{G}(t) \equiv \left(\frac{1}{N}\sum_{i=1}^N |\varphi_i(t)\varphi_i(0)| \right) \exp\left( -\frac{1}{2N} \sum_{i=1}^N \widetilde{\Theta}_i(t)^2 \right).
  \end{split}\label{eq:Gtildedef}
\end{equation}
Under the present assumptions this provides an accurate estimate of the correlation function as $N\rightarrow \infty$,
\begin{equation}
  \begin{split}
    \avg{  \widetilde{G}(t) } &\approx Z_{1;0,1} \prod_{i=1}^{N}\left[
    \frac{1}{\sqrt{4 \pi E t}}
    \int d \UWTheta_i(t)\ e^{-\UWTheta_i(t)^2 \kappa/(2 t) - \frac{1}{2N}\UWTheta_i(t)^2}  \right] \\
      &= Z_{1;0,1} \paren{1 + \frac{2 E t}{N}}^{-N/2} \\
      &= Z_{1;0,1} e^{-E t}\left[ 1 + \frac{(Et)^2}{N} + O\left(N^{-2}\right) \right] .
  \end{split}\label{eq:Gtildeav}
\end{equation}
The variance of $\widetilde{G}$ in the $N\rightarrow\infty$ limit can be computed similarly,
\begin{equation}
  \begin{split}
    \text{Var}\left(  \widetilde{G}(t) \right) &\approx
    -\avg{  \widetilde{G}(t) }^2 + Z_{1;0,1}^2 \prod_{t=0}^{L-1}\left[
    \frac{1}{\sqrt{4 \pi E t}}
    \int d \UWTheta_i(t)\ e^{-\UWTheta_i(t)^2 \kappa/(2 t) - \frac{1}{N}\UWTheta_i(t)^2} \right]\\
      &=  Z_{1;0,1}^2 \paren{1 + \frac{4 E t}{N}}^{-N/2} - Z_{1;0,1}^2 \paren{1 + \frac{2 E t}{N}}^{-N}  \\
      &= Z_{1;0,1}^2 e^{-2 E t}\left[ \frac{2 (E t)^2}{N} + O\left(N^{-2}\right) \right].
  \end{split}\label{eq:Gtildevar}
\end{equation}
The correlation function computed from normally distributed unwrapped phases therefore has a StN ratio
\begin{equation}
  \begin{split}
    \text{StN}\left( \widetilde{G}(t) \right) \approx \frac{\sqrt{N}}{\sqrt{2} Et}\left[ 1 + O(N^{-1/2})\right].
  \end{split}\label{eq:StNMbartilde}
\end{equation}
Eq.~\eqref{eq:StNMbartilde} demonstrates that normally distributed unwrapped phases provide correlation function estimates whose StN ratios decrease polynomially as $t^{-1}$ rather than exponentially as $e^{-E t}$ as the spacetime volume $t$ containing nonzero $U(1)$ charge is increased.

\subsection{Unwrapped characteristic function and cumulant expansion}\label{sec:cumulant}

For field configurations violating the small fluctuation assumptions of Eq.~\eqref{eq:small}, as demonstrated to occur even in free-field theory in Figs.~\ref{fig:phasehist}-\ref{fig:hardunwrapping}, it is necessary to construct an estimator for $\avg{\cos\Theta}$ from the unwrapped phase that does not depend on assumptions about the distribution of $\Theta$.
The constraint that the winding numbers of the unwrapped phase are integer values, $W[\widetilde{\theta}] = \theta$,
can be interpreted as a statement that the characteristic functions of the wrapped and unwrapped phase differences agree at every integer,
\begin{equation}
  \begin{split}
    \Phi_\Theta(n) = \avg{e^{i\Theta n}} = \avg{e^{i\widetilde{\Theta}n}} = \Phi_{\widetilde{\Theta}}(n),\hspace{20pt} n\in\mathbb{Z}.
  \end{split}\label{eq:charUW}
\end{equation}
For noninteger $n$ the wrapped and unwrapped characteristic functions can differ.
By constraining the unwrapped characteristic function with results not limited to integer $n$, winding number information present in the unwrapped phase  can be incorporated.
Once the unwrapped phase characteristic function is fit to numerical results by some method, the mean cosine of the (wrapped or unwrapped) phase is given by evaluating the resultant fit function at $n=1$,
\begin{equation}
  \begin{split}
    \Phi_{\widetilde{\Theta}}(1) = \avg{\cos(\Theta)} \approx e^{-E t}.
  \end{split}\label{eq:avecos}
\end{equation}

Cumulant expansion methods similar to those explored in Refs.~\cite{Endres:2011jm,Endres:2011er,Endres:2011mm,Lee:2011sm,Nicholson:2012xt,Grabowska:2012ik,Wagman:2016bam}  can be used to estimate $\Phi_{\widetilde{\Theta}}(1)$ with systematic uncertainties whose size can be assessed by varying the truncation order of the expansion. 
For a generic complex random variable $z$ with characteristic function $\Phi_z(k) = \avg{e^{ikz}}$, cumulants can be defined as the coefficients of a Taylor series for $\ln(\Phi_z)$,
\begin{equation}
  \begin{split}
    \Phi_z(k) = \avg{e^{ikz}} \equiv \exp\left[ \sum_{n=1}^\infty \frac{(ik)^n}{n!}\kappa_n(z) \right].
  \end{split}\label{eq:charfundef}
\end{equation}
Equivalent expansions can be constructed that perform a dual expansion in cumulants of the real and imaginary parts of $z$.
The cumulants appearing in Eq.~\eqref{eq:charfundef} can be related to the moments of $z$ by comparing Taylor series expansions for the exponentials in Eq.~\eqref{eq:charfundef},
\begin{equation}
  \begin{split}
    \kappa_n(z) = \avg{z^n} - \sum_{m=1}^{n-1} {n-1 \choose m-1} \kappa_m(z)\avg{z^{n-m}}.
  \end{split}\label{eq:cumulantmom}
\end{equation}
Noting that scalar field propagators are given in terms of the field's log-magnitude and unwrapped phase by
\begin{equation}
  \begin{split}
    C(t) = \varphi(t)\varphi^*(0) = e^{\R + i \widetilde{\Theta}},
  \end{split}\label{eq:CUW}
\end{equation}
an estimator for the scalar boson mass can be defined by 
\begin{equation}
  \begin{split}
    \widetilde{E}^{(n_{max})} &= -\sum_{n=1}^{n_{max}} \frac{1}{n!} \; \partial_t \kappa_n(\R+i\widetilde{\Theta}) \\
    &= -\sum_{n=1}^{n_{max}} \frac{1}{n!} \; \partial_t \kappa_n(\text{ln}|C| + i \text{Arg}(C)).
  \end{split}\label{eq:cumulantEMdef}
\end{equation}
 In the limits $n_{max}\rightarrow \infty$ and $N\rightarrow \infty$, Eq.~\eqref{eq:cumulantEMdef} should approach the scalar boson mass or the ground-state energy for two-point correlation functions in general LQFTs.
Consistency between unwrapped and wrapped phase cumulant expansions requires that $W[\widetilde{\theta}] = \theta$ but is otherwise independent of the particular choice of phase unwrapping algorithm used to define $\widetilde{\Theta}$.

The leading contributions to Eq.~\eqref{eq:cumulantEMdef} are
\begin{equation}
  \begin{split}
    \kappa_1(\R) = \avg{\R},\hspace{20pt}
    \kappa_2(\R) = \avg{\R^2}- \avg{\R}^2,\hspace{20pt} \kappa_2(\widetilde{\Theta}) = \avg{\widetilde{\Theta}^2},
  \end{split}\label{eq:leadingcumulants}
\end{equation}
since $\kappa_1(\UWTheta)$ and the covariance of $\R$ and $\UWTheta$ are guaranteed to vanish by $\UWTheta \rightarrow -\UWTheta$ symmetry.
In general LQFTs the magnitude and phase might make very different contributions to the effective mass, and so an arbitrary hierarchy is possible between odd cumulant contributions only involving the magnitude and even cumulant contributions that also involve the phase.
In particular, $\kappa_2(\widetilde{\Theta})$ dominates $\kappa_1(\R)$ for free complex scalar field theory and the leading contribution to the cumulant effective mass above is
\begin{equation}
  \begin{split}
    \widetilde{E}^{(2)} &= -\partial_t \kappa_1(\R) - \frac{1}{2}\partial_t \kappa_2(\R) + \frac{1}{2}\partial_t \kappa_2(\widetilde{\Theta}) + \dots,
  \end{split}\label{eq:leadingcumuEM}
\end{equation}
where the ellipsis denotes contributions from $\kappa_n(\R+i\widetilde{\Theta})$ with $n\geq 3$.
All omitted contributions with $n\geq 3$ would vanish in the infinite statistics $N\rightarrow \infty$ limit if $\R$ and $\widetilde{\Theta}$ were exactly normally distributed and independent.
For distributions with finite moments, contributions from $n\geq 3$ provide subdominant corrections that will be small for approximately normally distributed $\R$ and $\widetilde{\Theta}$.
The size of these contributions can be assessed in practice by comparing results for $\widetilde{E}^{(n_{max})}$ with multiple truncation points $n_{max}$ and systematic uncertainties can be assigned based off sensitivity of $\widetilde{E}^{(n_{max})}$ to the truncation point.
Since terms with odd $n$ have vanishing phase contributions by $\Theta \rightarrow -\Theta$ symmetry, the convergence pattern of $E^{(n_{max})}$ should be expected to strongly depend on whether $n_{max}$ is even or odd and be comparatively smooth as a function of even (odd) cumulant number $n_{max} = 2,4,6,\dots$ ($n_{max} = 3,5,7,\dots$).

An analogous expansion to Eq.~\eqref{eq:cumulantEMdef} could be defined for the wrapped phase; however, the $\mathcal{P}(\Theta)$ approaches a uniform distribution at large times and large cumulants will make sizable contributions to the wrapped analog of Eq.~\eqref{eq:cumulantEMdef}.
Much faster convergence is expected for Eq.~\eqref{eq:cumulantEMdef} if $\widetilde{\theta}(t)$ differs from $\theta(t)$ by nonzero winding numbers and $\mathcal{P}(\widetilde{\Theta})$ is approximately normal.
Any unwrapping algorithm with $W[\widetilde{\theta}]=\theta$ will define a $\widetilde{\Theta}$ such that Eq.~\eqref{eq:cumulantEMdef} is a consistent estimator in the $n_{max}\rightarrow \infty$ limit, but different algorithms may have different convergence rates.

Estimators for correlation functions including cumulant expansions of unwrapped phases are constructed by generalizing Eq.~\eqref{eq:Gtildedef} as
\begin{equation}
  \begin{split}
    \widetilde{G}^{(n_{max})}(t) \equiv  \exp\left[ \sum_{n=1}^{n_{max}} \frac{1}{n!}\kappa_n\left( \mathcal{R}(t) + i \widetilde{\Theta}(t) \right) \right].
  \end{split}\label{eq:Gtildedefn}
\end{equation}
Despite the vanishing of all cumulants with $n\geq 3$ under the assumption of uncorrelated $\R$ and $\UWTheta$ and in the $N\rightarrow \infty$ limit of an exactly normal unwrapped phase distribution, the statistical uncertainties of these higher cumulants increase with increasing $n$.
For large $n$, the variance of the $n$th cumulant will be dominated by the variance of the $n$th moment.
The large moment $n$ and large statistical ensemble size $N$ behavior of cumulant expansion contributions $\kappa_n / n!$  is therefore determined by the statistical behavior of
\begin{equation}
  \begin{split}
    \frac{1}{(2n)!}\avg{\widetilde{\Theta}^{2n}} = \frac{1}{(2n)!}\avg{ \frac{1}{N}\sum_{i=1}^N \widetilde{\Theta}_i^{2n}}.
  \end{split}\label{eq:momdef}
\end{equation}
For normally distributed unwrapped phases, these sample moments have expectation values
\begin{equation}
  \begin{split}
    \frac{1}{(2n)!}\avg{\widetilde{\Theta}^{2n}} \approx \frac{(2n-1)!!}{(2n)!} \left( \frac{t}{\kappa} \right)^n.
  \end{split}\label{eq:normalscaling}
\end{equation}
The variance of $\widetilde{\Theta}$ can be calculated straightforwardly from Eq.~\eqref{eq:normalscaling} and leads to StN behavior for large moments given by
\begin{equation}
  \begin{split}
    \text{StN}\left( \frac{1}{(2n!)}\widetilde{\Theta}^{2n} \right) \approx \sqrt{N} 2^{-n + 1/4} \left[ 1 + O(n^{-1}) + O(N^{-1}) \right].
  \end{split}\label{eq:momStN}
\end{equation}
Cancellations between contributions from different moments could lead to a smaller variance for $\widetilde{E}^{(n_{max})}$ and $\widetilde{G}^{(n_{max})}$ than that of $\widetilde{\Theta}^{2n}/(2n)!$, but exponentially precise cancellations that would be required to avoid $2^{-n}$ StN degradation are not expected to occur at finite $N$.
This suggests that $\widetilde{E}^{(n_{max})}$ and $\widetilde{G}^{(n_{max})}$ have StN ratios proportional to $\sqrt{N}2^{-n}$ as in Eq.~\eqref{eq:momStN}.
This suggests that even under the assumptions of Eq.~\eqref{eq:small}, the construction of a complete solution to the sign problem using phase unwrapping and the cumulant expansion still requires an extrapolation $n_{max}\rightarrow\infty$ where $N$ must be taken exponentially large in $n_{max}$ to remove all truncation errors at fixed statistical precision.

The appearance of such an exponentially hard extrapolation should be expected: phase unwrapping is applicable to generic LQFT correlation functions and the sign problem has been demonstrated to be NP-hard for some quantum systems by Troyer and Wiese~\cite{Troyer:2004ge}.
For LQFTs including LQCD, observations of the ubiquity of (complex-)log-normally distributed correlation functions~\cite{Hamber:1983vu,Guagnelli:1990jb,Endres:2011jm,DeGrand:2012ik,Drut:2015uua,Porter:2016vry,Wagman:2016bam,Rammelmuller:2017vqn} suggest that useful results might be obtained using modest $n_{max}$ despite the exponential difficulty of extrapolating to $n_{max}\rightarrow \infty$.
Understanding the size of truncation errors in practical calculations and systematic limitations of this method will likely require specific studies for particular LQFTs of interest.

Generic correlation functions in $(0+1)D$ complex scalar field theory can be analyzed similarly to the scalar boson propagators above.
The wrapped phase for a general correlation function $G_{Q,2P}$ only depends on its $U(1)$ charge $Q$ and not on $P$ and is denoted by 
\begin{equation}
  \begin{split}
  \Theta_Q \equiv \Theta_{Q,2P} = W[Q \Theta]  
  \end{split}\label{eq:ThQdef}
\end{equation}
The unwrapped phase can similarly be chosen to be independent of $P$ and defined by\footnote{Note that this is not the most general choice. Since large phase jumps are associated with regions of small magnitude by Eq.~\eqref{eq:ArgGdef}, a phase unwrapping scheme that depends on the magnitude and therefore on $P$ may have advantages.}
\begin{equation}
  \begin{split}
   \widetilde{\Theta}_Q   &\equiv \text{Arg}\left( \mathcal{O}_{Q,2P}(t) \mathcal{O}^*_{Q,2P}(0) \right) \\
    &= \text{Arg} \left(  e^{i Q \Theta } \right).
  \end{split}\label{eq:uwQ}
\end{equation}
Note that because Arg is a nonlinear function, $\widetilde{\Theta}_Q$ is not simply related to $\widetilde{\Theta}$ despite the identity $W[Q\widetilde{\Theta}] = \Theta_Q$.
In particular $Q \widetilde{\Theta}$ is sensitive to branch cut crossings of the variable $Q\Theta$ with principle domain $(-\pi Q, \pi Q]$ rather than branch cut crossings of $\Theta_{Q} = W[Q\Theta]$ with principle domain $(-\pi,\pi]$.
With $\widetilde{\Theta}$ defined by Eq.~\eqref{eq:nu1Ddef}, $Q\widetilde{\Theta}$ will include jumps of $2\pi Q$ at branch cut crossings of $\Theta$ rather than jumps of $2\pi$ at branch cut crossings of $\Theta_Q$ and is therefore not equal to $\widetilde{\Theta}_Q$.
With a consistent phase unwrapping of $\Theta_{Q,2P} \in (-\pi,\pi]$, correlation functions and ground-state energies can be estimated with the cumulant expansions 
\begin{equation}
  \begin{split}
    \widetilde{G}_{Q,2P}^{(n_{max})} = \exp\left[ \sum_{n=1}^{n_{max}} \frac{1}{n!}\kappa_n \left( (|Q| + 2P)\R + i \widetilde{\Theta}_Q \right) \right], \\
    \widetilde{E}_{Q,2P}^{(n_{max})} = -\sum_{n=1}^{n_{max}} \frac{1}{n!}\partial_t \kappa_n \left( (|Q| + 2P)\R + i \widetilde{\Theta}_Q \right).
  \end{split}\label{eq:EQPdef}
\end{equation}
For any phase unwrapping algorithm satisfying $W[\widetilde{\Theta}_Q] = \Theta_{Q,2P}$, these provide unbiased estimators for correlation functions and effective masses in the dual limit $N\rightarrow \infty$ and $n_{max} \rightarrow \infty$.
In general charge sectors, $\Theta_Q$ is wrapped normally distributed under the assumptions of Eq.~\eqref{eq:small} and $\widetilde{\Theta}_Q$ can be consistently defined to be normally distributed with variance $1/\kappa_Q$ chosen to reproduce the ground-state energy charge $Q$ sector.
In analogy to $\widetilde{\Theta}$, the correct ground-state energy $E_Q \equiv E_{Q,0}$ is reproduced if the variance of $\widetilde{\Theta}$ is taken to be $1/\kappa_Q \approx 2 E_Q$.
The StN results of Eq.~\eqref{eq:StNMbartilde} can therefore be applied to $\widetilde{G}_{Q,2P}^{(n_{max})}$ if $\kappa$ is replaced by $\kappa_Q$ to give
\begin{equation}
  \begin{split}
    \text{StN}\left( \widetilde{G}_{Q,2P}^{(2)} \right) &\approx \frac{\sqrt{N}}{2E_Q t}\left[ 1 + O(N^{-1}) \right].
  \end{split}\label{eq:EQPStN}
\end{equation}
The moment analysis of Eq.~\eqref{eq:momStN} can be applied to $\widetilde{\Theta}_Q$, and suggests that StN ratios for $\widetilde{G}_{Q,2P}^{(n_{max})}$ decrease polynomially with increasing $E_Q  t$ but exponentially with increasing $n_{max}$.

The avoidance of exponential StN degradation with increasing $E_Q t$ at fixed order in the cumulant expansion can also be understood in the language of sign problems.
Integration over phase fluctuations making nonpositive-definite contributions to path integrals is replaced by calculation of the moments of the unwrapped phase,
\begin{equation}
  \begin{split}
    \avg{\widetilde{\Theta}_Q^n} = \frac{1}{Z}\int_\varphi e^{-S} \widetilde{\Theta}_Q^{n} ,
  \end{split}\label{eq:GQPmagphaseUW}
\end{equation}
that vanish for odd $n$ by unitarity and are path integrals of positive-definite quantities without sign problems or phase fluctuations for even $n$.
A sign problem can reemerge beyond leading order in the cumulant expansion from linear combinations of positive-definite moments that enter the cumulant expansion with opposite signs.
In particular if $\widetilde{E}^{(2)}$ approximates $E$ poorly, then the sum of cumulants could be $O(e^{-E_Q t})$ while the individual cumulant contributions are $O(1)$ and the full sign problem could reemerge at full strength.

\begin{figure}
\centering
\includegraphics[width=\textwidth]{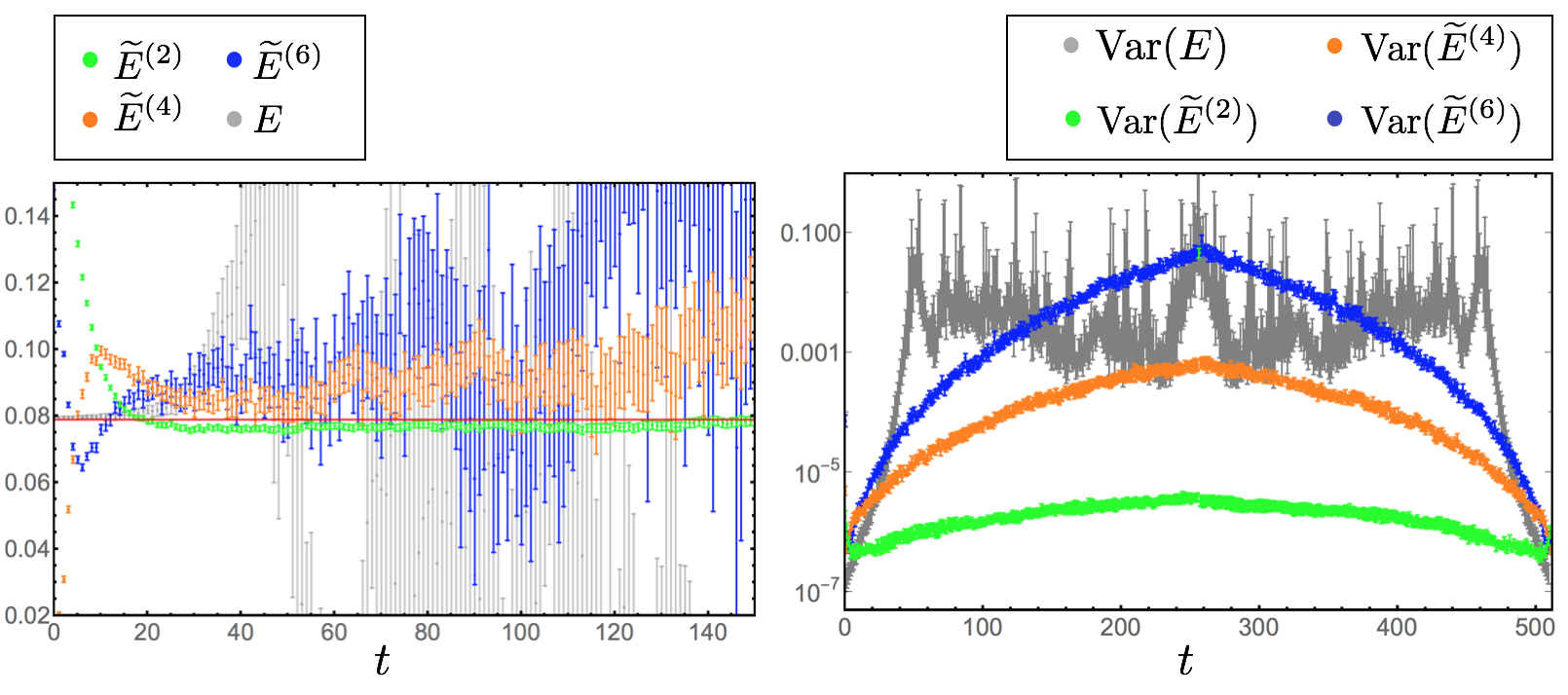}
\caption{ The left plot shows the scalar boson mass $\widetilde{E}^{(n_{max})}$ on ensemble $C_0$ obtained using cumulant expansions of the propagator log-magnitude and unwrapped phase truncated at order $n_{max} = 2,\ 4,\ 6$. The $L\rightarrow\infty$ analytic result is shown as a red line and the standard effective mass is shown in gray. The right plot shows the variances of these effective masses. Gaussian-weighted integration with $\sigma = 1.41$ is used to calculate the unwrapped phase.}
\label{fig:unwrapStN}
\end{figure}

\begin{figure}
\centering
\includegraphics[width=\textwidth]{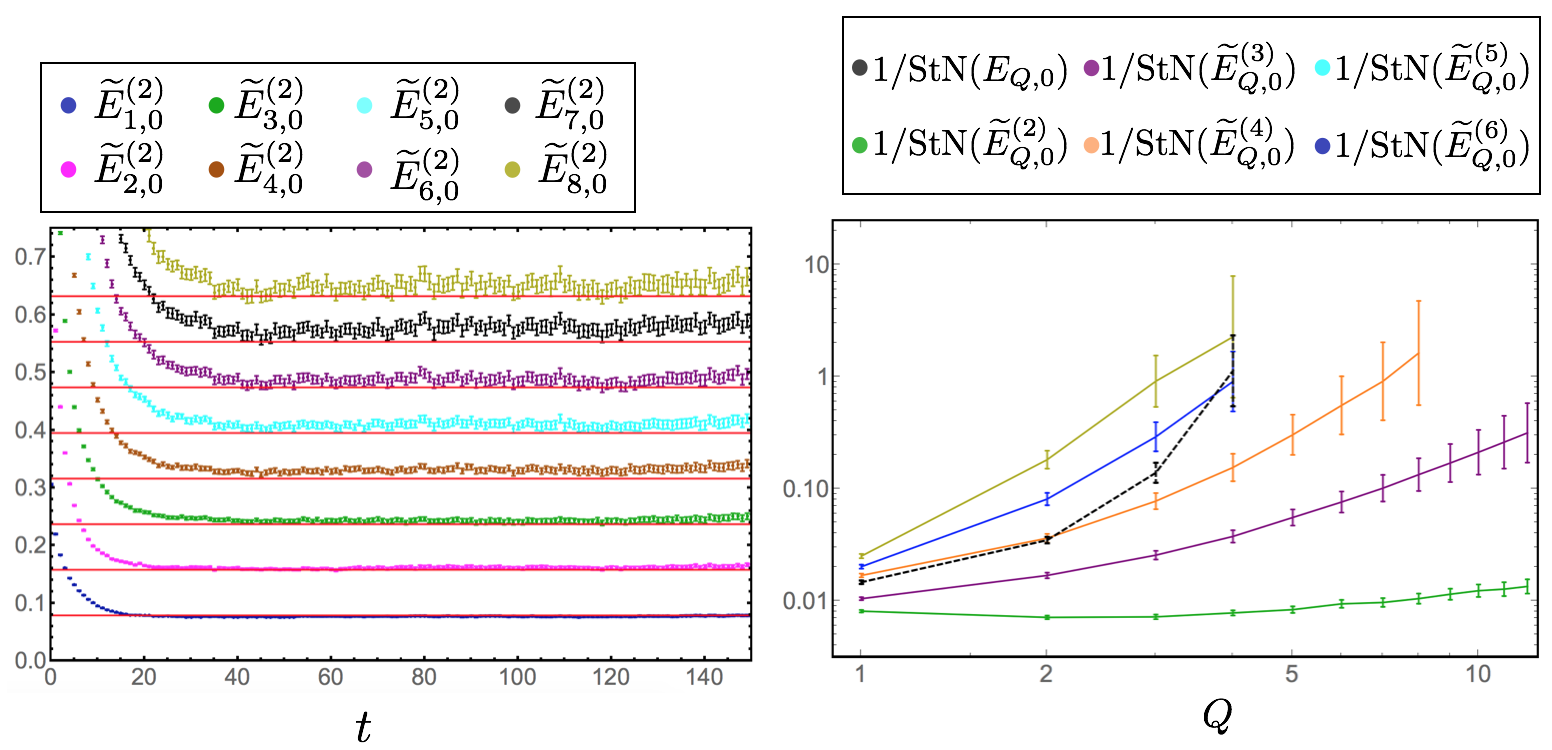}
\caption{ The left plot shows the ground-state energies $\widetilde{E}_{Q,0}^{(2)}$ of charge sectors $Q=1,\dots,8$ for ensemble $C_0$ that involve second-order truncations of cumulant expansions of the log-magnitudes and unwrapped phases of the correlation functions $G_{Q,0}$. Results for $E^{(2)}_{1,0}$ are identical to those shown in Fig.~\ref{fig:unwrapStN}. The right plot shows the average inverse StN of these ground-state energy measurements for a time region $t = 10\rightarrow 20$ as a function of $Q$ for various cumulant expansion truncation orders. Gaussian-weighted integration with $\sigma = 1.41$ is used to calculate the unwrapped phase.}
\label{fig:unwrapStNQ}
\end{figure}

\section{One-dimensional phase unwrapping}\label{sec:unwrap1D}

Phase unwrapping was shown analytically above to remove exponential StN degradation at fixed order in a cumulant expansion under the assumptions of Eq.~\eqref{eq:small}.
Relaxing these assumptions, the probability distribution of phase fluctuations becomes more complicated and an unwrapped phase distribution satisfying $W[\widetilde{\theta}] = \theta$ cannot be easily found analytically.
Numerical MC simulations are used in this section to analyze the accuracy and precision of cumulant expansions involving the log-magnitude and unwrapped phase in $(0+1)D$ scalar field theory without the small fluctuation assumptions of Eq.~\eqref{eq:small}.

\subsection{Numerical phase unwrapping schemes}\label{sec:unwrapnum}

For a field defined on a discrete lattice of points, the unwrapped phase is not uniquely defined without further assumptions that for instance could be based on a discretized definition of smoothness.
Precisely defining this smoothness assumption is essential for constructing a numerical phase unwrapping algorithm.
The assumptions $|\partial_t \widetilde{\theta}| < \pi$ and $W[\widetilde{\theta}]=\theta$
lead uniquely to the path unwrapping algorithm of Eq.~\eqref{eq:nu1Ddef}.
This section employs this phase unwrapping algorithm and two variations with alternative smoothness criteria that enforce smoothness on larger distances than a single lattice spacing.
\begin{enumerate}
\item Single-point integration: $\widetilde{\theta}(t)$ is determined by demanding
\begin{equation}
\abs{\widetilde{\theta}(t) - \widetilde{\theta}(t-1)} < \pi
\end{equation}
as in Eq.~\eqref{eq:nu1Ddef}. This technique assumes a finely sampled lattice, such that the probability density of phase jumps near $\pi$ is vanishing.

\item Windowed integration, with window $w$: $\widetilde{\theta}(t)$ is determined by demanding
\begin{equation}
\abs{\widetilde{\theta}(t) - \frac{1}{\text{min}(w,t)} \sum_{t' = \text{max}(t- w, 0)}^{t-1} \widetilde{\theta}(t')} < \pi.
\end{equation}
This technique more robustly handles large phase jumps by considering the average of previously unwrapped phases. This locally may allow the unwrapped phase jump magnitude to exceed $\pi$, but in such a way that global fluctuation is reduced. When $w = 1$, this reduces to single-point integration.

\item Gaussian-weighted integration, with width $\sigma$: $\widetilde{\theta}(t)$ is determined by demanding
\begin{equation}
\abs{\widetilde{\theta}(t) - \sum_{t' = 0}^{t-1} \mathcal{N} e^{-(t'-t)^2/(2 \sigma^2)} \widetilde{\theta}(t')} < \pi.
\end{equation}
The normalization $\mathcal{N}$ is fixed by
$\sum_{t' = 0}^{t-1} \mathcal{N} e^{-(t'-t)^2/(2 \sigma^2)} = 1$. This technique allows one to smoothly interpolate between integer window sizes by providing a non-integer tunable parameter. When $\sigma \sim w$, we expect the two techniques to perform similarly.
\end{enumerate}
Only phase unwrapping algorithms satisfying $W[\widetilde{\theta}] = \theta$
are considered, since this condition guarantees that the   unwrapped cumulant expansion reproduces the exact correlation function in the dual limit of infinite truncation order and infinite statistics. 

A time-reversal-symmetric integration path is used in which one of the forward integration techniques above is applied to determine the unwrapped phase in $[0,L/2]$ and the corresponding reverse integration technique is applied to determine the unwrapped phase in $[L/2+1,L]$.
This symmetric integration path has the advantage of beginning with regions closest to the source where phase gradients are smallest and the probability of unwrapping ambiguities due to physical fluctuations violating $|\partial_t \widetilde{\theta}| < \pi$ is correspondingly smallest.
The $t=0$ phase $\widetilde{\theta}(0) = \theta(0)$ is used as an initial condition for unwrapping, although this is irrelevant for correlation functions since they only involve phase differences $\widetilde{\theta}(t) - \widetilde{\theta}(0)$.
With this scheme the unwrapped phase is discontinuous at $L/2$ if the wrapped phase is associated with a $q\neq 0$ field configuration with nonzero $U(1)$ winding number.
Since the same point $t=0$ is used for the correlation function source and the initial unwrapping point in this scheme, the unwrapped phase should be separately calculated for each correlation function in a MC ensemble when multiple source points are used with each field configuration. 
As discussed above, the phase differences $Q\theta(t) - Q\theta(0)$ for each charge sector $Q$ also need to be unwrapped individually since the unwrapped phase is a nonlinear function of the wrapped phase.

\subsection{Complex scalar field MC ensembles}
\label{subsec:cho-ensembles}

Nineteen different choices of the $(0+1)D$ complex scalar field theory parameters $M^2$ and $\lambda$ indicated in Table~\ref{tbl:cho-ensembles} are employed to generate a variety of free and interacting MC ensembles.
Free-field ensembles $A_0$, $B_0$, and $C_0$ are generated with parameters $M^2= 0.1,\ 0.025,$ and $0.00625$ and serve as toy models for LQFTs with coarse, moderate, and fine lattice spacings, respectively. In lattice units, the free-field correlation lengths defined by $\xi = 1/E$ are
\begin{equation}
  \begin{split}
    \xi_{A_0} = 3.175, \hspace{20pt}   \xi_{B_0} = 6.331, \hspace{20pt}  \xi_{C_0} = 12.652,
  \end{split}\label{eq:corrdef} 
\end{equation} 
where the $L\rightarrow \infty$ approximation Eq.~\eqref{eq:lattpole} to the free scalar boson mass $E$ has been used. 
For each choice of $M^2$, the length of the lattice has been rescaled to $L = 128,\ 256,$ and $512$ respectively to enforce $M L = 128\sqrt{0.1} \approx 40.48$ and maintain a roughly constant temporal extent in units of the free-field correlation length. Two additional free-field ensembles $D_0$ and $E_0$
are generated with finer lattice spacing to explore lattice spacing dependence.
In addition to the free-field ensembles, a variety of interacting scalar field theory ensembles are generated.
Quartic self-interactions are used that correspond to potentials
\begin{equation}
  \begin{split}
    V(|\varphi|) = \lambda |\varphi|^4,
  \end{split}\label{eq:phi4}
\end{equation}
with a variety of couplings $\lambda$ indicated in Table~\ref{tbl:cho-ensembles}.
Repulsive couplings $\lambda > 0$ are necessary for the action to be bounded from below and for the path integral representing the thermal partition function to converge.
With $\lambda > 0$, the partition function is well defined for $M^2 < 0$.
In higher dimensions, this corresponds to a phase of complex scalar field theory where the $U(1)$ global symmetry is spontaneously broken.
In $(0+1)D$ at finite $L$ the correlation length is finite in the $M^2 < 0$ phase but much larger than in the $M^2 > 0$ phase with the same $\lambda$ and $|M^2|$.
Ensembles $A_n^\pm$, $B_n^\pm$ and $C_n^\pm$ describe interacting scalar field theories with the same $|M^2|$ as $A_0$, $B_0$, and $C_0$ with the sign of $M^2$ indicated by a superscript and two different values of $\lambda$ shown in Table~\ref{tbl:cho-ensembles} denoted by subscripts $n=1,2$.
Ensembles with different $|M^2|$ but the same $\lambda$ subscript correspond to choices of $\lambda$ that keep $\lambda L / |M^2|$ fixed.
Two additional negative $M^2$ ensembles $D_1^-$ and $E_1^-$, corresponding to $D_0$ and $E_0$, are also generated for a detailed study of lattice spacing dependence
in the interacting case.

To interpret calculations at different $M^2$ and $L$ as having a fixed physical correlation length and varying lattice spacing $a$ for a $(0+1)D$ field of mass dimension $[\varphi] = -1/2$, the dimensionless parameters used in the MC calculations should be interpreted as $(Ma)^2$, $L/a$, and $a^3 \lambda$.
This scaling is obtained if the dimensionless parameter $a^3 \lambda$ is chosen for calculations at different $(aM)^2$ and $L/a$ such that $\lambda L/M^2$ is held fixed.
In appropriately rescaled units, the spectrum $Ea$ obtained at different $(Ma)^2$, $L/a$, and $a^3\lambda$ but fixed $\lambda L/M^2$ will differ by $O(\lambda)$ in small-$\lambda$ perturbation theory.
It is tempting to interpret this as a renormalization condition that permits quantitative comparison of results at different $Ma$ and $L/a$; however, $(0+1)D$ complex scalar field theory is nonrenormalizable and an infinite number of renormalization conditions need to be imposed to consistently match results for the spectrum of LQFTs with different parameters to all orders in $\lambda$.
Spectral results from LQFT calculations with different parameters but fixed $\lambda L/M^2$ should approximately agree for $\lambda \ll 1$ and $L/M^2 \gg 1$ but will differ in general by nonuniversal corrections.
This nonuniversality is an artifact of working in $(0+1)D$ and arises even at $\lambda = 0$ where it can be understood as arising from higher-derivative operators in a Symanzik-improved lattice action~\cite{Symanzik:1983dc}.
Since there is no universal continuum limit for $(0+1)D$ scalar field theory, no attempt is made to employ non-perturbative renormalization conditions to match observables between ensembles with different parameters.
In the $ML \approx 40.48$ MC ensembles considered here, the two different values of $(\lambda L / M^2)_n$ held fixed among $A_n^\pm$, $B_n^\pm$ and $C_n^\pm$  correspond to $(\lambda L / |M^2|)_1 = 16$ and  $(\lambda L / |M^2|)_2 = 32$ respectively.

\begin{table}
\centering
\begin{tabular}{c c c c}
 & $A_0$ & $A_1^\pm$ & $A_2^\pm$ \\
\hline
\hline
$L$ &128&& \\
$M^2$ &$+0.1$&$\pm$&$\pm$ \\
$\lambda$ &
0&
0.0125&
0.025 \\
\hline
\end{tabular}
\begin{tabular}{c c c c}
 & $B_0$ & $B_1^\pm$ & $B_2^\pm$ \\
\hline
\hline
$L$ &256&& \\
$M^2$ &$+0.025$&$\pm$&$\pm$ \\
$10^3 \lambda$ &
0&
1.5625&
3.125 \\
\hline
\end{tabular}
\begin{tabular}{c c c c}
 & $C_0$ & $C_1^\pm$ & $C_2^\pm$ \\
\hline
\hline
$L$ &512&& \\
$M^2$ &$+0.00625$&$\pm$&$\pm$ \\
$10^3 \lambda$ &
0&
0.1953125&
0.390625 \\
\hline
\end{tabular}

\begin{tabular}{c c c}
 & $D_0$ & $D_1^-$ \\
\hline
\hline
$L$ &1024& \\
$M^2$ &$+0.0015625$&$-$ \\
$10^6 \lambda$ &
0&
24.4140625 \\
\hline
\end{tabular}
\begin{tabular}{c c c}
 & $E_0$ & $E_1^-$ \\
\hline
\hline
$L$ &2048& \\
$M^2$ &$+0.000390625$&$-$ \\
$10^6 \lambda$ &
0&
3.0517578125 \\
\hline
\end{tabular}
\caption{Ensembles used for complex scalar investigation, segmented by size. For the free-field cases $A_0$, $B_0$, $C_0$, $D_0$, and $E_0$, a consistent positive $M^2$ is chosen to match physical lattice extent. For the interacting cases of series $A$, $B$, and $C$, both $\pm M^2$ are used, while for series $D$ and $E$ only $-M^2$ is used. Each ensemble is updated via a Metropolis sweep over the odd/even lattice sites $N_{\text{skip}} = 10000$ times
between each measurement. Before saving lattice measurements, $N_{\text{therm}} = 50$ iterations of the complete measurement cycle are performed for thermalization. Following thermalization, $N_{\text{meas}} = 5000$ measurement cycles are performed with the values of $\varphi(t)$ saved each time.}
\label{tbl:cho-ensembles}
\end{table}

In addition to MC ensembles generated using the standard action in Eq.~\eqref{eq:actiondecomp} with the parameter choices described above, ensembles are also generated using the analytically phase-integrated dual form of the theory given in Eq.~\eqref{eq:Pprimedef}-\eqref{eq:GQ2Pdual}.
These dual ensembles only involve MC sampling over the magnitude of the scalar field, and as demonstrated analytically above they have no sign problem and a much milder (though still exponentially severe) StN problem with increasing $t$.
It is verified below that these dual ensembles lead to more precise calculations of ground-state energies $E_{Q,0}$ in charge sectors $Q=1,\dots,4$.
These results are used as a precise check on the accuracy of results obtained using the standard ensembles with and without phase unwrapping.
Results show small but statistically significant differences between the low-lying energy levels of ensembles with different parameters but equal $\lambda L / M^2$.
More details on free-field consistency checks and autocorrelation times can be found in Appendix~\ref{app:MC}.

\subsection{\texorpdfstring{$1D$}{1D} phase unwrapping results}
\label{subsec:cho-results}

Where the phase varies smoothly, a nearest-neighbor unwrapping scheme accurately captures the variation in phase across the ensemble.
Close to large phase jumps with $|\partial_t \theta| > \pi/2$, winding number assignment can vary depending on the phase unwrapping algorithm.
Different algorithms will lead to distributions that broaden more or less quickly with $t$ and therefore different low-order cumulant expansion truncation errors.
Finding an unwrapping scheme with fast convergence in the cumulant expansion amounts to choosing an algorithm for winding number assignment in neighborhoods of large phase jumps that appropriately tunes the unwrapped phase variance growth controlling $\widetilde{E}^{(2)}_{Q,0}$.

\begin{figure}
\centering
\includegraphics[width=0.8\textwidth]{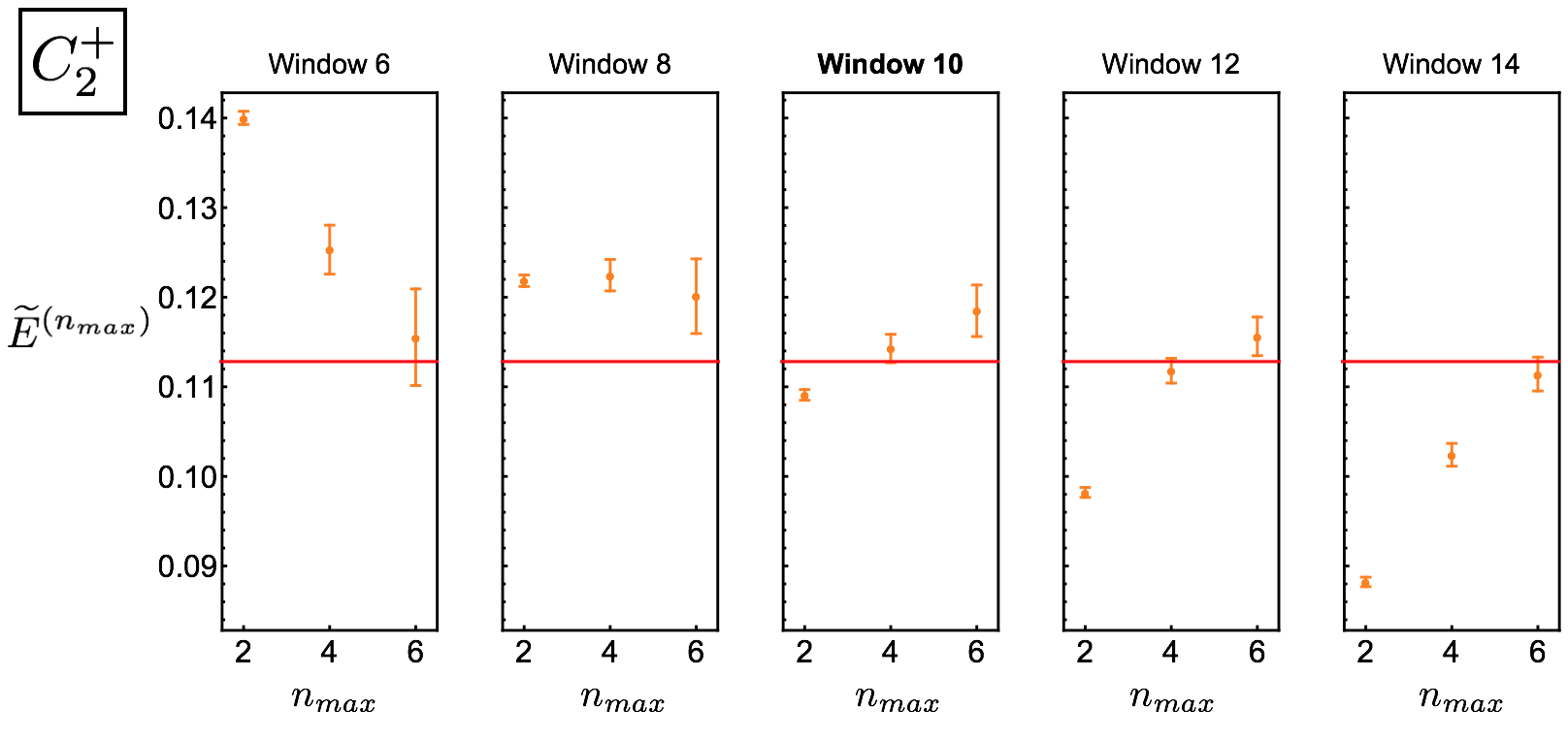}
\includegraphics[width=0.8\textwidth]{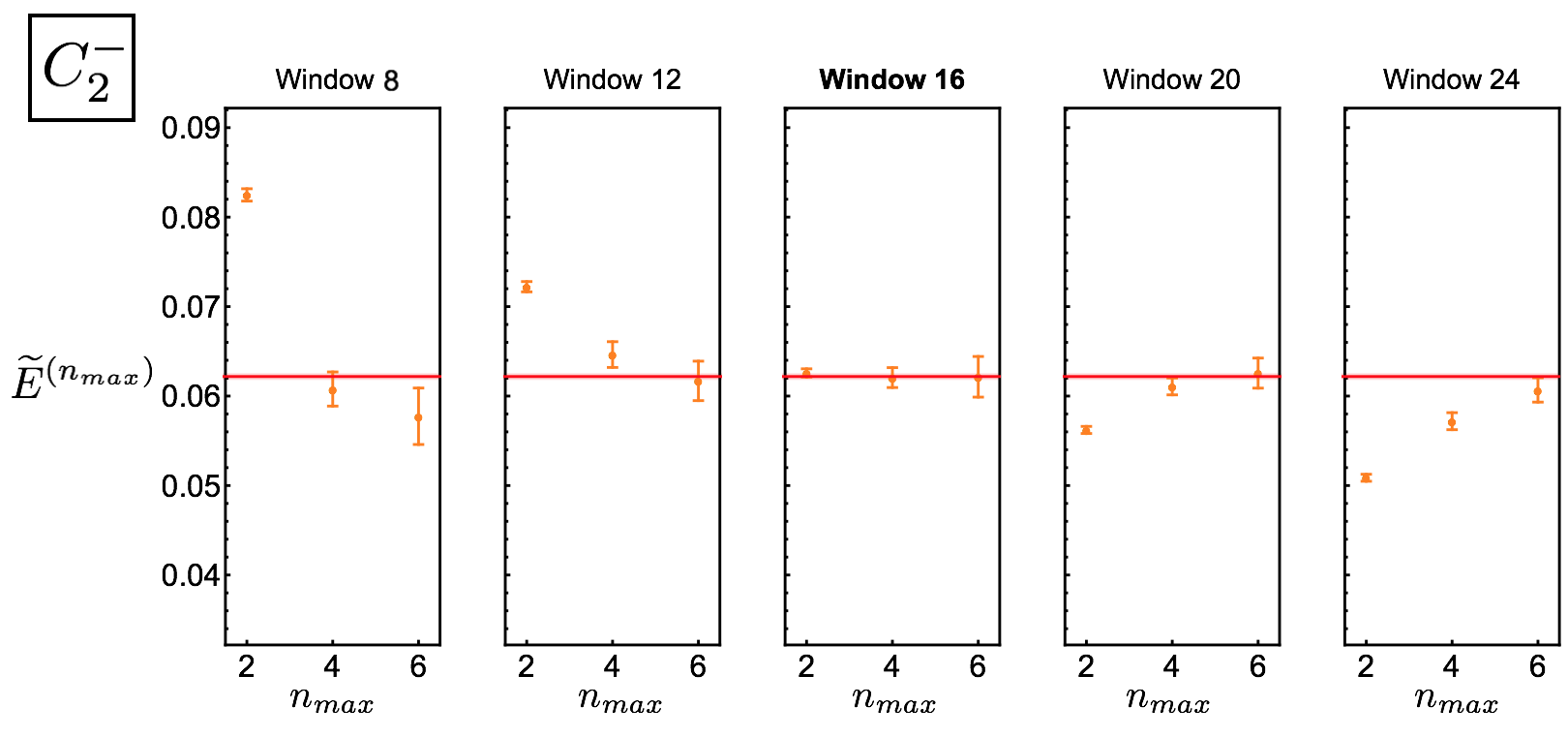}
\caption{Effective mass measurements comparing estimation using cumulants up to order $n_{max} = 2,4,6$ with several window sizes for ensembles $C_2^+$ and $C_2^-$. The red band indicates the dual variables estimate of the correct mass. The choice $w=16$ provides an accurate estimate for $C_2^-$ with $n_{max} = 2$ and little variation with increasing $n_{max}$, while other choices have much larger truncation errors at $n_{max} = 2$ that are reduced by increasing the truncation order. The choice $w=10$ provides the most accurate estimate for $C_2^+$ with $n_{max} = 2$, although statistically significant truncation errors are still visible. For all window choices shown, results with $n_{max} = 6$ are statistically consistent with dual variable calculations at the $1\sigma -2 \sigma$ level.
}
\label{fig:meff-vs-cumu-pos-neg-mass}
\end{figure}

It is empirically found that the single-point integration phase unwrapping method described in Sec.~\ref{sec:unwrapnum} that enforces $|\widetilde{\theta}(t) - \widetilde{\theta}(t-1)| < \pi$ gives poor results for the scalar boson mass across all ensembles.
Results do not markedly improve at finer lattice spacing.
Statistical precision is generally good for correlation functions and ground-state energies estimated from cumulant expansions truncated at low orders with qualitatively similar $t$ scalings to those shown for Gaussian-weighted integration of $C_0$ in Figs.~\ref{fig:unwrapStN}-\ref{fig:unwrapStNQ}.
However, the truncation errors of second- and third-order results for $\widetilde{E}^{(n)}$ are large, sometimes an order of magnitude larger than both the statistical uncertainties and central values of standard ensemble average estimates of $E$.
Truncation errors decrease at higher orders in the expansion, generally with a pattern of visible decreases at even orders that are sensitive to the shape of the phase distribution, but statistical errors increase dramatically.

\begin{figure}
\vspace{0.5cm}
\includegraphics[width=\textwidth]{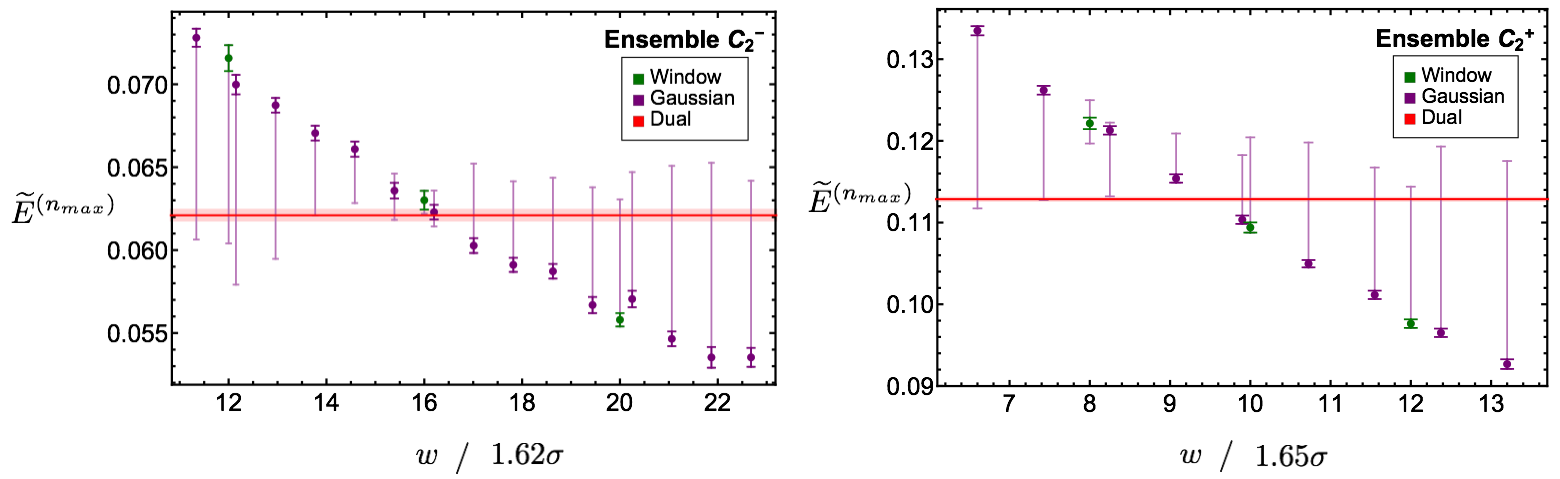}
\caption{The scalar boson mass determined with a variety of phase unwrapping algorithm parameters and truncated cumulant expansion for ensembles $C_2^\pm$.  The green points show $\widetilde{E}^{(2)}$ obtained using windowed integration with window sizes $w$ shown on the horizontal axis. Dark green error bars on these points show $68\%$ confidence intervals including statistical uncertainties. The dark purple points and error bars show $\widetilde{E}^{(2)}$ and its statistical uncertainties obtained using Gaussian integration. The Gaussian widths $\sigma$ used are proportional to the window size $w$ with $w = 1.65\sigma$ for $C_2^-$ (left) and $w = 1.62\sigma$ for $C_2^+$ (right) empirically found to provide agreement between windowed and Gaussian integration. The lighter purple error bars on both Gaussian and windowed unwrapping points show the extent of the variation in central values of $\widetilde{E}^{(2)}$, $\widetilde{E}^{(4)}$, and $\widetilde{E}^{(6)}$ and demonstrate that results tend to converge towards dual ensemble results as $n_{max}$ is increased and also that the size of truncation errors is sensitive to the unwrapping algorithm parameters used. The red bands show dual ensemble results and statistical uncertainties for comparison.
  }
\label{fig:gauss-window-sizes}
\end{figure}

\begin{figure}
\centering
\includegraphics[width=.8\textwidth]
{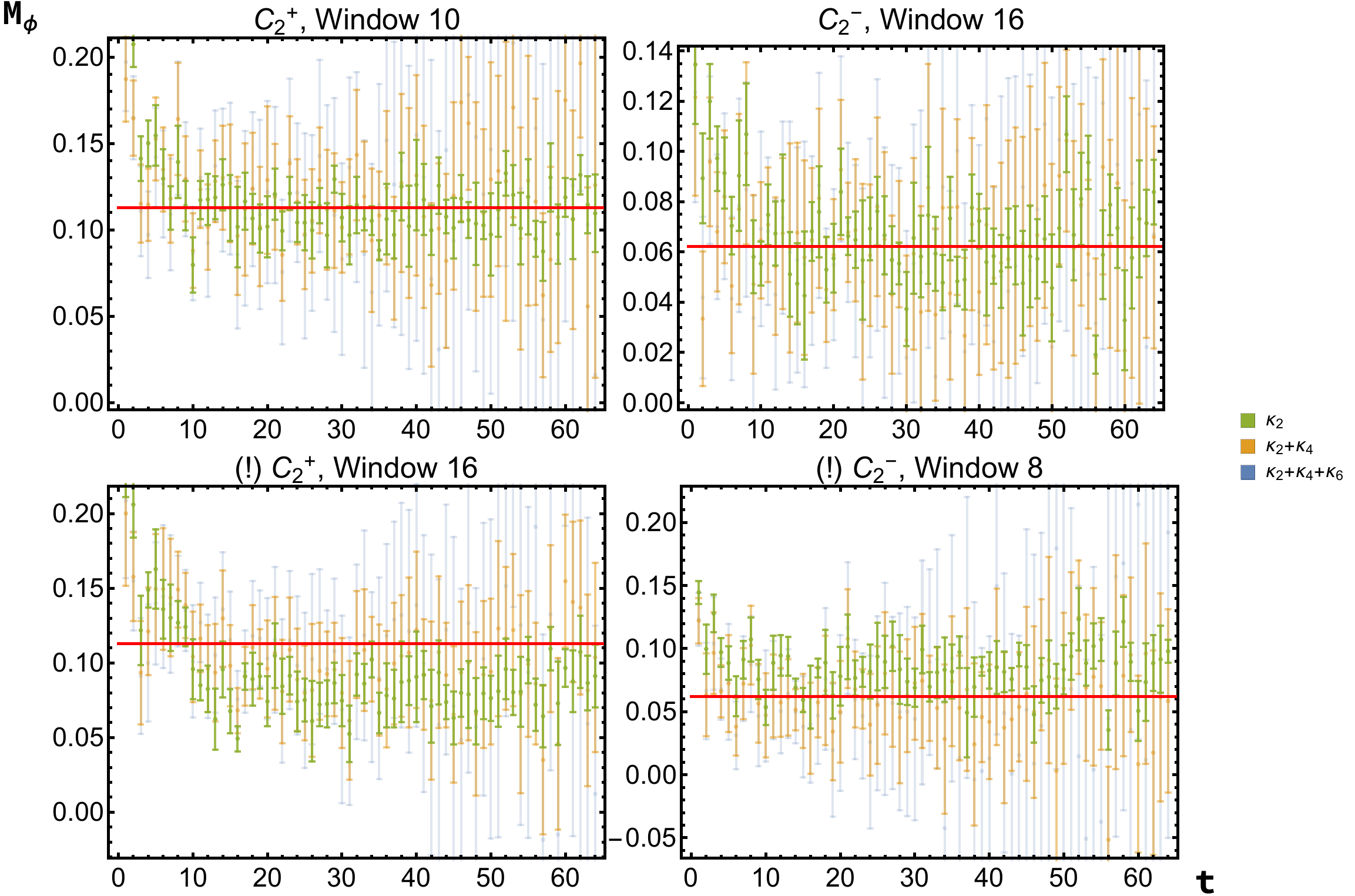}
\caption{Effective mass plots for unwrapped phase cumulant expansion mass calculations for ensembles $C_2^\pm$. The upper plots use the optimal window sizes for $C_2^+$ and $C_2^-$ indicated in Fig.~\ref{fig:meff-vs-cumu-pos-neg-mass} and show little variation with truncation order at all $t$. The lower plots use suboptimal window sizes that include significant truncation errors with $n_{max} = 2$ and smaller truncation errors with larger $n_{max}$. The red band indicates the dual variables estimate of the scalar mass and its uncertainties.}
\label{fig:meff-vs-time-pos-neg-mass}
\end{figure}

The windowed and Gaussian-weighted integration methods that enforce smoothness on scales larger than the lattice scale provide estimates for correlation functions and energies with much smaller truncation errors than single-point integration.
The statistical precision of results at various orders in the cumulant expansion is roughly independent of the choice of smearing scale (the window size $w$ or Gaussian width $\sigma$ used when calculating winding numbers) but the central values of low-order results depend sensitively on the smearing scale.
A representative demonstration of this tendency is shown for $\widetilde{E}^{(n)}$ for ensembles $C_2^+$ and $C_2^-$ in Figs.~\ref{fig:meff-vs-cumu-pos-neg-mass}-\ref{fig:gauss-window-sizes}.
Results with Gaussian-weighted integration tend to match results with windowed integration up to a single $O(1)$ constant of proportionality between $w$ and $\sigma$. 
Gaussian-weighted integration can be tuned to interpolate between integer-valued window sizes in this way.
An empirical condition relating the phase unwrapping smearing scale to the correlation length or another cost function penalizing large truncation errors could be used to self-consistently define an optimal smearing scale, but it is difficult to assess the systematic errors of optimized estimates without sacrificing precision by going to higher orders in the cumulant expansion.

\begin{figure}
  \centering
\includegraphics[width=\textwidth]{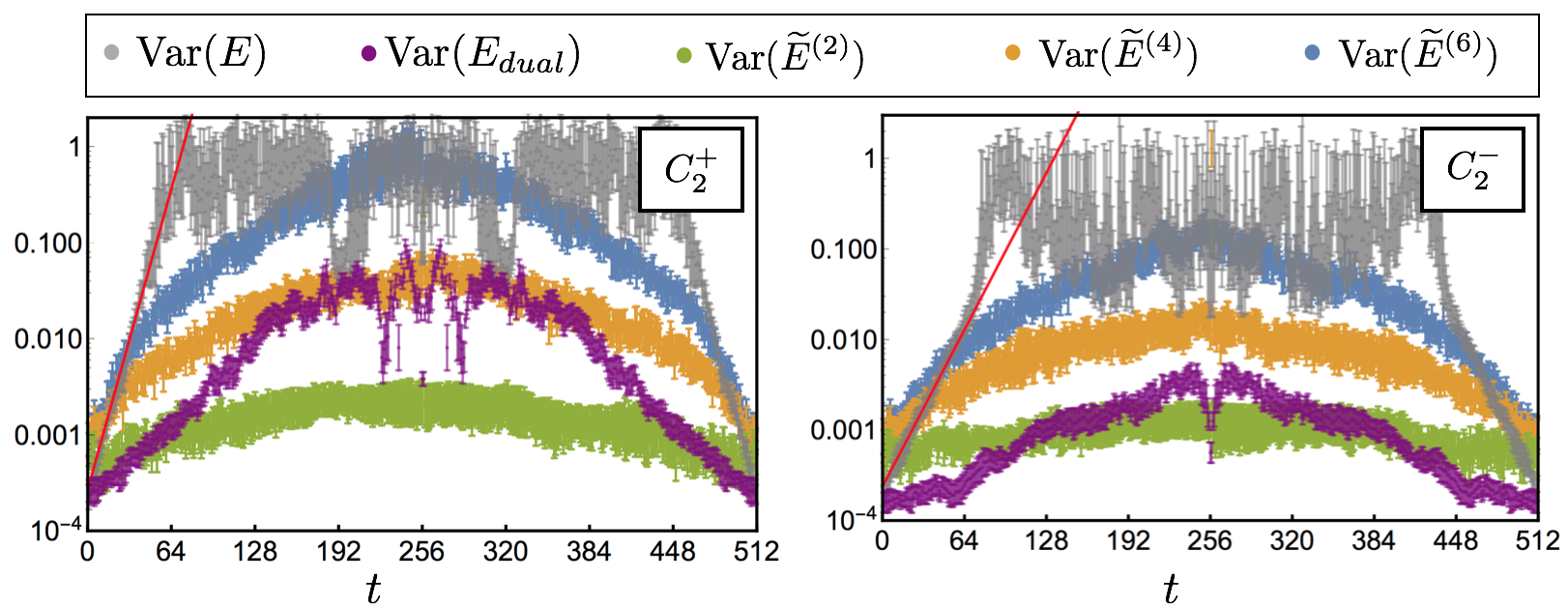}
\caption{Statistical variance in ground-state energy estimate versus correlator time separation for ensembles $C_2^\pm$. The gray overlay plots the variance for the standard effective mass estimator, demonstrating the exponentially decaying StN problem where the variance estimate remains reliable. The red line indicates the theoretical Parisi-Lepage StN decay $\mathcal{N} e^{-Et}$ with $E$ given by the precise dual variables estimate and normalization $\mathcal{N}$ determined by a fit to the first $L/8$ values. The purple points show the variance of the effective mass in the dual lattice variable ensemble and demonstrate exponential variance growth that is significantly less severe than the standard effective mass. The $n_{max} = 2$ estimate with phase unwrapping has even less severe variance growth and becomes more precise than the dual variable estimate at large $t$. Phase unwrapped cumulant effective masses with $n_{max} = 2,\ 4,\ 6$ show variance growth with downward curvature on the logarithmic scale shown that is consistent with polynomial variance growth, though it is difficult to robustly distinguish high-order polynomial from exponential variance growth numerically.}
\label{fig:cho-stn}
\end{figure}

\begin{figure}[p]
  \vspace*{2cm}
  \centering
  \includegraphics[width=\textwidth]{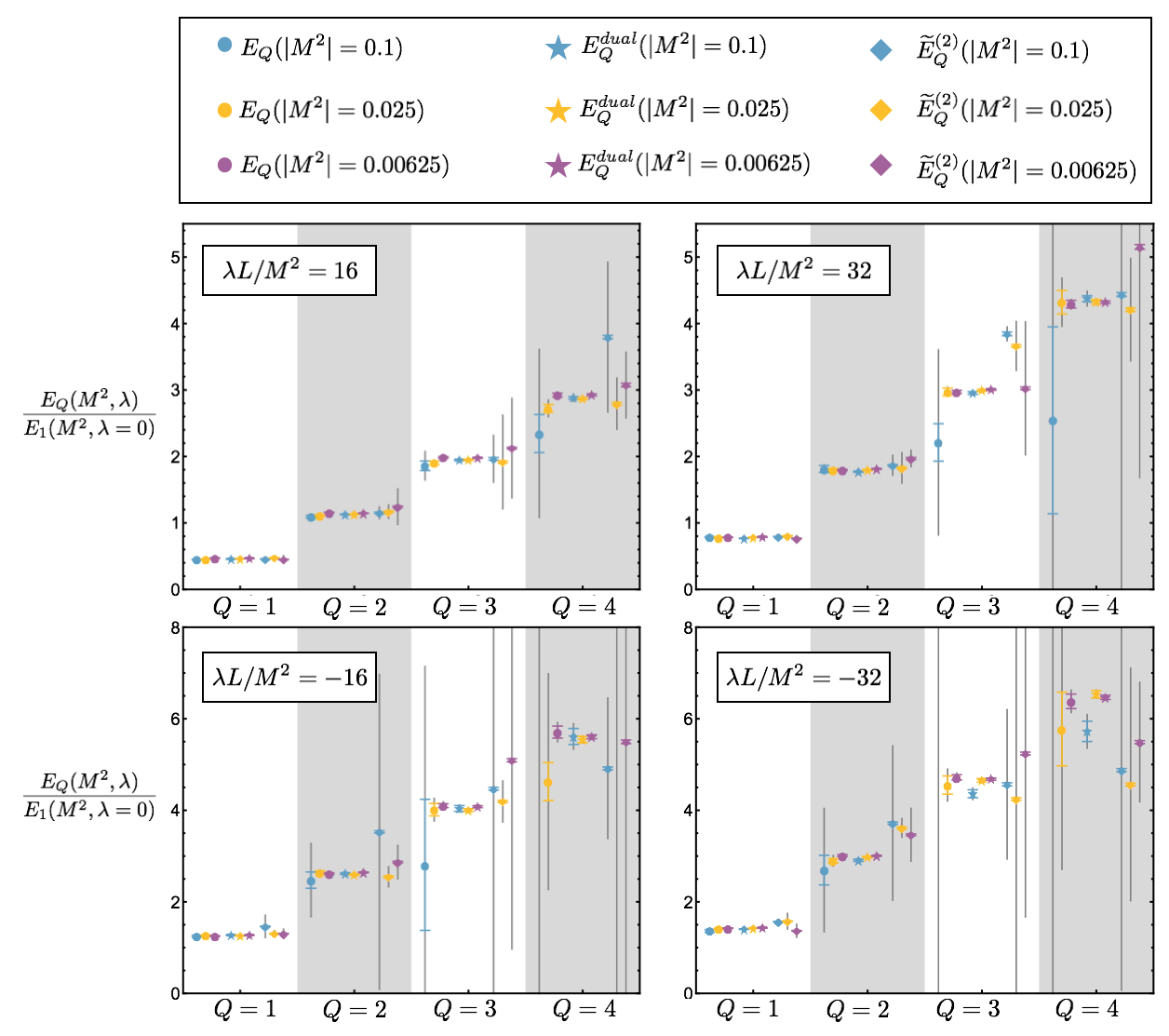}
\caption{Ground-state energies $E_Q \equiv E_{Q,0}$ for charge sectors $Q=1,\dots,4$ in interacting complex scalar field theory with positive $M^2$ (top) and negative $M^2$ (bottom) as well as two choices of
$|\lambda L / M^2| = 16$ (left) and $|\lambda L / M^2| = 32$ (right).
Colored error bars indicate 68\% confidence intervals including statistical errors only, while
thin gray error bars include systematic uncertainties associated with variation of fitting window range and cumulant expansion truncation errors added to statistical errors in quadrature.
Systematic uncertainties associated with fitting window range variation are estimated as one half the
difference between maximum and minimum central values for fit windows shifted by one and two time slices.
Systematic uncertainties associated with truncation errors are estimated by the maximum difference
between the central value of $\widetilde{E}^{(2)}_{Q,0}$ and the central values of $\widetilde{E}^{(4)}_{Q,0}$
and $\widetilde{E}^{(6)}_{Q,0}$.
In several cases the $Q = 4, M^2 = 0.1$ standard estimator does not reliably plateau and systematic errors cover the plot range with no estimate for the central value shown.
The vertical axis shows ratios of interacting energies $E_{Q,0}(\lambda,M,L)$ to the noninteracting scalar boson mass $E$ to facilitate comparison with noninteracting energies $E_{Q,0}(\lambda=0,M,L) = Q E$.
}
\label{fig:cho-spectrum-phi4}
  \vfill
\end{figure}

Empirically, window sizes tuned to reproduce the correlation length in each charge sector $w_Q \sim \xi_Q = 1/E_{Q,0}$ tend to give accurate results for $E_{Q,0}$ at low orders in the cumulant expansion.
Results for the ground-state energies in $Q=1,\dots,4$ charge sectors obtained with optimally tuned Gaussian integration phase unwrapping for the free-field ensemble $C_0$ are summarized in Figs.~\ref{fig:unwrapStN}-\ref{fig:unwrapStNQ}.
It is noteworthy that second-order truncated cumulant expansion energy estimates $\widetilde{E}^{(2)}_{Q,0}$ have negligible StN loss with increasing $t$ at fixed $Q$ and with increasing $Q$ at fixed $t$. 
Close agreement between $\widetilde{E}^{(2)}_{Q,0}$ and the exact results for $E_{Q,0}$ is obtained but requires tuning a smearing parameter in the phase unwrapping algorithm, see Sec.~\ref{sec:unwrap1D} for further discussion.
Higher-order truncations $\widetilde{E}^{(n_{max})}_{Q,0}$ have StN ratios that noticeably decrease with increasing $t$ and with increasing $Q$.
This StN decrease shows less curvature on the log-log scale in Fig.~\ref{fig:unwrapStNQ} than the exponential StN decrease of $E_{Q,0}$, and numerical results are consistent with constant StN at second-order and increasingly high-order polynomial StN degradation at increasingly high cumulant expansion truncation order.

The cumulant expansion tends to converge from above or below depending on whether the smearing scale is tuned to be larger or smaller than the physical correlation length.
A heuristic explanation for these observations is that unwrapping with an overly small smearing scale is overly sensitive to short-distance fluctuations and erroneously adds winding numbers while unwrapping with an overly large smearing scale penalizes diffusive motion away from physically uncorrelated points and leads to underbroadening.
Fig.~\ref{fig:meff-vs-time-pos-neg-mass} demonstrates that this overbroadening or underbroadening is a time-independent feature when estimating the effective mass.
In either case, truncation errors are reduced by going to higher order in the cumulant expansion at the cost of decreased statistical precision.

The StN behavior of phase unwrapped ensembles using optimally tuned smearing parameters is shown in Fig.~\ref{fig:cho-stn}.
There is very little StN degradation in $\widetilde{E}^{(2)}$.
Standard ensemble average correlation functions show exponential StN degradation with the expected $O(e^{-E_{Q,0}t})$ scaling, while dual variable correlation functions show much more mild but likely still exponential StN scaling.
For the largest source/sink separations, the dual estimate precision grows to become worse than the precision of $\widetilde{E}^{(2)}$.
The limiting factors on the accuracy of low-order results in the cumulant expansion extracted with optimally tuned phase unwrapping are the systematic uncertainties associated with truncation errors and phase unwrapping parameter tuning, not statistical precision.
The precision of $\widetilde{E}^{(2)}_{Q,0}$ and systematic truncation errors are both clearly visible in results for the ground-state energies of charge sectors $Q=1,\dots,4$ in interacting complex scalar field MC ensembles in Fig.~\ref{fig:cho-spectrum-phi4}.
Truncation errors are estimated from the maximum difference between the central value of $\widetilde{E}^{(2)}_{Q,0}$ and the
central values of $\widetilde{E}^{(4)}_{Q,0}$ and $\widetilde{E}^{(6)}_{Q,0}$. After including this difference as a systematic uncertainty
added in quadrature with the statistical errors,
$\widetilde{E}^{(2)}_{Q,0}$ results are consistent with precise results from the dual ensembles.
Systematic truncation uncertainties determined in this way are significantly larger than statistical uncertainties.
The combination of large truncation errors in $\widetilde{E}^{(2)}_{Q,0}$ and significant variance growth with increasing $n_{max}$ prevents phase unwrapped results from providing precise and accurate results for the spectrum of $(0+1)D$ complex scalar field theory.

\begin{figure}
\includegraphics[width=.9\textwidth]{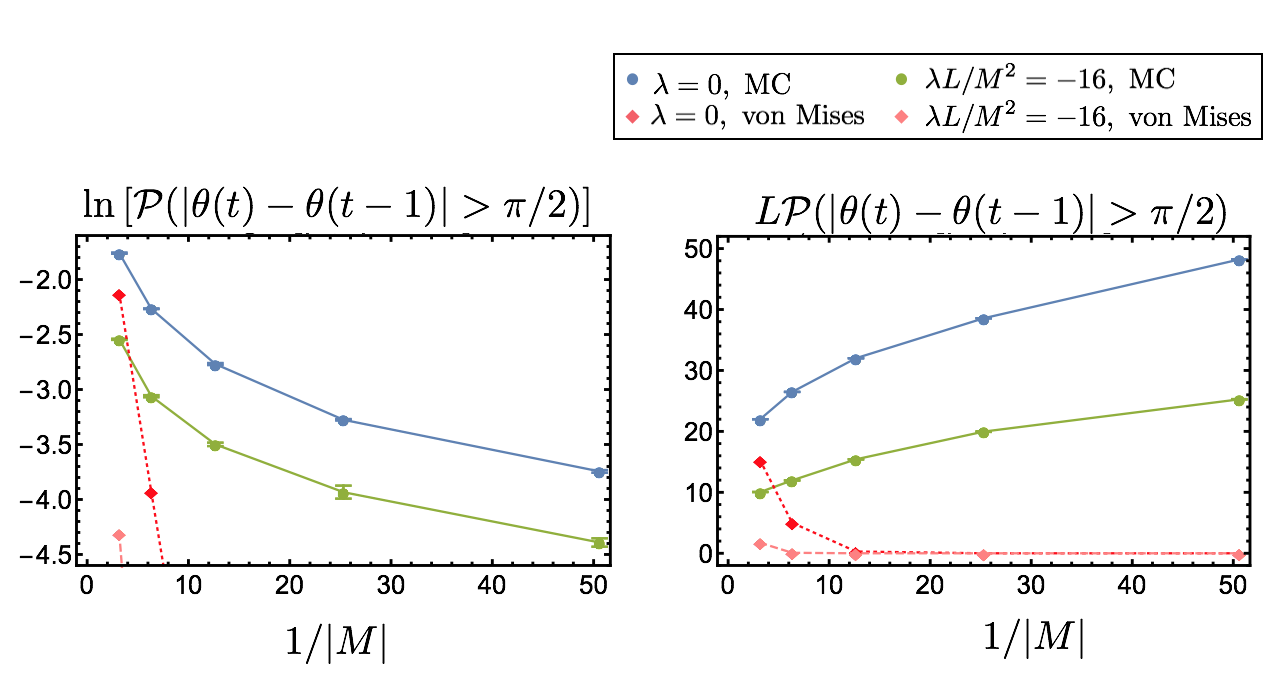}
\caption{The left plot shows the probability of large phase jumps defined by $|\theta(t) - \theta(t-1)| > \pi/2$ for a variety of ensembles. The blue curve shows results for free-field ensembles $A_0$, $B_0$, $C_0$, $D_0$, and $E_0$ as a function of $1/|M|$ and therefore approximately as a function of the correlation length. The green curve shows analogous results for interacting scalar field ensembles $A_1^-$, $B_1^-$, $C_1^-$, $D_1^-$ and $E_1^-$ with $M^2 < 0$ and fixed $-\lambda L/M^2 = 16$ corresponding to fixed coupling strength in units of the $O(\lambda^0)$ tree-level correlation length. Results for phase differences between all nearest-neighbor sites on the lattice are averaged, and error bars on each point indicate $68\%$ confidence intervals calculated using bootstrap techniques. The dotted red (dashed pink) curve shows the predictions of Eq.~\eqref{eq:Panalytic} corresponding to von Mises distributed phase differences with $\kappa \approx 1/(2E)$ calculated for the free (interacting) ensembles. The right plot shows the same probabilities multiplied by the lattice size $L$ to represent the probability that a field configuration will have a large phase jump.
}
\label{fig:phase-jumps}
\end{figure}

As the lattice spacing is taken much smaller than the physical correlation length, phase differences between neighboring lattices become smaller on average.
Under the fixed scalar field magnitude assumption, this probability can be calculated using the von Mises distribution derived exactly  for $\partial_t \theta$ in Eq.~\eqref{eq:VM} to be
\begin{equation}
  \begin{split}
    \mathcal{P}(|\partial_t \theta| > \pi - \varepsilon) = \frac{2}{I_0(\kappa)} \int_{\pi - \varepsilon}^\pi \frac{d \Delta}{2\pi} e^{\kappa \cos(\Delta)}.
  \end{split}\label{eq:Panalytic}
\end{equation}
Under small fluctuation assumptions and neglecting excited-states,  $\kappa \approx 1/(2E) $ as in Eq.~\eqref{eq:kappa} and $\kappa$ therefore becomes large as the correlation length becomes large in lattice units.
The probability in Eq.~\eqref{eq:Panalytic} vanishes rapidly as $\kappa \rightarrow\infty$ with $\varepsilon >0$, and so one may expect the probability of large phase jumps in a MC ensemble to vanish as $M^2 \rightarrow 0$.
However, there is some non-negligible probability that $|\varphi|$ fluctuates to become arbitrarily small even at very small lattice spacing; for example, this occurs due to nearly coincident zero crossings of the real and imaginary parts of $\varphi$ as they fluctuate from one sign to the other as shown in Fig.~\ref{fig:hardunwrapping}.
The distribution of $\partial_t \theta$ in LQFT MC ensembles is given by marginalizing over $\kappa$, and nontrivial correlations between the magnitude and phase could lead to significant departures from the expectations of Eq.~\eqref{eq:Panalytic}.
Such departures are seen in Fig.~\ref{fig:phase-jumps}, where the probability of jumps larger than $\varepsilon = \pi/2$ appears to vanish as $M^2 \rightarrow 0 $ much more slowly than predicted by Eq.~\eqref{eq:Panalytic}.
Similar scaling is found in free-field theory, interacting field theory with $M^2>0$, and somewhat surprisingly also in the $M^2 < 0$ regime where the magnitude typically fluctuates about local minima where $\kappa$ in Eq.~\eqref{eq:Panalytic} is nonzero.
The expected number of large phase jumps per field configuration $L \times \mathcal{P}(|\partial_t \theta|  > \pi/2)$ is empirically observed to grow as $|M^2|$ is decreased and $L$ is increased to hold $|M|L$ fixed, suggesting that there is never a physically relevant regime that is likely to be free of large phase jumps and the phase unwrapping ambiguities associated with them.

The result that the number of large phase jumps per configuration grows faster than the number of sites grows as the lattice spacing is decreased is particularly troubling because of an accumulation-of-errors (or differences) problem arising in $1D$ phase unwrapping.
If there is a link connecting $t_{jump}$ and $t_{jump}+1$ with a large phase difference $|\theta(t_{jump}+1) - \theta(t_{jump})| \gtrapprox \pi$, then different phase unwrapping algorithms tend to assign different winding numbers following $t_{jump}$.
With the forward integration schemes described above, this differing winding number at $t_{jump}$ will lead to differences of $2\pi$ in the phase unwrapped by different unwrapping schemes at all $t \geq t_{jump}$.
This accumulation-of-errors problem means that large phase differences between nearest neighbor lattice sites, which might be considered lattice artifacts, lead to scheme-dependent variation of size $(2\pi)^n$ in contributions of a MC correlation function to the $n$th moment of the unwrapped phase that do not disappear as the lattice spacing is reduced.
Further studies are necessary to understand whether this scaling is an artifact of the nonuniversality of $(0+1)D$ complex scalar field theory or a feature that persists in $1D$ unwrapping of momentum-projected correlation functions of renormalizable LQFTs.

While these results on the prevalence of large phase jumps and accumulation of errors in phase unwrapping suggest a pessimistic outlook for $(0+1)D$ complex scalar field theory, other applications of phase unwrapping provide encouraging results demonstrating that the $1D$ accumulation of errors problem becomes more tractable in higher dimensions.
It was realized in the 1980s that accumulation of phase unwrapping errors along a $1D$ integration path is a generic problem in the presence of undersampling~\cite{Itoh:1982} but can be avoided in alternative algorithms for $2D$ phase unwrapping~\cite{Goldstein:1988,Huntley:1989}.
The basic source of greater robustness in higher dimensions is that in $1D$ only one integration path\footnote{It is not expected that additional information from phase unwrapping integration paths that wind around the circle of a finite volume can be used to resolve the accumulation of errors issue.} can be used to connect two points $t$ and $t^\prime$, while in two- and higher-dimensions multiple paths can be used to connect the same points.
Assuming that $\Theta(x,y) = \text{arg} [G(x,y)]$ where $G(x,y)$ is an analytic function, $1D$ unwrapping  will provide identical results for phase integration from $y$ to $x$ that do not depend on the choice of $1D$ integration contour, see e.g. Ref~\cite{Ghiglia:98}.
Under this analyticity assumption, path dependence that arises in numerical data must be the result of numerical noise and sampling several $1D$ unwrapping paths adds error correction through redundancy. 
Applications in 3D have been found to be even more robust to noise than applications in $2D$, suggesting that phase unwrapping generically becomes more robust as the number of dimensions is increased~\cite{Huntley:01,Hooper:07,Abdul-Rahman:09}.
A simple argument supporting this idea is that phase unwrapping makes smoothness assumptions informed by nearest-neighbor phases, and as the number of dimensions increases the number of nearest neighbors that can be used to inform a phase unwrapping algorithm also increases.

Successful applications of numerically robust phase unwrapping algorithms in higher dimensions crucially rest on the assumption that wrapped phases are discrete samples of the complex logarithm of an underlying analytic function.
In this case
noise may locally
produce regions in which phase unwrapping along differing paths
produces results that differ by multiples of $2\pi$, but unwrapping
along paths avoiding these regions is guaranteed by the underlying analyticity to produce identical results.
If the wrapped phase is sampled with sufficiently high resolution and low noise, then the density of points with large phase jumps leading to unwrapping ambiguities is guaranteed to be vanishingly small.
Field configurations in LQFTs are not analytic and are not expected to approach smooth or even continuous functions as the continuum limit is approached.
Instead, field configurations may approach distributions including isolated singularities that will lead to nonlocal unwrapping ambiguities.
It is not clear without further studies of multidimensional LQFTs whether appropriately smeared configurations calculated on finely discretized lattices will be smooth enough for phase unwrapping algorithms to determine unwrapped phases without ambiguities arising from large phase jumps.

Even if large phase jumps are unavoidable in multidimensional LQFTs, more robust multidimensional phase unwrapping algorithms can still be used to avoid the $1D$ unwrapping accumulation of errors problem encountered here.
By applying multidimensional phase unwrapping algorithms to correlation functions in coordinate space, $2\pi$ ambiguities from large phase jumps leading to the $1D$ accumulation of errors problem could be localized to isolated neighborhoods of spacetime.
This might improve the convergence of the cumulant expansion and reduce StN degradation at higher orders.
Precise but approximate results at low orders in the cumulant expansion could also be used as starting points for subsequent calculations of differences between exact and approximate correlation functions that might be more efficient than calculations of exact correlation functions alone. 

\section{Conclusions}\label{sec:conclusions}

In $(0+1)D$ complex scalar field theory, phase fluctuations distinguish correlation functions in $Q\neq 0$ charge sectors from vacuum sector correlation functions.
These phase fluctuations result in sign problems for the path integrals representing correlation functions, even though the vacuum sector partition function is positive-definite.
A method for avoiding $(0+1)D$ scalar field sign problems is introduced that relies on numerically integrating time series of phase differences at a range of source/sink separation using phase unwrapping techniques developed for signal processing and a variety of engineering applications.
The nonzero moments of the unwrapped phase distribution can be computed with positive-definite path integrals without sign problems.
A cumulant expansion involving moments of correlation function log-magnitudes and unwrapped phases can be used to reproduce the spectrum of $(0+1)D$ complex scalar field theory.
The numerical results presented here include large systematic truncation errors at low orders in the cumulant expansion and decreased precision as well as the reemergence of a mild StN problem at higher orders.
It is argued that the large truncation errors arise from isolated large phase jumps that lead to errors in the $n$th moment proportional to $(2\pi)^n$ at all subsequent times.
This accumulation-of-errors problem makes results using a cumulant expansion of the unwrapped phase numerically sensitive to the presence of large phase jumps.
Numerical MC studies suggest that in $(0+1)D$ scalar field theory the probability of  having one or more large phase jumps per lattice extent grows as the $M^2\rightarrow 0$ limit is taken to increase the correlation length, and this accumulation-of-errors problem leads to large systematic errors even at very fine lattice spacing.
This may be due to the non-renormalizability of $(0+1)D$ scalar field theory and these investigations should be extended to renormalizable field theories to better understand this issue.
The appearance of heavy-tailed phase derivative distributions in free-field theory as well as in LQCD baryon correlation functions~\cite{Wagman:2016bam} suggests that these problematic large phase jumps are present in physically relevant LQFTs and are possibly generic features of correlation functions with phase fluctuations.
If heavy-tailed phase differences are generic features of LQFT, then high moments of the unwrapped phase sensitive to the tails of the distribution may be noisy and convergence of the cumulant expansion may be slow.
Leading-order cumulant expansion results using appropriately tuned phase unwrapping algorithms provide precise approximations to correlation functions that avoid sign or StN problems, but robust applications of phase unwrapping in multidimensional LQFTs will require a solution to the $1D$ accumulation of errors problem and perhaps alternative methods of including corrections from noisy higher-order terms.

\vspace*{5mm}

{\bf Acknowledgments:}
	We would like to thank Adam Bene Watts, Dorota Grabowska, David Kaplan, Christopher Monahan, Andrew Pochinsky, Martin Savage, Phiala Shanahan, and Daniel Trewartha for helpful discussions.
         This work was partially supported by the U. S. Department of Energy through Early Career Research Award No. de-sc0010495 and Grant No. de-sc0011090 and by the SciDAC4 Grant No. de-sc0018121.
        MLW was supported by a MIT Pappalardo Fellowship.

%


\bibliography{refs}

\vfill
\pagebreak

\appendix

\section{Dual Lattice Variable Phase Integration}\label{app:dual}

The action for a $(0+1)D$ complex scalar field with an arbitrary $U(1)$ invariant, spacetime translation invariant potential energy function $V(|\varphi|)$ can be decomposed into magnitude and phase contributions as
\begin{equation}
  \begin{split}
    S(\varphi) &= \sum_{t=0}^{L - 1} \left\lbrace |\varphi(t)|\left[ -|\varphi(t-1)|e^{i\theta(t) - i\theta(t-1)} +  2|\varphi(t)| - |\varphi(t+1)|e^{i\theta(t) - i\theta(t+1)}  \right] + V(|\varphi(t)|) \right\rbrace \\
    &= \sum_{t=0}^{L - 1} \left\lbrace 2|\varphi(t)|^2 + V(|\varphi(t)|) - \kappa(t)\cos(\theta(t) - \theta(t-1)) \right\rbrace
  \end{split}\label{eqA:actiondecomp}
\end{equation}
where translation invariance has been used to shift field arguments. 
The partition function for the interacting theory can be similarly decomposed as,
\begin{equation}
  \begin{split}
    Z &= \int_0^\infty \prod_{t= 0}^{L-1} \left[ d|\varphi(t)|\; |\varphi(t)| \; e^{-2|\varphi(t)|^2 -  V(|\varphi(t)|)} \right] \int_{-\pi}^\pi \prod_{t= 0}^{L-1} \left[ \frac{1}{\pi }d \theta(t) \; e^{\kappa(t)\cos(\theta(t) - \theta(t-1))} \right]
  \end{split}\label{eqA:Zdecomp}
\end{equation}
The phase integral can be evaluated analytically be introducing dual lattice variables representing the differences between phases at adjacent lattice sites,
\begin{equation}
  \begin{split}
    \Delta(t) &\equiv \theta(t) - \theta(t-1).
  \end{split}\label{eq:Detladef}
\end{equation}
This transformation is related to dual lattice variable methods that have a long history in lattice gauge theory~\cite{Ukawa:1979yv} and  can be viewed as a $1D$ analog of the $O(N)$ model dual lattice variable transformation introduced in Ref.~\cite{Endres:2006xu}.
To simplify the change of variables, we first use $2\pi$-periodicity to rotate the integration domains for the sequence of $\theta(t)$ integrals,
\begin{equation}
\int_{-\pi}^\pi d\theta(0) \prod_{t=1}^{L-1} \sq{\int_{-\pi}^\pi d\theta(t)}
\qquad \rightarrow \qquad
\int_{-\pi}^\pi d\theta(0) \prod_{t=1}^{L-1} \sq{\int_{\theta(t-1)-\pi}^{\theta(t-1)+\pi} d\theta(t)}.
\end{equation}
We then change variables from $\theta(t)$ to $\Delta(t)$ for all $t \geq 1$, with trivial Jacobian,
\begin{equation}
  \begin{split}
    \det \left(\frac{\partial(\theta(0),\Delta(1),\dots,\Delta(L - 1))}{\partial(\theta(0),\theta(1),\dots,\theta(L-1))}\right)  = \det \begin{pmatrix} 1&0&0&0&\cdots \\ -1&1&0&0&\cdots \\ 0&-1&1&0& \cdots \\ \vdots&\vdots&\vdots&\vdots&\ddots \end{pmatrix} = 1
  \end{split}.
\label{eq:Jacobian}
\end{equation}
Having first rotated the integration ranges, we are left with simple integration bounds for the newly-introduced dual variables,
\begin{equation}
\int_{-\pi}^\pi d\theta(0) \prod_{t=1}^{L-1} \sq{\int_{\theta(t-1)-\pi}^{\theta(t-1)+\pi} d\theta(t)}
= \int_{-\pi}^\pi d\theta(0) \prod_{t=1}^{L-1}
\sq{\int_{-\pi}^{\pi} d\Delta(t)}.
\end{equation}
To completely decouple the integrals, we would like to
introduce the final variable $\Delta(0) = \theta(0) - \theta(L-1)$.
The presence of PBCs slightly complicates this transformation by introducing the constraint
\begin{equation}
\Delta(0) \equiv \theta(0) - \theta(L-1)  - \sum_{t=1}^{L-1} \Delta(t)
\end{equation}
which can be implemented by a $\delta$-function
\begin{equation}
1 = \int_{-\infty}^{\infty} d\Delta(0) \delta \paren{\sum_{t=0}^{L-1} \Delta(t)}.
\end{equation}
The compact nature of the phase variables requires this final integral to run over all reals to satisfy PBCs in all winding number sectors. To treat all dual variables on equal footing, we instead handle this sum over winding number sectors directly,
\begin{equation}
\int_{-\infty}^{\infty} d\Delta(0) \delta \paren{\sum_{t=0}^{L-1} \Delta(t)}
= \sum_{w \in \mathbb{Z}} \int_{-\pi}^{\pi} d\Delta(0) \delta \paren{\sum_{t=0}^{L-1} \Delta(t) + 2 \pi w}.
\end{equation}

This path integral change of variables from $\theta$ to $\Delta$ turns the phase integrals turns into a product of decoupled integrals.
This representation allows the integral over phases to be explicitly evaluated as
\begin{equation}
  \begin{split}
    Z &=  \int_0^\infty \prod_{t= 0}^{L-1} \left[ d|\varphi(t)|\; |\varphi(t)| \; e^{-2|\varphi(t)|^2 - V(|\varphi(t)|)} \right] \\
    &\hspace{20pt} \times \int_{-\pi}^\pi \frac{1}{\pi}d\theta(0) \int_{-\pi}^{\pi} \prod_{t= 0}^{L-1} \left[ \frac{1}{\pi} d \Delta(t) \; e^{\kappa(t)\cos(\Delta(t))} \right] \sum_{w \in \mathbb{Z}} \delta\left( \sum_{t=0}^{L-1} \Delta(t) + 2\pi w \right)\\
    &=  2 \int_0^\infty \prod_{t= 0}^{L-1} \left[ d|\varphi(t)|\; |\varphi(t)| \; e^{-2|\varphi(t)|^2 - V(|\varphi(t)|)} \right] \\
    &\hspace{20pt} \times  \int_{-\infty}^\infty dq \; \sum_{w \in \mathbb{Z}} e^{2\pi i q w} \int_{-\pi}^{\pi} \prod_{t= 0}^{L-1} \left[ \frac{1}{\pi} d \Delta(t) \; e^{\kappa(t)\cos(\Delta(t)) + i q \Delta(t) } \right] \\
    &= 2 \sum_{q\in\mathbb{Z}}    \int_0^\infty \prod_{t= 0}^{L-1} \left[ d|\varphi(t)|\; |\varphi(t)| \; e^{-2|\varphi(t)|^2 - V(|\varphi(t)|)} \; 2\;  I_{\abs{q}}\left( \kappa(t) \right) \right]
  \end{split}\label{eq:CHOZdecomp}
\end{equation}
where we have used an integral representation to factorize the $\delta$-function
and explicitly integrated the $\Delta(t)$ to produce modified Bessel functions
of the first kind, $I_{|q|}(z)$. The remaining integrals over $|\varphi(t)|$ cannot be evaluated in closed form for arbitrary $V(|\varphi|)$; however, since $I_{|q|}(z) \geq 0$ for $z\geq 0$, the form of the partition function given in the final
line of Eq.~\eqref{eq:CHOZdecomp} defines a
positive-definite probability density for $q, \varphi$:
\begin{equation}
\begin{split}
    1 &= \sum_{q\in\mathbb{Z}} \int \mathcal{D}|\varphi|\; \mathcal{P}(q,|\varphi|) \\
    &\equiv \frac{2}{Z} \sum_{q\in\mathbb{Z}}
    \int 
    \prod_{t=0}^{L-1} d|\varphi(t)| \curly{ |\varphi(t)| e^{-2|\varphi(t)|^2 - V(|\varphi(t)|)} \;
    2\; I_{\abs{q}}\left( \kappa(t) \right)
    }.
\end{split}
\label{eq:CHOP1}
\end{equation}
The probability distribution $\mathcal{P}(q,|\varphi|)$ is a positive-definite, normalizable function that can be used for MC sampling of $|\varphi|$ and $q$ as described below.

The integrals over phase variables can similarly be performed analytically for scalar field correlation functions. The general correlation function can first be written in terms of the new dual variables,
\begin{equation}
  \mathcal{O}_{Q,2P}(t)\mathcal{O}_{Q,2P}^*(0) = |\varphi(t)|^{|Q|+2P}|\varphi(0)|^{|Q|+2P} e^{i Q \sum_{t^\prime = 1}^t \Delta(t^\prime)}.
\end{equation}
Inserting this observable into the path integration and explicitly evaluating gives
\begin{equation}
  \begin{split}
    G_{Q,2P}(t) &= \frac{1}{Z} \int_0^\infty \prod_{t= 0}^{L-1} \left[ d|\varphi(t)|\; |\varphi(t)| \; e^{-2|\varphi(t)|^2 -  V(|\varphi(t)|)} \right] |\varphi(t)|^{|Q|+2P}|\varphi(0)|^{|Q|+2P} \\
    &\hspace{20pt} \times \int_{-\pi}^\pi \frac{1}{\pi } d\theta(0) \int_{-\pi}^{\pi} \prod_{t= 0}^{L-1} \left[ \frac{1}{\pi} d \Delta(t) \; e^{\kappa(t)\cos(\Delta(t))} \right] \sum_{w \in \mathbb{Z}} \delta\left( \sum_{t=0}^{L-1} \Delta(t) + 2\pi w \right) e^{i Q \sum_{t^\prime = 1}^t \Delta(t^\prime)} \\
    &= \frac{2}{Z} \int_0^\infty \prod_{t= 0}^{L-1} \left[ d|\varphi(t)|\; |\varphi(t)| \; e^{-2|\varphi(t)|^2 -  V(|\varphi(t)|)} \right] |\varphi(t)|^{|Q|+2P}|\varphi(0)|^{|Q|+2P} \\
    &\hspace{20pt} \times \int_{-\infty}^\infty dq \; \sum_{w \in \mathbb{Z}} e^{2\pi i q w} \int_{-\pi}^{\pi} \prod_{t= 0}^{L-1} \left[ \frac{1}{\pi} d \Delta(t) \; e^{\kappa(t)\cos(\Delta(t)) + i q \Delta(t)  } \right]e^{i Q \sum_{t^\prime = 1}^t \Delta(t^\prime)} \\
    &= \sum_{q\in\mathbb{Z}}  \int \mathcal{D}|\varphi|\; \mathcal{P}(q,|\varphi|)
    \; |\varphi(t)|^{|Q|+2P}|\varphi(0)|^{|Q|+2P}
    \prod_{t^\prime=1}^{t}\left[ \frac{I_{\abs{Q+q}}\left( 2|\varphi(t^\prime)||\varphi(t^\prime-1)| \right)}{I_{\abs{q}}\left( 2|\varphi(t^\prime)||\varphi(t^\prime-1)| \right)} \right].
  \end{split}\label{eq:CHOpropdecomp}
\end{equation}
The integrand is again positive-definite and can be interpreted as an integration measure without a sign problem, in contrast to Eq.~\eqref{eq:Gmagphase}.
It is also possible to calculate correlation functions by MC sampling field configurations according to $\mathcal{P}(q,|\varphi|)$ and then including the ensemble average of the product of Bessel functions in Eq.~\eqref{eq:CHOpropdecomp} as a reweighting factor.

Care must be taken in defining MC updates of $q$.
For instance, a Metropolis scheme in which updates $q\rightarrow q^\prime$ are proposed and then accepted with probability  $\text{min}\left( 1,\; e^{-S_{\eff} (q,|\varphi|) + S_{\eff}(q^\prime,|\varphi|)}  \right)$ with
\begin{equation}
  \begin{split}
  S_{\eff} (q,|\varphi|) = \sum_{t=0}^{L-1} 2|\varphi(t)|^2 + V(|\varphi(t)|) - \ln|\varphi(t)| - \ln\left[ I_{|q|}(\kappa(t))\right],
  \end{split}\label{eq:Sqdef}
\end{equation}
will experience ``topological freezing''. The minimum action $q=0$ sector is sampled effectively but $q \neq 0$ sectors make $O(e^{-L})$ suppressed contributions  to the partition function, because they involve products of $L$ small factors $\prod_{t=0}^{L-1} \frac{I_{|q|}}{I_0}$, so are scarcely or never present in a finite-$N$ MC ensemble.
This is problematic for MC calculations of correlation functions, because $q \neq 0$ contributions can provide significant contributions to correlation functions with nonzero $U(1)$ charge.
Considering Eq.~\eqref{eq:CHOpropdecomp} for the case of the scalar field propagator $G = G_{1,0}$, the $q=-1$ sector makes a contribution at large $t \sim L$ involving the exponentially large product $\prod_{t^\prime = 1}^L \frac{I_0}{I_1} \sim e^{L}$.
This situation of exponentially rare MC configurations making exponentially large contributions to observables suggests this MC scheme has an overlap problem where the distribution being importance sampled has poor overlap with the region of configuration space making dominant contributions to observables of interest.

Figure~\ref{fig:cho-standard-autocorrs} plots the integrated
autocorrelation time for $\ob = \braket{|\varphi|^2}$ for all
standard ensembles used in this work.
Figure~\ref{fig:cho-pi-autocorrs} analogously plots the 
integrated autocorrelation time for the dual variable ensembles. 
Autocorrelation times between corresponding standard and dual
ensembles are similar. There is significant autocorrelation only
on the finest lattice ($M^2 = 0.00625$), and as such all analyses
of methods we introduce applied binning with bin size $\geq 10$
on the finest lattice to more accurately estimate errors in the presence
of this autocorrelation. Larger bin sizes were used in some cases to
further improve the $\chi^2$/DoF estimates.
Tables~\ref{tbl:free-fit-data}-\ref{tbl:l16-neg-fit-data} 
indicate the bin and  window sizes used for all fits to energy
levels required to produce the various spectrum plots in
this work.

Instead, MC sampling over $q$ can be replaced by explicit summation over a finite subset of winding numbers that make dominant contributions to particular observables.
MC sampling can be performed using the modified probability distribution
\begin{equation}
  \begin{split}
    \mathcal{D}|\varphi| \mathcal{P}_0(|\varphi|) = \frac{2}{Z_0}   \prod_{t=0}^{L-1} \left[  |\varphi(t)| d|\varphi(t)| e^{-2|\varphi(t)|^2 - V(|\varphi(t)|)} \;  2\;  I_{0}\left( \kappa(t) \right) \right],
  \end{split}\label{eqA:Pprimedef}
\end{equation}
where $Z_0$ represents the $q=0$ contribution to the partition function and is defined to ensure that $\int \mathcal{D}|\varphi| \mathcal{P}_0(|\varphi|) = 1$.
Importance sampling with respect to $\mathcal{P}_0(|\varphi|)$ can be performed with local Metropolis update steps of $|\varphi(t)|$ and an accept-reject probability determined by changes in the action $S_{\eff}(q=0,|\varphi|)$.
Correlation functions $G_{Q,2P}$ can be calculated from field configurations importance sampled according to $\mathcal{P}_0(|\varphi|)$ by explicit summation over all $q$,
\begin{equation} \begin{aligned}
    G_{Q,2P}(t) &= \sum_{q \in \mathbb{Z}} \int \mathcal{D}|\varphi| \mathcal{P}_0(|\varphi|) \;  |\varphi(t)|^{|Q|+2P} |\varphi(0)|^{|Q| + 2P} \\
    &\hspace{20pt} \times \prod_{t^\prime = 1}^t \frac{I_{|Q + q|}(\kappa(t))}{I_0(\kappa(t))}  \prod_{t^\prime = t+1}^L \frac{I_{|q|}(\kappa(t))}{I_0(\kappa(t))}.
  \end{aligned}\end{equation}
Given a finite MC ensemble of scalar field magnitude $|\varphi_i|$, $i=1,\dots,N$ sampled from  Eq.~\eqref{eqA:Pprimedef}, correlation functions can be estimated from the corresponding ensemble averages
\begin{equation}\begin{aligned}
    \overline{G}_{Q,2P}^{dual}(t) &= \frac{1}{N} \sum_{i=1}^{N} 
    \sum_{q \in \mathbb{Z}} \Bigg\{
    |\varphi_i(t)|^{|Q|+2P} |\varphi_i(0)|^{|Q| + 2P} \\
    &\hspace{20pt} \times \prod_{t^\prime = 1}^t \frac{I_{|Q + q|}(\kappa_i(t))}{I_0(\kappa_i(t))}  \prod_{t^\prime = t+1}^L \frac{I_{|q|}(\kappa_i(t))}{I_0(\kappa_i(t))} \Bigg\},
  \end{aligned}\label{eqA:GQ2Pdual}
\end{equation}
where $\overline{G}_{Q,2P}^{dual}$ denotes ensemble estimates of $G_{Q,2P}$ in this dual variables approach.
Significant contributions to Eq.~\eqref{eqA:GQ2Pdual} arise for $q = -Q,\dots,+Q$ but topological charge sectors with $|q| > |Q|$ make subdominant contributions that rapidly converge to zero and allow the sum over topological charge sector to be truncated in practical calculations.

\section{Unwrapped Phase Definition}\label{app:unwrapping}

The complex exponential function is not injective because $e^{z} = e^{z + 2\pi i}$.
The complex logarithm function, intuitively describing the inverse of the complex exponential function, therefore requires care to define.
The principal value of the complex logarithm of an analytic function $f$ of a complex variable $z$ defined such that $-\pi < \text{Im}\ln f(z) \leq \pi$  is given by
\begin{equation}
  \begin{split}
    \ln f(z) &\equiv \ln(|f(z)|e^{i\theta(z)}) \equiv \ln |f(z)| + i \text{arg}(e^{i\theta(z)}) \equiv \ln |f(z)| + i \theta(t), 
  \end{split}\label{eq:logdef}
\end{equation}
where $-\pi < \theta \leq \pi$.
The principal-valued logarithm is not a continuous function on the punctured complex plane $\mathbb{C} \setminus \{0\}$ because
\begin{equation}
  \begin{split}
    \lim_{\theta\rightarrow \pi^-}\text{arg}(e^{i\theta})= \pi \neq -\pi = \lim_{\theta\rightarrow \pi^+}\text{arg}(e^{i\theta}) .
  \end{split}\label{eq:logdisc}
\end{equation}
The standard method of defining a continuous logarithm function involves analytic continuation and construction of a Riemann surface including infinitely many copies of the punctured complex plane glued together at branch cuts $\theta = \pi + 2\pi \nu$.
Instead, one can assume the existence of a single-valued, analytic function $\text{Ln}: \mathbb{C}\setminus \{0\} \rightarrow \mathbb{C}$ satisfying
\begin{equation}
  \begin{split}
    \frac{d}{dz}\text{Ln}(f(z)) \equiv \frac{1}{f(z)}\frac{df}{dz}, \hspace{20pt} \text{Ln}(1) \equiv 0.
  \end{split}\label{eq:Lnderivdef}
\end{equation}
Applying the fundamental theorem of calculus for integration along a curve $\gamma : [a,b] \rightarrow \mathbb{C}\setminus \{0\}$ gives
\begin{equation}
  \begin{split}
    \text{Ln}(f(b)) - \text{Ln}(f(a)) = \int_{\gamma(z)} \frac{d}{dz}\text{Ln}(f(z)) dz = \int_{\gamma(z)} \frac{\frac{df}{dz}}{f(z)}.
  \end{split}\label{eq:Lndef}
\end{equation}
Note that this construction implicitly depends on the function $f$
under consideration.
To gain intuition for Eq.~\eqref{eq:Lndef} first consider the case $f(z) = z$.
By analyticity of $\text{Ln}(z)$ in the domain $\mathbb{C}\setminus \{0\}$, any line integral along a  curve $\gamma(z)$  that is assumed to be in the trivial homotopy class of $\mathbb{C}\setminus \{0\}$  can be deformed into an integral along a piecewise continuous path $\gamma = \gamma_R \cup \gamma_\theta$ composed of a purely radial path $\gamma_R : |a|e^{i\text{arg}(a)} \rightarrow |b|e^{i\text{arg}(a)}$ and a purely angular path $\gamma_\theta : |b|e^{i\text{arg}(a)} \rightarrow |b|e^{i\text{arg}(b)}$.
Since a radial path $\gamma_R$ does not cross any branch cuts of $\ln(z)$, the integral along $\gamma_R$ can be evaluated as
\begin{equation}
  \begin{split}
    \int_{\gamma_R(z)} \frac{dz}{z} = \int_{|a|e^{i\text{arg}(a)}}^{|b|e^{i\text{arg}(a)}} \frac{dz}{z} = \ln|a| - \ln|b|.
  \end{split}\label{eq:Lnreal}
\end{equation}
If the angular segment of the path crosses the branch cut of $\ln(z)$ placed at $z = \pi$, then a further path deformation should be made so that the angular path stops a distance $\varepsilon$ before the branch cut, runs radially along the branch cut from radius $|b|$ to the origin, encircles the origin in a circle of radius $\varepsilon$, runs radially back to radius $|b|$ along the opposite side of the branch cut, and then continues along the remainder of the angular path.
In the $\varepsilon\rightarrow 0$ limit the contributions from the radial paths vanish and contribution from the path encircling the origin can be evaluated by Cauchy's theorem, 
\begin{equation}
  \begin{split}
    \int_{\gamma_\theta(z)} \frac{dz}{z} = \fint_{|b|e^{i\text{arg}(a)}}^{|b|e^{i\text{arg}(b)}} \frac{dz}{z} + \nu \oint \frac{dz}{z} = 2\pi i \nu +  i \fint_{\text{arg}(a)}^{\text{arg}(b)} d\theta = 2\pi i \nu + i (\text{arg}(b) - \text{arg}(a)),
  \end{split}\label{eq:Lnim}
\end{equation}
where $\nu = +1$ if $\gamma$ includes a positive branch cut crossing (since $\gamma$ is simply connected this occurs if and only if $a$ is in quadrant II and $b$ is in quadrant III), $\nu= -1$ if $\gamma$ includes a negative branch cut crossing (if $b$ is in quadrant II and $a$ is in quadrant III), and zero otherwise.
Choosing instead $a = 1$ such that $\text{Ln}(a) = 0$, $b = z = |z|e^{i\theta}$ arbitrary, and $\gamma$ a counter-clockwise path that does not encircle the origin gives
\begin{equation}
  \begin{split}
    \text{Ln}(z) &= \int_{\gamma_R(z)} \frac{dz}{z} + \int_{\gamma_\theta(z)} \frac{dz}{z} = \int_1^{|z|} \frac{dz}{z} + \fint_{|z|}^{|z|e^{i\text{arg}(z)}} \frac{dz}{z} + \nu \oint \frac{dz}{z} \\
    &= \ln |z| + \text{arg}(z) + 2\pi i \nu,
  \end{split}\label{eq:Lnln}
\end{equation}
where $\nu = +1$ if $\text{Im}z < 0$ and $\nu=0$ otherwise.
The imaginary part of this expression defines a function that agrees with $\text{arg}(z)$ modulo $2\pi$,
\begin{equation}
  \begin{split}
    \text{Arg}(z) \equiv \text{Im}\text{Ln}(z) = \text{arg}(z) + 2\pi i \nu.
  \end{split}\label{eqA:Argzdef}
\end{equation}
For an arbitrary analytic function $f(z)$, $\text{Arg}(f(z))$ can be computed by integrating $\text{Im}\frac{d}{dz}\text{Ln}(f(z))$ along the angular piece of $\gamma: [0, z] \rightarrow \mathbb{C} \setminus \{0\}$,
\begin{equation}
  \begin{split}
    \text{Arg}(f(|z|e^{i\theta})) = \int_0^\theta \text{Im}\left[ \frac{\frac{df}{d\theta^\prime} }{f(|z|e^{i\theta^\prime})} \right]d\theta^\prime.
  \end{split}\label{eqA:Argdef}
\end{equation}
Assuming $f(|z|e^{i\theta^\prime}) \neq 0$, the integral in Eq.~\eqref{eqA:Argdef} is well-defined and $\text{Arg}(f(|z|e^{i\theta}))$ is continuous and analytic provided that $f$ is continuous and analytic~\cite{Kitahara:2015}.

\section{Monte Carlo Ensembles}\label{app:MC}

\begin{figure}
\centering
\begin{minipage}{.53\textwidth}
\centering
\includegraphics[width=\linewidth]{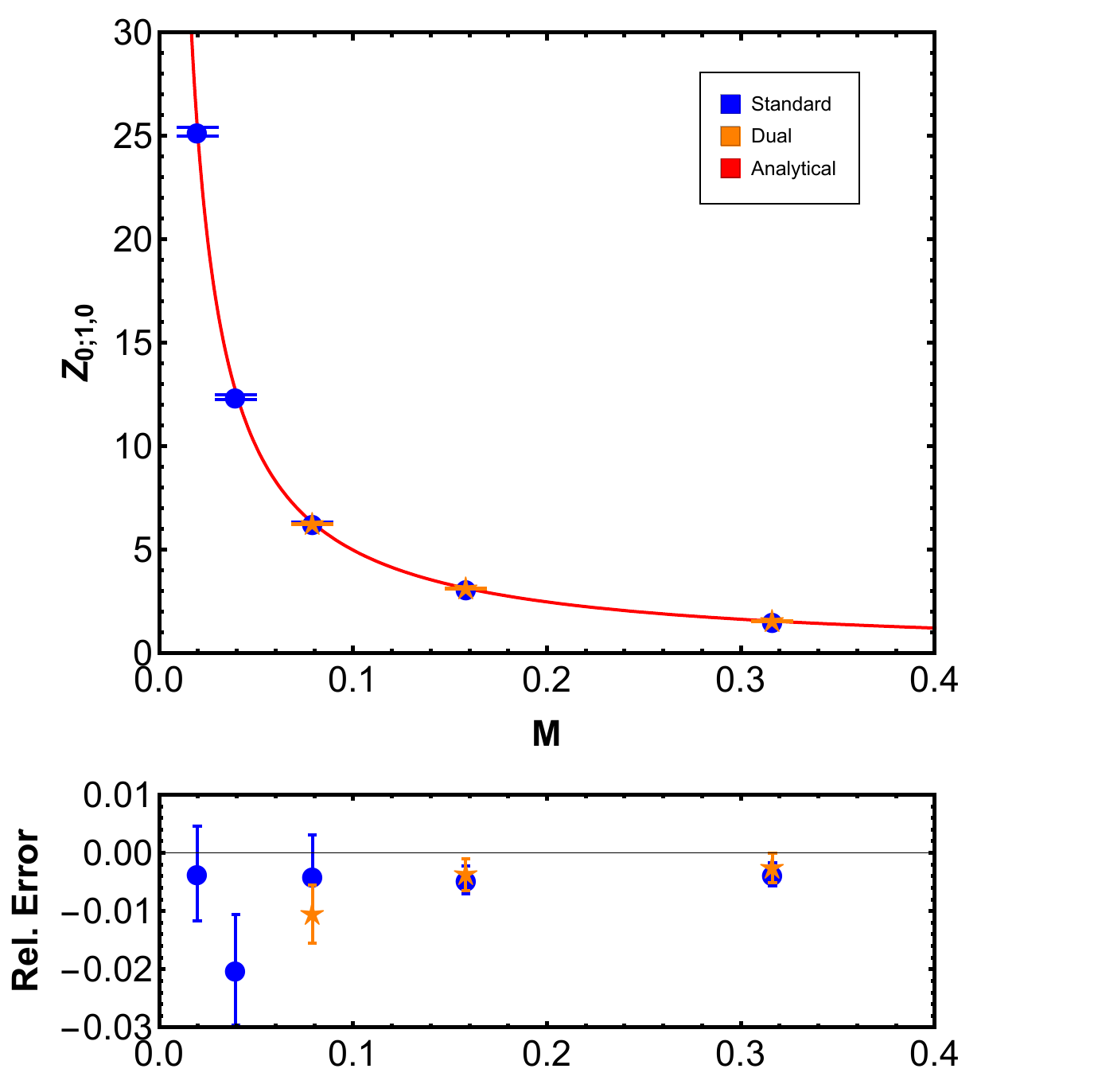}
\end{minipage}
\begin{minipage}{.42\textwidth}
\centering
\includegraphics[width=\linewidth]{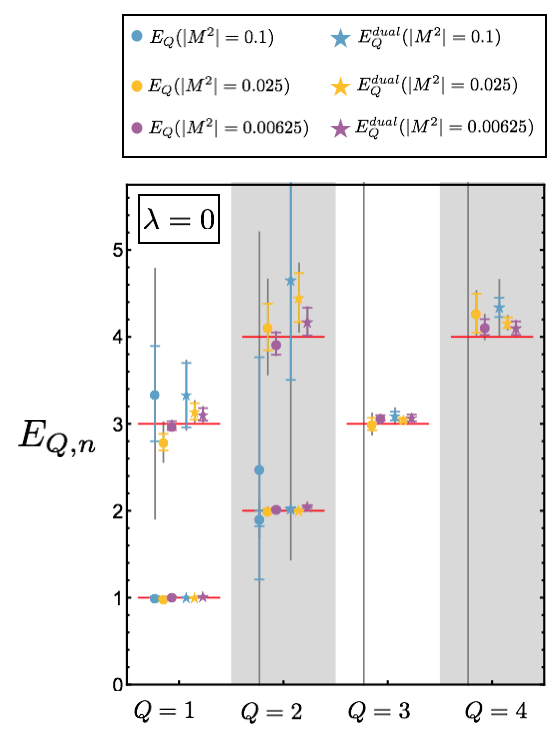}
\end{minipage}
\caption{In the left plots, we compare free-field $Z_{0;1,0} = \braket{|\varphi|^2}$ estimated using both the standard and phase-integrated ensembles versus the analytical value given in Eq.~\eqref{eq:lattpole}. The three main ensembles $A_0$, $B_0$, and $C_0$, agree with the analytical prediction to the percent level. The auxiliary $D_0$ and $E_0$ ensembles agree to the few percent level. In the right plot, GEVP methods are used to determine the lowest six energy levels in the spectrum of the free theory, rescaled into physical units. The coarsest ensemble $A_0$ exhibits large statistical and systematic uncertainties in fitting, and for $Q=3,4$ no plateau could be reliably fit (indicated by vertical gray lines). Where reliable estimates are possible, the data agree with analytical predictions.}
\label{fig:cho-phisq}
\end{figure}

Ensembles are generated via Metropolis sweeps over the sites in a red-black alternating pattern for efficient execution,
with $N_{\text{skip}}/2$ odd and $N_{\text{skip}}/2$ even updates between each measurement.
Both the standard $1D$ complex scalar field action defined in Eq.~\eqref{eq:actiondecomp} with the potential Eq.~\eqref{eq:phi4} and the analytically phase-integrated dual form of the theory given in Eq.~\eqref{eq:Pprimedef}-\eqref{eq:GQ2Pdual} are used to perform MC calculations using identical values of the parameters $M^2$, $L$, and $\lambda$ given in Table.~\ref{tbl:cho-ensembles}.
The phase unwrapping techniques based on smoothed numerical integration of the wrapped phase described above are applied to all correlation functions generated using the standard complex scalar action.
The cumulant expansion is then used to estimate correlation functions from sample moments of the corresponding unwrapped phases and log-magnitudes, and a generalized eigenvalue problem (GEVP) is solved to numerically extract the low-lying spectrum of the theory from the resulting correlation function estimates~\cite{Luscher:1990ck}.

We perform some checks for ensemble consistency. Eq.~\eqref{eq:lattpole} gives the noninteracting expectation for $Z_{1;0,1} = \braket{|\varphi|^2}$. We can reliably estimate this overlap on our noninteracting lattices and compare against the theoretically predicted value. The left plot of Figure~\ref{fig:cho-phisq} compares the standard ensemble and dual variable ensemble estimates versus the theoretical prediction, finding agreement to the percent level for the three main ensembles $A_0$, $B_0$, and $C_0$, while the auxiliary ensembles (used only for investigation of lattice spacing effects) match the prediction at the few percent level. Eqs.~\eqref{eq:GQP} and \eqref{eq:GQPspec} describe the noninteracting spectrum in terms of $Q=0$ ground state energy $E = E_0 = 2 \text{arcsinh} \paren{M/2}$. There are six low-lying states (energy $E_i \leq 4E_0$), with two states in each of the $Q=0$ and $Q=1$ channels, and one state in each of the $Q=2$ and $Q=3$ channels. Figure~\ref{fig:cho-phisq} further demonstrates that our free-field ensembles correctly reproduce this low-lying spectrum to within statistical and systematic fitting errors.

\begin{figure}
\includegraphics[height=5cm]{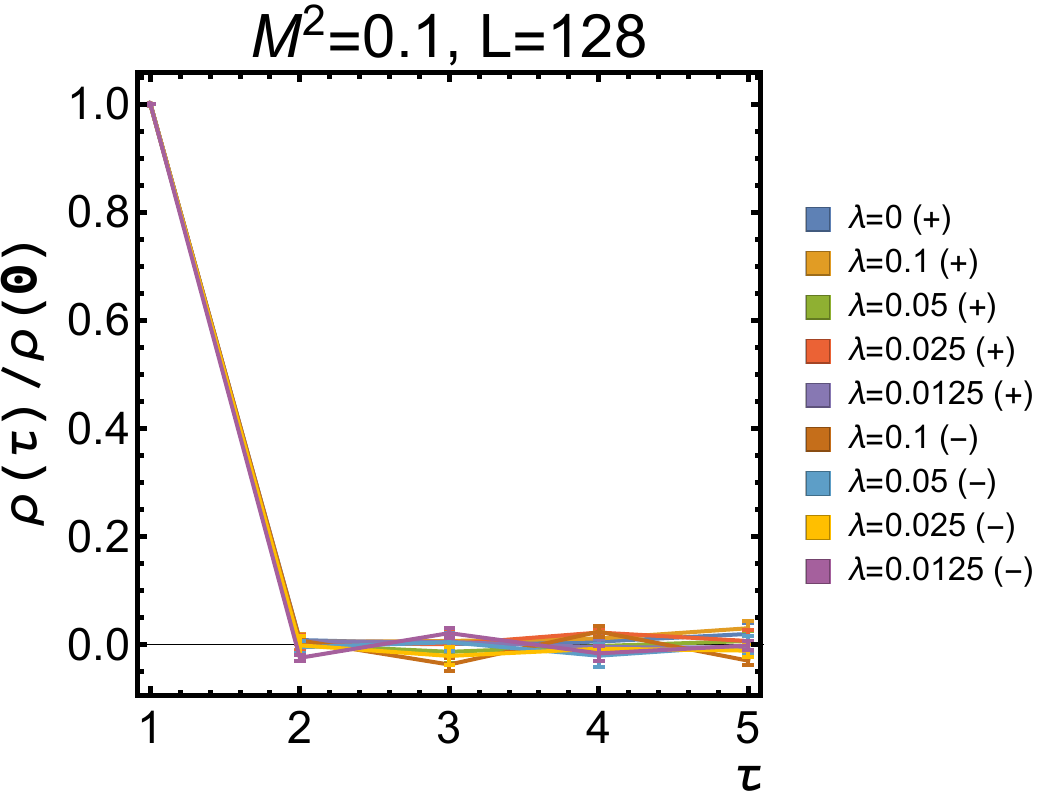}
\includegraphics[height=5cm]{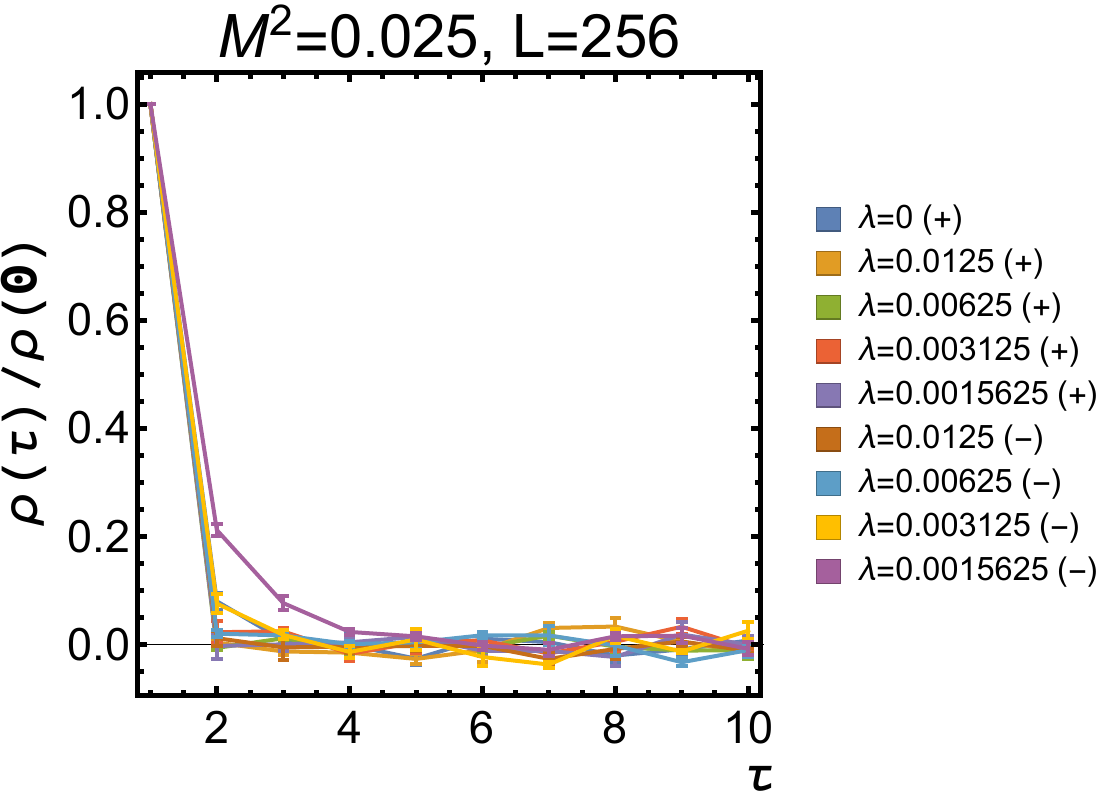}
\includegraphics[height=5cm]{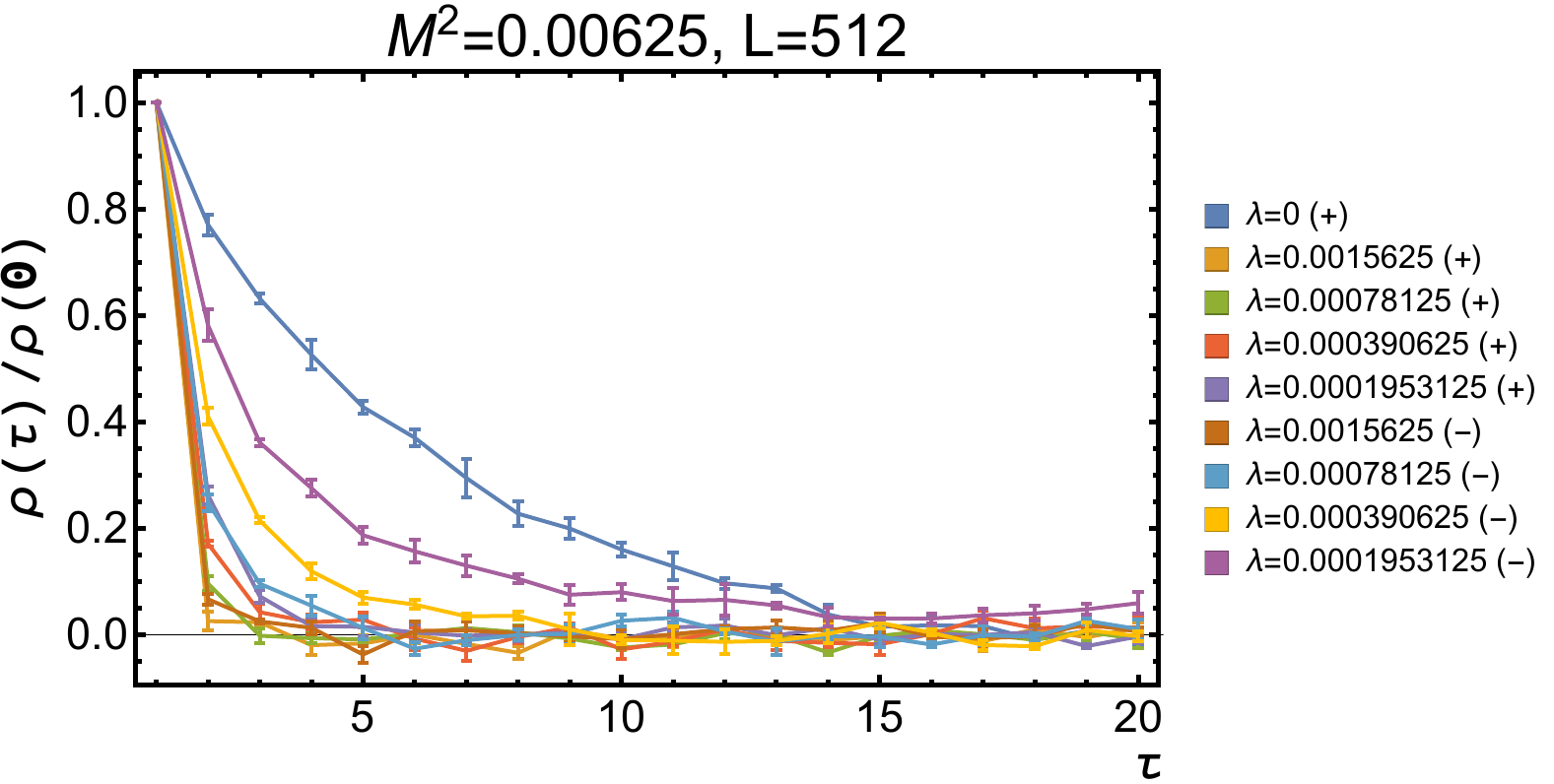}
\caption{Integrated autocorrelation times $\rho(\tau)/\rho(0)$ versus MCMC stream separation $\tau$ for the $\braket{|\phi|^2}$ observables on all standard ensembles. There is significant autocorrelation up to roughly $\author \approx 10$ for the finest lattice ($M^2 = 0.00625$), and significantly less autocorrelation on the two coarser lattices.}
\label{fig:cho-standard-autocorrs}
\vspace{0.5cm}
\includegraphics[height=5cm]{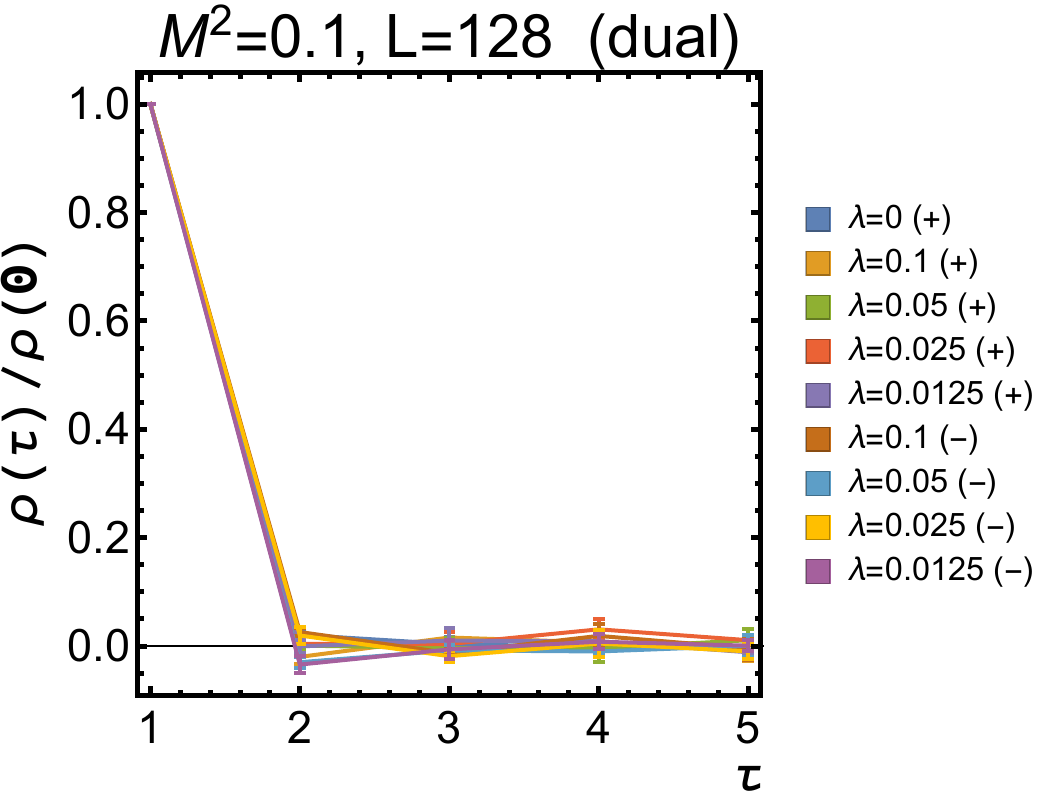}
\includegraphics[height=5cm]{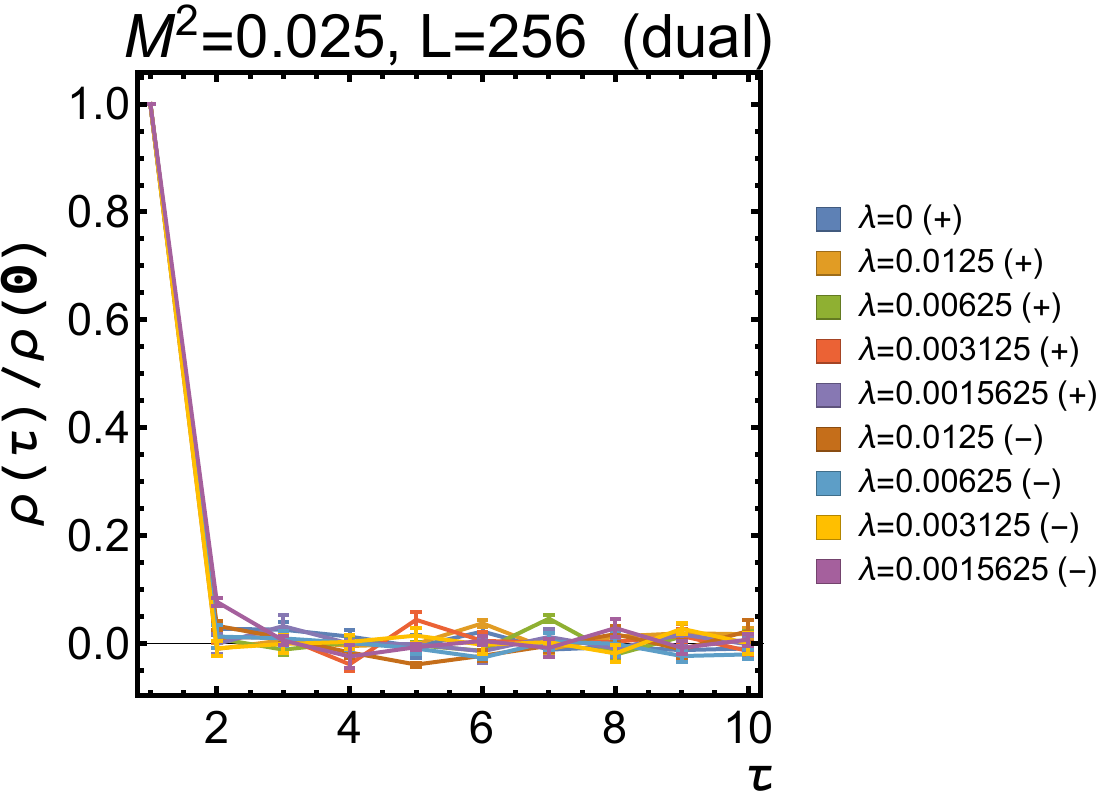}
\includegraphics[height=5cm]{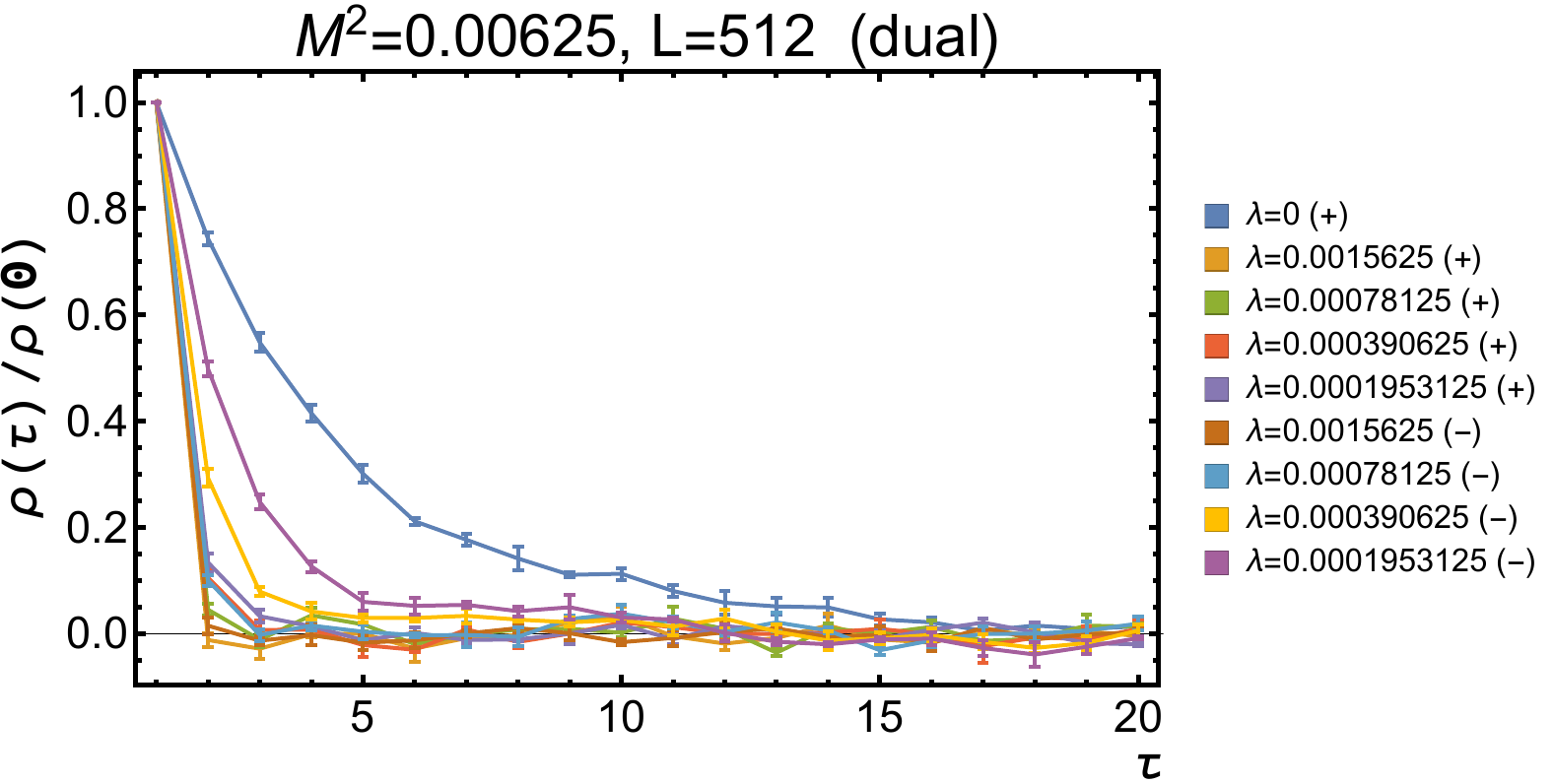}
\caption{Integrated autocorrelation times $\rho(\tau)/\rho(0)$ versus MCMC stream separation $\tau$ for the $\braket{|\phi|^2}$ observables on all dual variable ensembles. Similarly to the standard ensembles, there is significant autocorrelation up to roughly $\author \approx 10$ for the finest lattice ($M^2 = 0.00625$), and significantly less autocorrelation on the two coarser lattices.}
\label{fig:cho-pi-autocorrs}
\end{figure}

Figure~\ref{fig:cho-standard-autocorrs} plots the integrated
autocorrelation time for $\ob = \braket{|\varphi|^2}$ for all
standard ensembles used in this work.
Figure~\ref{fig:cho-pi-autocorrs} analogously plots the 
integrated autocorrelation time for the dual variable ensembles. 
Autocorrelation times between corresponding standard and dual
ensembles are similar. There is significant autocorrelation only
on the finest lattice ($M^2 = 0.00625$), and as such all analyses
applied binning with bin size $\geq 10$ on the finest lattice to
more accurately estimate errors in the presence of this
autocorrelation. Larger bin sizes were used in some cases to
further improve the $\chi^2$/DoF estimates.
Tables~\ref{tbl:free-fit-data}-\ref{tbl:l16-neg-fit-data} 
indicate the bin and  window sizes used for all fits to energy
levels required to produce the various spectrum plots in
this work.

\begin{table}
\begin{tabular}{c c c c c c c c c}
Ensemble & Eigenvalue & Central value & Error (stat.) & Error (syst.) & $\chi^2$/DoF & Window & Bin Size \\
\hline
\hline
$A_0$ & $Q=1$ & 0.317 & 0.005 & 0.001 & 0.466 & [1,8] & 1\\
	& $Q=1^*$ & 0.972 & 0.159 & 0.329 & 1.289 & [1,2] & \\
    & $Q=2$ & 0.603 & 0.026 & 0.066 & 1.605 & [1,3] & \\
    & $Q=2^*$ & 0.854 & 0.494 & 0.812 & 0.696 & [1,2] & \\
    & $Q=3$ &--&--&--&--&--& \\
    & $Q=4$ &--&--&--&--&--& \\
$B_0$ & $Q=1$ & 0.157 & 0.001 & 0.000 & 0.333 & [1,16] & 1\\
	& $Q=1^*$ & 0.444 & 0.013 & 0.024 & 0.396 & [1,8] & \\
    & $Q=2$ & 0.316 & 0.004 & 0.003 & 0.459 & [1,8] & \\
    & $Q=2^*$ & 0.662 & 0.037 & 0.060 & 0.301 & [1,5] & \\
    & $Q=3$ & 0.476 & 0.011 & 0.022 & 1.571 & [1,8] & \\
    & $Q=4$ & 0.664 & 0.036 & 0.012 & 1.003 & [1,3] & \\
$C_0$ & $Q=1$ & 0.080 & 0.000 & 0.000 & 2.117 & [1,15] & 1\\
	& $Q=1^*$ & 0.237 & 0.003 & 0.003 & 1.989 & [1,6] & \\
    & $Q=2$ & 0.160 & 0.001 & 0.000 & 0.491 & [1,15] & \\
    & $Q=2^*$ & 0.315 & 0.011 & 0.006 & 1.376 & [1,6] & \\
    & $Q=3$ & 0.241 & 0.002 & 0.002 & 1.783 & [1,4] & \\
    & $Q=4$ & 0.326 & 0.007 & 0.006 & 2.093 & [1,2] & \\
\hline
\end{tabular}
\vspace{1cm}

\begin{tabular}{c c c c c c c c c}
Ensemble & Eigenvalue & Central value & Error (stat.) & Error (syst.) & $\chi^2$/DoF & Window & Bin Size \\
\hline
\hline
$A_0^\bigstar$ & $Q=1$ & 0.316 & 0.001 & 0.000 & 0.515 & [1,6] & 5\\
	& $Q=1^*$ & 1.051 & 0.124 & 0.048 & 0.946 & [1,2] & \\
    & $Q=2$ & 0.636 & 0.005 & 0.04 & 0.795 & [1,6] & \\
    & $Q=2^*$ & 1.505 & 0.360 & 0.900 & 1.056 & [1,2] & \\
    & $Q=3$ & 0.974 & 0.016 & 0.020 & 0.716 & [1,4] & \\
    & $Q=4$ & 1.375 & 0.037 & 0.094 & 0.941 & [1,4] & \\
$B_0^\bigstar$ & $Q=1$ & 0.159 & 0.001 & 0.000 & 0.736 & [1,10] & 10\\
	& $Q=1^*$ & 0.495 & 0.016 & 0.010 & 0.343 & [1,4] & \\
    & $Q=2$ & 0.316 & 0.002 & 0.001 & 0.480 & [1,10] & \\
    & $Q=2^*$ & 0.711 & 0.043 & 0.052 & 0.213 & [1,3] & \\
    & $Q=3$ & 0.481 & 0.005 & 0.003 & 0.948 & [1,5] & \\
    & $Q=4$ & 0.656 & 0.010 & 0.008 & 0.711 & [1,5] & \\
$C_0^\bigstar$ & $Q=1$ & 0.080 & 0.000 & 0.000 & 1.493 & [1,15] & 20\\
	& $Q=1^*$ & 0.246 & 0.006 & 0.004 & 0.503 & [1,6] & \\
    & $Q=2$ & 0.162 & 0.001 & 0.000 & 0.963 & [1,15] & \\
    & $Q=2^*$ & 0.330 & 0.013 & 0.005 & 0.290 & [1,6] & \\
    & $Q=3$ & 0.242 & 0.003 & 0.001 & 0.406 & [1,4] & \\
    & $Q=4$ & 0.324 & 0.006 & 0.003 & 0.212 & [1,4] & \\
\hline
\end{tabular}
\caption{Free spectrum fitting information. Charge eigenvalues decorated with an asterisk indicate first excitations.
  Ensembles reported with a $\bigstar$ symbol indicate the dual-variable method, while those without indicate the standard method.}
\label{tbl:free-fit-data}
\end{table}

Systematic errors presented in the fit tables indicate the
variation in central value as the fit window is offset by up to
two lattice points. Systematic errors for the phase unwrapping
technique additionally include variation in the central value
as higher-order cumulants are included. In this work, these
errors include variation up to cumulant order 6 (the second
subleading order in phase variations), as higher-order
cumulants include too much noise to be reliably estimated.

\begin{table}
\begin{tabular}{c c c c c c c c c}
Ensemble & Eigenvalue & Central value & Error (stat.) & Error (syst.) & $\chi^2$/DoF & Window & Bin Size \\
\hline
\hline
$A_2^+$ & $Q=1$ & 0.430 & 0.009 & 0.016 & 0.480 & [1,8] & 1\\
    & $Q=2$ & 0.848 & 0.102 & 0.412 & 0.469 & [1,3] & \\
    & $Q=3$ &--&--&--&--&--& \\
    & $Q=4$ &--&--&--&--&--& \\
$B_2^+$ & $Q=1$ & 0.222 & 0.002 & 0.001 & 0.689 & [1,8] & 1\\
    & $Q=2$ & 0.457 & 0.007 & 0.017 & 1.420 & [1,5] & \\
    & $Q=3$ & 0.719 & 0.031 & 0.045 & 0.550 & [1,3] & \\
    & $Q=4$ & 0.912 & 0.127 & 0.467 & 0.699 & [1,3] & \\
$C_2^+$ & $Q=1$ & 0.113 & 0.000 & 0.000 & 1.354 & [1,16] & 1\\
    & $Q=2$ & 0.238 & 0.002 & 0.001 & 0.878 & [1,10] & \\
    & $Q=3$ & 0.373 & 0.005 & 0.006 & 0.536 & [1,5] & \\
    & $Q=4$ & 0.504 & 0.013 & 0.015 & 0.664 & [1,5] & \\
\hline
\end{tabular}
\vspace{1cm}

\begin{tabular}{c c c c c c c c c}
Ensemble & Eigenvalue & Central value & Error (stat.) & Error (syst.) & $\chi^2$/DoF & Window & Bin Size \\
\hline
\hline
$A_2^{+\bigstar} $ & $Q=1$ & 0.442 & 0.002 & 0.001 & 0.523 & [1,8] & 5\\
    & $Q=2$ & 0.916 & 0.008 & 0.003 & 1.464 & [1,4] & \\
    & $Q=3$ & 1.373 & 0.027 & 0.028 & 1.081 & [1,2] & \\
    & $Q=4$ & 1.803 & 0.070 & 0.092 & 0.716 & [1,2] & \\
$B_2^{+\bigstar} $ & $Q=1$ & 0.225 & 0.001 & 0.000 & 1.804 & [1,15] & 10\\
    & $Q=2$ & 0.470 & 0.002 & 0.001 & 1.893 & [1,10] & \\
    & $Q=3$ & 0.735 & 0.006 & 0.001 & 1.336 & [1,8] & \\
    & $Q=4$ & 1.033 & 0.013 & 0.007 & 2.003 & [1,8] & \\
$C_2^{+\bigstar} $ & $Q=1$ & 0.113 & 0.000 & 0.000 & 1.306 & [1,24] & 20\\
    & $Q=2$ & 0.237 & 0.001 & 0.000 & 1.148 & [1,16] & \\
    & $Q=3$ & 0.370 & 0.002 & 0.001 & 1.613 & [1,10] & \\
    & $Q=4$ & 0.511 & 0.004 & 0.002 & 1.091 & [1,6] & \\
\hline
\end{tabular}
\vspace{1cm}

\begin{tabular}{c c c c c c c c c}
Ensemble & Eigenvalue & Central value & Error (stat.) & Error (syst.) & $\chi^2$/DoF & Window & Bin Size \\
\hline
\hline
$A_2^{+\blacklozenge} $
	& $Q=1$ & 0.493 & 0.003 & 0.016 & 1.226 & [8,32] & 10\\
    & $Q=2$ & 1.172 & 0.007 & 0.531 & 0.737 & [8,24] & \\
    & $Q=3$ & 1.438 & 0.011 & 0.514 & 1.447 & [8,24] & \\
    & $Q=4$ & 1.536 & 0.011 & 1.514 & 1.574 & [8,24] & \\
$B_2^{+\blacklozenge} $
	& $Q=1$ & 0.249 & 0.001 & 0.027 & 1.922 & [16,64] & 10\\
    & $Q=2$ & 0.571 & 0.003 & 0.032 & 1.794 & [16,48] & \\
    & $Q=3$ & 0.671 & 0.004 & 0.753 & 1.482 & [16,48] & \\
    & $Q=4$ & 0.721 & 0.004 & 0.401 & 0.979 & [16,48] & \\
$C_2^{+\blacklozenge} $
	& $Q=1$ & 0.108 & 0.001 & 0.011 & 1.185 & [32,64] & 10\\
    & $Q=2$ & 0.274 & 0.001 & 0.046 & 1.179 & [32,64] & \\
    & $Q=3$ & 0.414 & 0.002 & 0.282 & 1.245 & [32,64] & \\
    & $Q=4$ & 0.434 & 0.002 & 0.103 & 1.362 & [32,64] & \\
\hline
\end{tabular}

\caption{Interacting spectrum fitting information for $\lambda L / M^2 = +32$. Ensembles reported with a $\bigstar$ symbol indicate the dual-variable method, ensembles reported with a $\blacklozenge$ symbol indicate the phase unwrapping method, and those without any symbol indicate the standard method.}
\label{tbl:l32-pos-fit-data}
\end{table}
\clearpage

\begin{table}
\begin{tabular}{c c c c c c c c c}
Ensemble & Eigenvalue & Central value & Error (stat.) & Error (syst.) & $\chi^2$/DoF & Window & Bin Size \\
\hline
\hline
$A_2^-$
	& $Q=1$ & 0.248 & 0.003 & 0.002 & 0.702 & [1,8] & 1\\
    & $Q=2$ & 0.571 & 0.016 & 0.009 & 0.433 & [1,4] & \\
    & $Q=3$ & 0.696 & 0.089 & 0.429 & 2.164 & [1,4] & \\
    & $Q=4$ & 0.801 & 0.443 & 1.007 & 0.305 & [1,2] & \\
$B_2^-$
	& $Q=1$ & 0.124 & 0.001 & 0.001 & 1.458 & [1,8] & 1\\
    & $Q=2$ & 0.285 & 0.003 & 0.005 & 0.937 & [1,8] & \\
    & $Q=3$ & 0.469 & 0.009 & 0.006 & 1.066 & [1,8] & \\
    & $Q=4$ & 0.682 & 0.028 & 0.050 & 1.340 & [1,4] & \\
$C_2^-$
	& $Q=1$ & 0.062 & 0.000 & 0.000 & 1.194 & [1,16] & 1\\
    & $Q=2$ & 0.142 & 0.001 & 0.001 & 2.057 & [1,16] & \\
    & $Q=3$ & 0.235 & 0.002 & 0.001 & 1.294 & [1,16] & \\
    & $Q=4$ & 0.339 & 0.005 & 0.003 & 0.796 & [1,8] & \\
\hline
\end{tabular}
\vspace{1cm}

\begin{tabular}{c c c c c c c c c}
Ensemble & Eigenvalue & Central value & Error (stat.) & Error (syst.) & $\chi^2$/DoF & Window & Bin Size \\
\hline
\hline
$A_2^{-\bigstar} $
	& $Q=1$ & 0.242 & 0.001 & 0.000 & 1.412 & [1,8] & 5\\
    & $Q=2$ & 0.558 & 0.003 & 0.003 & 1.981 & [1,8] & \\
    & $Q=3$ & 0.930 & 0.006 & 0.008 & 1.395 & [1,8] & \\
    & $Q=4$ & 1.378 & 0.014 & 0.032 & 1.007 & [1,8] & \\
$B_2^{-\bigstar} $
	& $Q=1$ & 0.123 & 0.000 & 0.000 & 0.527 & [1,10] & 10\\
    & $Q=2$ & 0.283 & 0.001 & 0.001 & 0.886 & [1,8] & \\
    & $Q=3$ & 0.473 & 0.002 & 0.000 & 0.469 & [1,6] & \\
    & $Q=4$ & 0.683 & 0.005 & 0.004 & 1.737 & [1,4] & \\
$C_2^{-\bigstar} $
	& $Q=1$ & 0.062 & 0.000 & 0.000 & 1.417 & [1,24] & 20\\
    & $Q=2$ & 0.143 & 0.000 & 0.001 & 2.386 & [1,12] & \\
    & $Q=3$ & 0.238 & 0.001 & 0.001 & 2.638 & [1,8] & \\
    & $Q=4$ & 0.341 & 0.002 & 0.002 & 1.436 & [1,6] & \\
\hline
\end{tabular}
\vspace{1cm}

\begin{tabular}{c c c c c c c c c}
Ensemble & Eigenvalue & Central value & Error (stat.) & Error (syst.) & $\chi^2$/DoF & Window & Bin Size \\
\hline
\hline
$A_2^{-\blacklozenge} $
	& $Q=1$ & 0.250 & 0.002 & 0.010 & 1.096 & [8,24] & 10\\
    & $Q=2$ & 0.588 & 0.003 & 0.047 & 1.832 & [8,24] & \\
    & $Q=3$ & 1.211 & 0.009 & 0.033 & 1.352 & [8,24] & \\
    & $Q=4$ & 1.397 & 0.009 & 2.369 & 0.818 & [8,24] & \\
$B_2^{-\blacklozenge} $
	& $Q=1$ & 0.127 & 0.001 & 0.004 & 1.947 & [16,48] & 10\\
    & $Q=2$ & 0.289 & 0.002 & 0.036 & 1.930 & [16,48] & \\
    & $Q=3$ & 0.579 & 0.003 & 0.058 & 1.303 & [16,48] & \\
    & $Q=4$ & 0.665 & 0.004 & 0.122 & 1.171 & [16,48] & \\
$C_2^{-\blacklozenge} $
	& $Q=1$ & 0.061 & 0.000 & 0.003 & 2.091 & [32,48] & 10\\
    & $Q=2$ & 0.156 & 0.001 & 0.010 & 0.713 & [32,48] & \\
    & $Q=3$ & 0.239 & 0.002 & 0.079 & 0.806 & [32,48] & \\
    & $Q=4$ & 0.407 & 0.003 & 0.274 & 1.097 & [32,48] & \\
\hline
\end{tabular}

\caption{Interacting spectrum fitting information for $\lambda L / M^2 = -32$. Ensembles reported with a $\bigstar$ symbol indicate the dual-variable method, ensembles reported with a $\blacklozenge$ symbol indicate the phase unwrapping method, and those without any symbol indicate the standard method.}
\label{tbl:l32-neg-fit-data}
\end{table}
\clearpage

\begin{table}
\begin{tabular}{c c c c c c c c c}
Ensemble & Eigenvalue & Central value & Error (stat.) & Error (syst.) & $\chi^2$/DoF & Window & Bin Size \\
\hline
\hline
$A_1^+$
	& $Q=1$ & 0.394 & 0.006 & 0.011 & 0.511 & [1,8] & 1\\
    & $Q=2$ & 0.779 & 0.056 & 0.247 & 0.942 & [1,4] & \\
    & $Q=3$ & 0.885 & 0.451 & 1.289 & 0.497 & [1,2] & \\
    & $Q=4$ & -- & -- & -- & -- & -- & \\
$B_1^+$
	& $Q=1$ & 0.200 & 0.001 & 0.001 & 0.449 & [1,8] & 1\\
    & $Q=2$ & 0.418 & 0.005 & 0.006 & 0.692 & [1,8] & \\
    & $Q=3$ & 0.634 & 0.021 & 0.033 & 0.612 & [1,4] & \\
    & $Q=4$ & 0.731 & 0.065 & 0.367 & 1.944 & [1,4] & \\
$C_1^+$
	& $Q=1$ & 0.099 & 0.000 & 0.000 & 1.083 & [1,16] & 1\\
    & $Q=2$ & 0.207 & 0.001 & 0.001 & 0.498 & [1,16] & \\
    & $Q=3$ & 0.324 & 0.003 & 0.006 & 1.807 & [1,16] & \\
    & $Q=4$ & 0.451 & 0.010 & 0.013 & 1.223 & [1,8] & \\
\hline
\end{tabular}
\vspace{1cm}

\begin{tabular}{c c c c c c c c c}
Ensemble & Eigenvalue & Central value & Error (stat.) & Error (syst.) & $\chi^2$/DoF & Window & Bin Size \\
\hline
\hline
$A_1^{+\bigstar} $
	& $Q=1$ & 0.398 & 0.002 & 0.001 & 0.883 & [1,8] & 5\\
    & $Q=2$ & 0.824 & 0.006 & 0.004 & 2.060 & [1,8] & \\
    & $Q=3$ & 1.272 & 0.020 & 0.015 & 0.377 & [1,4] & \\
    & $Q=4$ & 1.768 & 0.055 & 0.068 & 1.465 & [1,4] & \\
$B_1^{+\bigstar} $
	& $Q=1$ & 0.198 & 0.001 & 0.000 & 1.177 & [1,10] & 10\\
    & $Q=2$ & 0.410 & 0.002 & 0.000 & 1.679 & [1,8] & \\
    & $Q=3$ & 0.632 & 0.005 & 0.002 & 0.326 & [1,6] & \\
    & $Q=4$ & 0.876 & 0.011 & 0.005 & 0.540 & [1,4] & \\
$C_1^{+\bigstar} $
	& $Q=1$ & 0.100 & 0.000 & 0.000 & 1.069 & [1,24] & 20\\
    & $Q=2$ & 0.209 & 0.001 & 0.001 & 1.906 & [1,12] & \\
    & $Q=3$ & 0.322 & 0.002 & 0.001 & 0.717 & [1,8] & \\
    & $Q=4$ & 0.443 & 0.003 & 0.003 & 0.746 & [1,6] & \\
\hline
\end{tabular}
\vspace{1cm}

\begin{tabular}{c c c c c c c c c}
Ensemble & Eigenvalue & Central value & Error (stat.) & Error (syst.) & $\chi^2$/DoF & Window & Bin Size \\
\hline
\hline
$A_1^{+\blacklozenge} $
	& $Q=1$ & 0.460 & 0.003 & 0.077 & 1.890 & [8,32] & 10\\
    & $Q=2$ & 1.111 & 0.007 & 1.083 & 1.825 & [8,32] & \\
    & $Q=3$ & 1.408 & 0.009 & 1.448 & 1.125 & [8,32] & \\
    & $Q=4$ & 1.549 & 0.009 & 0.483 & 0.753 & [8,32] & \\
$B_1^{+\blacklozenge} $
	& $Q=1$ & 0.206 & 0.001 & 0.010 & 2.013 & [16,48] & 10\\
    & $Q=2$ & 0.403 & 0.003 & 0.034 & 1.306 & [16,48] & \\
    & $Q=3$ & 0.662 & 0.004 & 0.071 & 1.666 & [16,40] & \\
    & $Q=4$ & 1.345 & 0.007 & 4.357 & 1.047 & [16,40] & \\
$C_1^{+\blacklozenge} $
	& $Q=1$ & 0.103 & 0.001 & 0.008 & 1.602 & [32,64] & 10\\
    & $Q=2$ & 0.227 & 0.001 & 0.029 & 1.055 & [32,64] & \\
    & $Q=3$ & 0.403 & 0.003 & 0.326 & 2.010 & [32,64] & \\
    & $Q=4$ & 0.434 & 0.003 & 0.801 & 1.274 & [32,64] & \\
\hline
\end{tabular}

\caption{Interacting spectrum fitting information for $\lambda L / M^2 = +16$. Ensembles reported with a $\bigstar$ symbol indicate the dual-variable method, ensembles reported with a $\blacklozenge$ symbol indicate the phase unwrapping method, and those without any symbol indicate the standard method.}
\label{tbl:l16-pos-fit-data}
\end{table}
\clearpage

\begin{table}
\begin{tabular}{c c c c c c c c c}
Ensemble & Eigenvalue & Central value & Error (stat.) & Error (syst.) & $\chi^2$/DoF & Window & Bin Size \\
\hline
\hline
$A_1^-$
	& $Q=1$ & 0.141 & 0.001 & 0.002 & 1.919 & [1,8] & 1\\
    & $Q=2$ & 0.345 & 0.006 & 0.011 & 1.136 & [1,4] & \\
    & $Q=3$ & 0.586 & 0.023 & 0.064 & 1.423 & [1,4] & \\
    & $Q=4$ & 0.739 & 0.090 & 0.388 & 1.607 & [1,4] & \\
$B_1^-$
	& $Q=1$ & 0.072 & 0.000 & 0.001 & 1.664 & [1,8] & 1\\
    & $Q=2$ & 0.176 & 0.001 & 0.004 & 0.618 & [1,8] & \\
    & $Q=3$ & 0.301 & 0.004 & 0.006 & 0.468 & [1,8] & \\
    & $Q=4$ & 0.430 & 0.009 & 0.018 & 1.301 & [1,4] & \\
$C_1^-$
	& $Q=1$ & 0.037 & 0.000 & 0.000 & 2.926 & [1,10] & 1\\
    & $Q=2$ & 0.091 & 0.000 & 0.001 & 1.855 & [1,10] & \\
    & $Q=3$ & 0.157 & 0.001 & 0.002 & 1.284 & [1,8] & \\
    & $Q=4$ & 0.231 & 0.002 & 0.003 & 0.248 & [1,6] & \\
\hline
\end{tabular}
\vspace{1cm}

\begin{tabular}{c c c c c c c c c}
Ensemble & Eigenvalue & Central value & Error (stat.) & Error (syst.) & $\chi^2$/DoF & Window & Bin Size \\
\hline
\hline
$A_1^{-\bigstar} $
	& $Q=1$ & 0.144 & 0.000 & 0.000 & 2.531 & [1,8] & 5\\
    & $Q=2$ & 0.354 & 0.001 & 0.002 & 1.506 & [1,8] & \\
    & $Q=3$ & 0.612 & 0.003 & 0.005 & 1.366 & [1,8] & \\
    & $Q=4$ & 0.907 & 0.007 & 0.014 & 1.733 & [1,8] & \\
$B_1^{-\bigstar} $
	& $Q=1$ & 0.073 & 0.000 & 0.000 & 1.561 & [1,10] & 10\\
    & $Q=2$ & 0.179 & 0.001 & 0.001 & 1.636 & [1,8] & \\
    & $Q=3$ & 0.307 & 0.001 & 0.002 & 1.610 & [1,6] & \\
    & $Q=4$ & 0.453 & 0.003 & 0.004 & 0.848 & [1,4] & \\
$C_1^{-\bigstar} $
	& $Q=1$ & 0.037 & 0.000 & 0.000 & 1.676 & [4,24] & 20\\
    & $Q=2$ & 0.091 & 0.000 & 0.000 & 2.067 & [4,12] & \\
    & $Q=3$ & 0.156 & 0.001 & 0.001 & 1.211 & [2,8] & \\
    & $Q=4$ & 0.231 & 0.001 & 0.001 & 0.810 & [2,6] & \\
\hline
\end{tabular}
\vspace{1cm}

\begin{tabular}{c c c c c c c c c}
Ensemble & Eigenvalue & Central value & Error (stat.) & Error (syst.) & $\chi^2$/DoF & Window & Bin Size \\
\hline
\hline
$A_1^{-\blacklozenge} $
	& $Q=1$ & 0.144 & 0.001 & 0.006 & 1.614 & [12,24] & 10\\
    & $Q=2$ & 0.363 & 0.003 & 0.027 & 1.967 & [12,24] & \\
    & $Q=3$ & 0.619 & 0.006 & 0.111 & 0.874 & [12,24] & \\
    & $Q=4$ & 1.195 & 0.008 & 0.355 & 1.857 & [12,24] & \\
$B_1^{-\blacklozenge} $
	& $Q=1$ & 0.075 & 0.001 & 0.001 & 1.412 & [24,48] & 10\\
    & $Q=2$ & 0.185 & 0.002 & 0.016 & 0.795 & [24,48] & \\
    & $Q=3$ & 0.303 & 0.003 & 0.111 & 1.254 & [24,48] & \\
    & $Q=4$ & 0.441 & 0.003 & 0.061 & 1.585 & [24,48] & \\
$C_1^{-\blacklozenge} $
	& $Q=1$ & 0.036 & 0.000 & 0.005 & 2.258 & [40,70] & 10\\
    & $Q=2$ & 0.098 & 0.001 & 0.021 & 1.632 & [40,70] & \\
    & $Q=3$ & 0.168 & 0.001 & 0.059 & 2.488 & [40,70] & \\
    & $Q=4$ & 0.243 & 0.002 & 0.039 & 1.993 & [40,70] & \\
\hline
\end{tabular}

\caption{Interacting spectrum fitting information for $\lambda L / M^2 = -16$. Ensembles reported with a $\bigstar$ symbol indicate the dual-variable method, ensembles reported with a $\blacklozenge$ symbol indicate the phase unwrapping method, and those without any symbol indicate the standard method.}
\label{tbl:l16-neg-fit-data}
\end{table}
\clearpage

\vfill
\pagebreak

\end{document}